\documentclass[final,5p,sort&compress,twocolumn,floatfix]{elsarticle}

\usepackage{amsfonts,amsmath}
\usepackage{esvect}

\journal{Journal of Magnetism and Magnetic Materials}

\begin{document}

\title{Thermal effects in spintronic materials and devices: an experimentalist's guide}

\begin{frontmatter}

\author{B. L. Zink}\ead{barry.zink@du.edu}
\affiliation{organization={Department of Physics and Astronomy}, addressline={University of Denver}, city={Denver}, state={CO}, postcode={80208}, country={USA}}

\begin{abstract}
Research in spintronics often involves generation of heat in nanoscale magnetic systems.  This heat generation can be intentional, as when studying effects created by an external applied temperature difference, or unintentional, coming as a consequence of driving relatively large charge currents through tiny structures.  Understanding and controlling these thermal gradients can present challenges to experimentalists, which are related at some level to the fact that heat flow is much more difficult to isolate and manipulate than charge flow.  This paper aims to provide a simple, intuitive framework to understand the fundamental issues that arise in spintronic materials and devices involving thermal gradients.  The first goal is to provide simple tools to demonstrate how thermal gradients arise in systems with thin conducting films on bulk substrates.  The main results are that a thermal gradient pointing perpendicular to the plane of a thin film supported on a macroscopic substrate is very common, even while the largest temperature drop in the system will exist across the bulk substrate itself. These results point to the need to understand the range of thermoelectric and magnetothermoelectric effects that can generate voltage signals and other responses to thermal gradients.  I provide a brief review of these, along with relevant spin effects.  The review concludes with examples and comments on several important ongoing issues in spintronics where thermal gradients play key roles.  
\end{abstract}


\end{frontmatter}

\section{Introduction} 

Spintronics seeks to manipulate and use the spin degree of freedom of electrons to add new functionality and fundamental properties to electronic systems.\cite{HirohataJMMM2020,HoffmannPRAp2015,BaderAnnRevCMP2010,ZuticRMP04,WolfScience01}  Achieving this control almost always uses tools of micro- or nanoscale science and engineering to create physical systems from thin films with one or more dimensions less than $100$ nm.  One of the earliest examples, and perhaps most notable to date, is the ``giant magnetoresistance" (GMR) effect, where spin-dependent scattering of electrons flowing through a thin film heterostructure of two decoupled ferromagnetic layers causes a significant field-dependent resistance.\cite{ParkinPRL1990,UngurisPRL1991,PotterPRB1994}  This fundamental physics leads to sensitive magnetic field sensors that drove advances in magnetic hard disk information storage technologies that enabled the information age.\cite{NakataniMRSBull2018,ChappertNatMater2007,ParkinIEEE2003,MoserJPhysD2002}   The reciprocal effect of GMR, where angular momentum conservation can cause the induced spin polarization of the electrons to transfer sufficient angular momentum to switch the magnetization direction of a ferromagnetic thin film element, called spin transfer torque (STT), opened dramatic new possibilities for spintronics, allowing control of information by application of large local currents instead of distant external fields.\cite{RalphJMMM2008,StilesPRB2002,BergerPRB1996,SlonczewskiJMMM96}  

The STT effect also highlights an important complication that arises in a wide range of ongoing studies in spintronics and related fields.  The fundamental physics of this effect requires a large current density to pass through tiny nanoscale structures.  The large current flow almost always brings large Joule heating through charge scattering, meaning that a complete understanding of spintronic devices often requires knowledge of heat flow in complex nanoscale structures, and the effects or artifacts this heat flow can introduce.  This significant heat flow is fundamentally much more difficult to control or manage than charge flow, since there is no such thing as a heat insulator.  To be more quantitative, where typical materials used in spintronic systems can easily have electrical conductivity, $\sigma$, that differ by much more than ten orders of magnitude, their thermal conductivity varies at most by a factor of $\sim1000$.    Historically, researchers in spintronics have a somewhat widely varied approach to these heating effects, ranging from ignoring them entirely to intentionally using them to open new functionalities.  The latter approach, using existing thermal gradients or intentionally created thermal gradients to manipulate the spin degree of freedom in a magnetic system, has grown into its own very active sub-field of spin caloritronics.\cite{BackJPhysD2019,BoonaEES2014,BauerNatMat2012,BauerSSC10}

The goal of this review is first to provide a simple and accessible motivation for the importance of thermal effects in spintronic materials and devices, focusing on model systems and calculations that demonstrate the important fundamentals that often arise.  The models demonstrating the fundamentals of thermal gradient generation and direction will point to the importance of understanding the ``zoo" of thermoelectric and magneto-thermoelectric effects, and their more recently appreciated spin counterparts.  I will therefore also provide a short review of these effects, providing some essential materials parameters for some key spintronic constituents and highlighting selected literature where these effects arise or are demonstrated.  This will include the Spin Seebeck Effect, Spin-Dependent Seebeck Effect and other important spin caloritronic demonstrations, and the Anomalous Nernst Effect.  I will conclude by highlighting some important examples and current areas of interest in spintronics, where thermal effects will likely always remain an important consideration.  This will include some aspects of Spin Orbit Torque (SOT) switching.   As indicated in the title, I write this from the point of view of an experimentalist with interest and expertise in developing measurement methods for thermal effects in thin films and nanostructures, and in applications to fundamental materials physics of magnetic systems.  I hope that the resulting guide is useful for students or more senior researchers new to the field, and especially to those wishing to develop a simple framework to better understand how heat flow in nanoscale structures can impact their own work.

\section{Generation of Thermal Gradients (Intentional and Unintentional)}

To begin, consider a large current density, $J=I/A$, where $A$ is the cross-sectional area perpendicular to the electron flow, and $I$ is the charge current applied to a wire patterned from a metallic thin film, with electrical resistivity $\rho$, deposited on an electrically insulating substrate.  Such structures are common in spintronic systems and devices, and often have thickness, $t$, ranging from a few to a few hundred nanometers, and width, $w$ from about $100$ nanometers to dozens of microns.  Electron flow through such wires generates Joule heat, with dissipated power $P_{\mathrm{j}}=I^{2}R$, with resistance $R$ determined by the material composition and geometry of the wire, $R=\rho \ell/w t$, where $\ell$ is the length of the current path.  The heated volume is then $V=\ell w t$. We can easily relate $P_{\mathrm{j}}$ directly to current density, such that 
\begin{equation}
P_{\mathrm{j}}=I^{2}\left( \frac{\rho \ell}{w t}\right)=\frac{I^2}{(w t)^{2}}\left(\ell w t\right) \rho \leadsto \frac{P_{\mathrm{j}}}{V}= J^{2}\rho
\end{equation} .
It is fairly common in experiments such as STT and SOT switching to reach current density on order of $J\sim10^{11}\ \mathrm{A/m}^2$.  For our simple wire this could represent a current $I=1\ \mathrm{mA}$ applied to a $100\ \mathrm{nm}$ thick, $100\ \mathrm{nm}$ wide wire.  Depending on the material used for the wire, $\rho$ can easily be fairly large.  If we assume $\rho=30\ \mathrm{\mu \Omega\ cm}$ (in the range of typical ferromagnetic metal alloys), then the resulting power density in the wire that results from Joule heating is $>10^{15}\ \mathrm{W/m}^{3}$.  This power density exceeds that in the core of a nuclear reactor by many orders of magnitude. \cite{ReactorNote}
Here we are mostly concerned with the temperature gradients that arise from this large energy density, which as we will see, depend strongly on how this energy escapes the wire in the form of heat flow.  But even before developing a picture of this heat flow, the huge energy density seen here should warn us that very significant thermal gradients will often be present in spintronic devices.  

\begin{figure}
\includegraphics[width=3.38in]{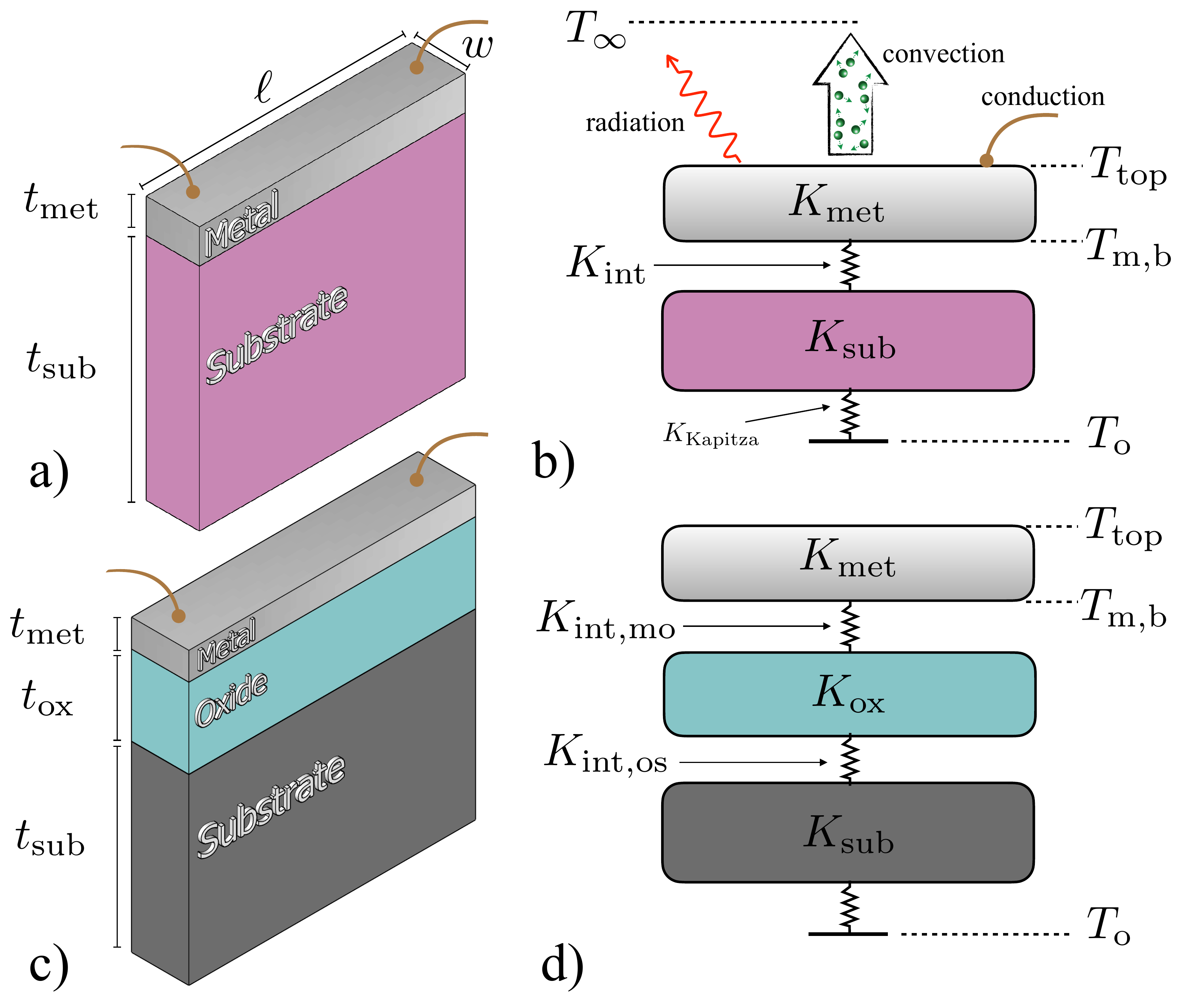}
\caption{\label{Fig1} Schematic views and simple thermal models for two situations common in spintronic devices. \textbf{a)} A slice of an insulating substrate with a deposited metal film, with thicknesses, $t_{\mathrm{met}}$ and $t_{\mathrm{sub}}$, and length $\ell$ and width $w$ indicated.  Electrical connections are shown schematically in brown. \textbf{b)}  Corresponding thermal model, showing contributions to heat transport from radiation, convection, conduction (through experimental wiring), and the sample heterostructure.  \textbf{c)} A similar schematic slice of an insulating substrate for the case of a metal film on top of a thin oxide layer, on top of a bulk substrate.  \textbf{d)} Corresponding thermal model with the additional conductances introduced by the oxide (heat loss from the top of the film is not shown for simplicity).   }
\end{figure}

We can gain a bit more understanding of the nature of these thermal gradients using very simple analytic models of the steady-state heat current that flows in various thin films or heterostructures supported by bulk substrates.  As shown in Fig.\ \ref{Fig1}, we will consider the temperature profile in a slice of two common structures.  Fig.\ \ref{Fig1}a) shows a schematic view of a slice, with width $w$ and length $\ell$ through a thick substrate, with thickness $t_{\mathrm{sub}}$, with a very thin metallic film, with thickness $t_{\mathrm{met}}$, deposited on top.   Since electrical connections to this film are often important to generate or measure the desired signal, and since we will see these often play an important role in the resulting thermal profile, we also schematically indicate the electrical connections to the top of the metal film.  For this structure, we can write the corresponding ``lumped element" thermal model seen in Fig.\ \ref{Fig1}b).  We assume this substrate will be mounted in an experimental platform with a constant base temperature, $T_{\mathrm{o}}$, and held in a surrounding environment at temperature $T_{\infty}$.  The top surface of the metal film can exchange heat energy with the surroundings potentially via all three of the typical mechanisms: radiation, convection, and conduction.  Radiation, the exchange of energy with the distant environment via thermally excited (blackbody) photons, is described by the Stefan-Boltzmann law, $P_{\mathrm{rad}}\propto A_{\mathrm{eff}}\sigma \epsilon T^{4}$, where $A_{\mathrm{eff}}$ is the effective area of the radiating surface, $\sigma$ is the Stefan-Boltzmann constant, and $\epsilon$ is the emissivity of the surface.  Despite the strong $T$-dependence, in many situations of interest this contribution is small and can often be ignored. However, some care must be taken when a surface has components with very different emissivities, especially since these can be poorly known for materials of interest.\cite{DaimonPRB2017}  If the sample is held in atmosphere or other gaseous environment, the heated film surface can also exchange energy via convection and conduction through this gas.  For the case of atmosphere surrounding a spintronic device, this contribution is often significant.  Finally, heat can flow away from the top surface of the film via conduction (transport of electrons and phonons) in the wires or probes used to make electrical connections to this film element.  This also can be a very significant contribution.  

The heated film can also exchange energy via conduction of heat downward toward the base temperature.  In this event the heat current flows through a series of thermal impedances $W$, or thermal conductances $K=1/W$.   The first such thermal conductance comes from the metal film itself, $K_{\mathrm{film}}$, related to the limitations of heat flow in a given material that we typically describe as the thermal conductivity of a material, $k$.  In analogy to the typical relation used to determine electrical resistance from electrical resistivity, we can write:
\begin{equation}
W=\frac{(1/k) \ell}{w t} \leadsto K=\frac{k A}{\ell},
\end{equation}
where the cross-sectional area perpendicular to heat flow is $A=w\cdot t$, and the length of the heat flow path is $\ell$.  In the case of the heat flowing through the metal film, the area is $w\cdot\ell$, and the path length $t_{\mathrm{met}}$ as indicated in Fig.\ \ref{Fig1}a), such that $K_{\mathrm{met}}=k_{\mathrm{met}} \ell w/t_{\mathrm{met}}$.  

The next thermal conductance comes from the interfacial thermal conductance between the metal and substrate.  The thermal boundary resistance is the reciprocal, and refers to the same physical picture.   One intuitive picture of this physical mechanism comes from imagining that phonons carrying heat across such an interface between dissimilar materials experience a change in their group velocity that is similar to the case for photons traveling across an interface between two materials with different index of refraction.  Interfacial thermal conductance is an important and active field of study in its own right, both from a theoretical (or numerical) and experimental point of view.  Most of this work is outside the scope of this paper, though I recommend those wanting to gain in-depth understanding of their own particular thermal profile in a spintronic device or material engage with excellent general reviews on nanoscale heat flow,\cite{CahillJAP2003,CahillAPR2014} and more specific reviews of interfacial thermal conductance.\cite{ChenRMP2022,MonachonARMR2016}  A recent paper\cite{AngelesPRM2021} has also experimentally investigated many important spintronic systems to determine the interfacial thermal conductance using the now fairly common approach of ultrafast time domain thermoreflectance (TDTR), which provides crucial information to improve understanding of the thermal gradients in these systems.  The interfacial thermal conductance, $G_{\mathrm{int}}$ is typically reported in units of $\mathrm{W/m^{2}}$, so to include in the thermal models that follow, we calculate $K_{\mathrm{int}}=G_{\mathrm{int}}A$, where $A$ is again cross-sectional area.   
Though in the few specific situations and geometries I calculate below, the interfacial effects do not dominate, I have included these both for physical accuracy regarding the nature of the heat flow, and for ease of adapting the simple analytical picture to situations where interfaces could play a more important role.  

The next thermal conductance comes from what is nearly always, and by many orders of magnitude, the largest component of the system, the bulk substrate that supports the thin film structure.  Before continuing, I point out that Figs.\ \ref{Fig1}a) and c), and any similar cartoons, \emph{dramatically} misrepresent the actual ratio between the thickness of a typical bulk substrate and typical thin film.  It is common for the substrate to be on order $10,000\times$ thicker than the film, such that a truly correct picture of the substrate thickness for the films represented in Fig.\ \ref{Fig1} by boxes that are perhaps $3\ \mathrm{mm}$ on the rendered page would be $\sim 30\ \mathrm{m}$ tall!  Keeping this simple fact in mind will be helpful as we move forward to find that, in almost any case imaginable, when a film is supported on a substrate, the substrate dominates the heat flow problem.  As with the film, $K_{\mathrm{sub}}$ is related to the substrate thermal conductivity, $k_{\mathrm{sub}}$, and the geometry, such that $K_{\mathrm{sub}}=k_{\mathrm{sub}}\ell w/t_{\mathrm{sub}}$.  Dividing by the much larger $t_{\mathrm{sub}}$ means that $K_{\mathrm{sub}}$ will almost always be the smallest thermal conductance, or largest thermal impedance, in the problem.  

The final thermal conductance shown in Fig.\ \ref{Fig1}b) represents a possible thermal impedance between the bottom of the substrate and the experimental platform.  Especially in low temperature physics, this impedance is typically called a Kapitza resistance, in analogy to the similar physical phenomenon originally observed between a liquid helium bath and a bulk object.\cite{PollackRMP1969}  This thermal impedance could be included in calculations to model any expected or encountered poor thermal link between the back of a sample and its heat sink.  In the few calculations I present below, I will typically assume this conductance is high enough not to contribute.  

Fig.\ \ref{Fig1}c) and d) present a similar schematic and model for the case where the metal film is deposited on a thin film of oxide, which was in-turn deposited on or grown from a bulk substrate.  This adds two thermal conductances, that arising from the interface between the metal and the oxide, $K_{\mathrm{int,mo}}$ and that from the bulk heat flow through the oxide $K_{\mathrm{ox}}= k_{\mathrm{ox}}\ell w/t_{\mathrm{ox}}$ (we also modify the name of interface contribution between oxide and substrate, accordingly, to $K_{\mathrm{int,os}}$).  This situation will represent a wide range of studies where thin metal films are patterned on insulating oxide or nitride layers on a silicon substrate.

\subsection{Simple Analytic Models: Heat flowing from back of substrate\label{BackHeat}}

We first consider a simple model of heat flow through the structures shown in Fig.\ \ref{Fig1} where the back of the substrate, $T_{\mathrm{o}}$, is kept at a temperature somewhat higher than the temperature of the environment $T_{\infty}$.  This could be the case for a range of experiments where a film on a substrate is heated from below, including those where the intention is to create a thermal gradient pointing in the plane of the substrate.  First we will investigate the situation when heat loss from the film surface occurs only from convection.  The power flowing out of the film, upwards from the film surface to the surrounding air can be reasonably approximated by:
\begin{equation}
P_{\mathrm{conv}}=h_{\mathrm{c}}A \Delta T = h_{\mathrm{c}}A (T_{\mathrm{top}}-T_{\infty}),
\end{equation}
where $h_{\mathrm{c}}$ is the convection heat transfer coefficient, $h_{\mathrm{c}}\cong25\ \mathrm{W/m}^{2}\mathrm{K}$ for slow moving air, which we assume to be in equilibrium with the sample environment at $T_{\infty}$.   
Power also flows into the metal film, toward the top surface via conduction from the heated substrate below.  This power flows through the entire heterostructure, and can be written:
\begin{equation}
P_{\mathrm{het}}=K_{\mathrm{het}}(T_{\mathrm{o}}-T_{\mathrm{top}}),
\end{equation}
where $K_{\mathrm{het}}=1/W_{\mathrm{het}}$ is the thermal conductance of the entire sample stack, with:
\begin{equation}
K_{\mathrm{het}}=\frac{1}{W_{\mathrm{het}}}=\frac{1}{W_{\mathrm{sub}}+W_{\mathrm{int}}+W_{\mathrm{met}}}
\label{KhetA}
\end{equation}
for the heterostructure of Fig.\ \ref{Fig1}a), and 
\begin{equation}
K_{\mathrm{het}}=\frac{1}{W_{\mathrm{sub}}+W_{\mathrm{int,so}}+W_{\mathrm{ox}}+W_{\mathrm{int,mo}}+W_{\mathrm{met}}}
\label{KhetB}
\end{equation} 
for the case with the additional oxide layer shown in Fig.\ \ref{Fig1}b).  
In steady-state, these incoming and outgoing powers must be equal:
\begin{equation}
K_{\mathrm{het}}(T_{\mathrm{o}}-T_{\mathrm{top}})=h_{\mathrm{c}}A \Delta T = h_{\mathrm{c}}A (T_{\mathrm{top}}-T_{\infty}).
\end{equation}
Solving for $T_{\mathrm{top}}$ gives:
\begin{equation}
T_{\mathrm{top}}=\frac{K_{\mathrm{het}}T_{\mathrm{o}}+h_{\mathrm{c}}A T_{\infty}}{h_{\mathrm{c}}A+K_{\mathrm{het}}}.
\label{TtopConv}
\end{equation}

For this particular geometry, where the cross-sectional area is the same for all the elements in the heterostructure and the convection loss, the area cancels and the expression for $T_{\mathrm{top}}$ can be simplified to:
\begin{equation}
T_{\mathrm{top}}=\frac{K_{\mathrm{het},\square}T_{\mathrm{o}}+ h_{\mathrm{c}}T_{\infty}}{K_{\mathrm{het},\square}+ h_{\mathrm{c}}},
\end{equation}
where
\begin{equation} 
K_{\mathrm{het},\square} =\frac{K_{\mathrm{het}}}{A}=\left(\frac{t_{\mathrm{film}}}{k_{\mathrm{film}}} +  \frac{1}{G_{\mathrm{int}}} +\frac{t_{\mathrm{sub}}}{k_{\mathrm{sub}}} \right) 
\end{equation}
is the heterostructure thermal conductance per area.  The cooling of the film surface via conduction when in atmosphere then only depends on the balance of the convection coefficient and the thermal conductance of the substrate/interface/film stack.  

A similar calculation is possible using the power leaving the top of the film surface via metallic wires or probes that may be used for voltage measurements or current probes in various experiments.  The heat current flowing upward away from the substrate in this case will almost certainly be non-uniform across the length of the sample, since these probes make contact at distinct locations.  The simple model we are constructing will not capture the effect of these complications, but can illustrate the overall magnitude of the out-of-plane component of the resulting thermal gradient.  As we will discuss further below, several groups have carefully documented experimental evidence for both this out-of-plane component and the in-plane non-uniformity.\cite{HuangPRL2011,MeierPRB2013,SchmidPRL2013,ShestakovPRB2015,MeierNatCom2015}  The power flowing out of the film, upwards from the top surface via conduction through these electrical connections is $P_{\mathrm{wires}}=K_{\mathrm{wires}}(T_{\mathrm{top}}-T_{\infty})$.  Note that the choice of the temperature where the resulting heat current is sink can depend on the details of a given experimental setup.  For this calculation, I will estimate this contribution by assuming two gold wire bonds, with circular cross sectional area $A_{\mathrm{wires}}$ and length $\ell_{\mathrm{wires}}$, are attached to the top of the metal film.  The thermal conductance of the wires is related to the geometry and thermal conductivity: $K_{\mathrm{wires}}=k_{\mathrm{Au}}A_{\mathrm{wires}}/\ell_{\mathrm{wires}}$.   

We can now model an experiment in vacuum (so convection does not contribute), with the wire bonds attached, where similar steady-state balance of heat powers flowing into and out of the top metal surface yields:
\begin{equation}
K_{\mathrm{het}}(T_{\mathrm{o}}-T_{\mathrm{top}})=K_{\mathrm{wires}}(T_{\mathrm{top}}-T_{\infty}).
\end{equation}
Solving for $T_{\mathrm{top}}$ for the wire bond conduction case then gives:
\begin{equation}
T_{\mathrm{top}}=\frac{K_{\mathrm{het}}T_{\mathrm{o}}+K_{\mathrm{wires}} T_{\infty}}{K_{\mathrm{wires}}+K_{\mathrm{het}}}.
\label{TtopBond}
\end{equation}

\begin{figure}
\includegraphics[width=3.38in]{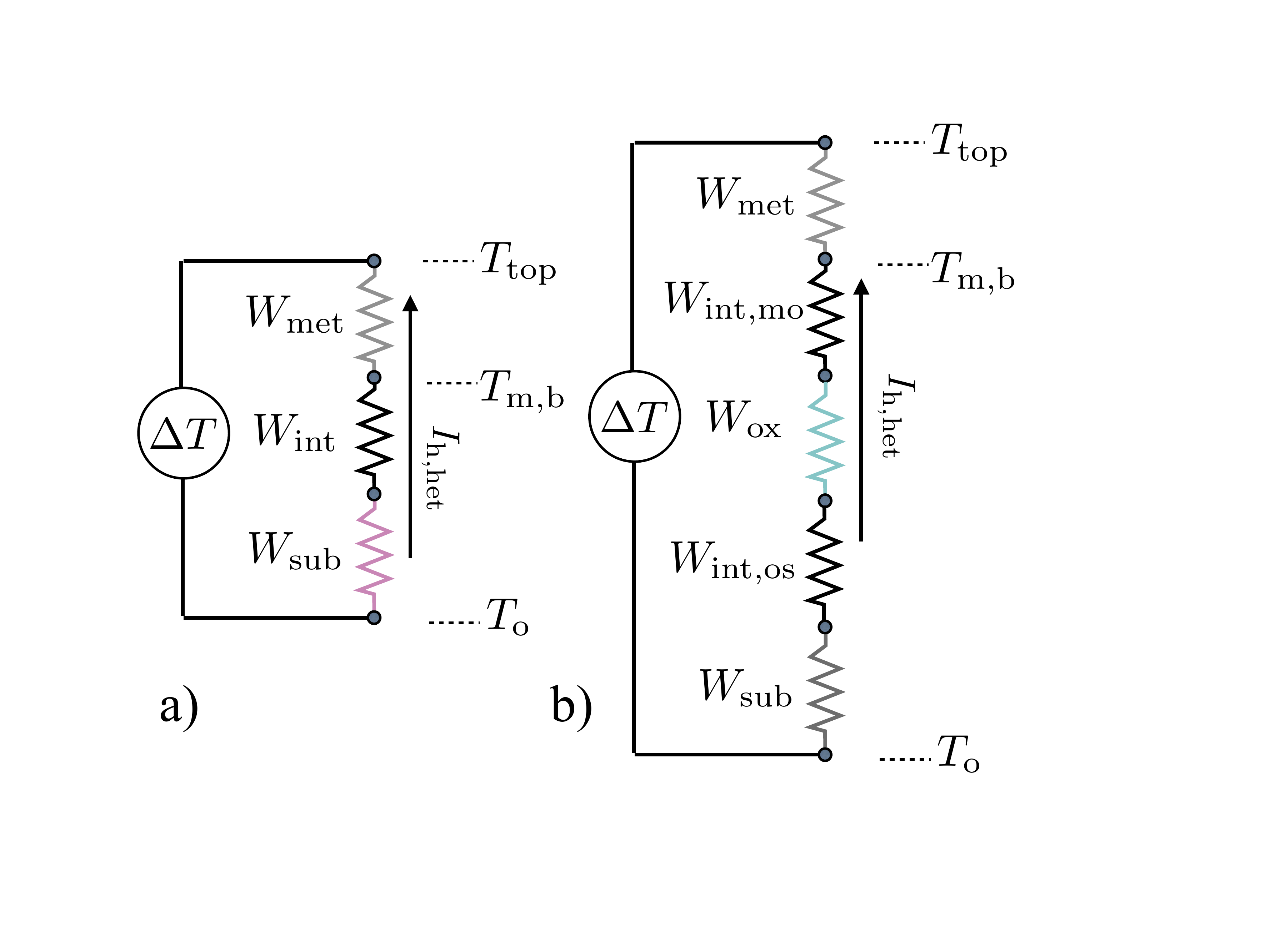}
\caption{\label{Fig2} Thermal circuit models for heterostructure branch of the two simple thermal models shown in Fig.\ \ref{Fig1}. In each, the total heat current flowing in the heterostructure is $I_{\mathrm{h,het}}$, set by $\Delta T$ and the total thermal impedance of these circuits.  \textbf{a)}  Thermal circuit for a metal film on a substrate. \textbf{b)} Thermal circuit for a film on an oxide film on a bulk substrate.    }
\end{figure}

Once $T_{\mathrm{top}}$ is determined in either case, the temperature profile in the sample heterostructure can be calculated using the thermal circuits shown in Fig.\ \ref{Fig2}.  The total temperature difference across the sample structure, $\Delta T=T_{\mathrm{top}}-T_{\mathrm{o}}$ drives a total heat current through the heterostructure, $I_{\mathrm{h,het}}=\Delta T/W_{\mathrm{het}}$.  The temperature drop, the thermal analog to the voltage drop, on each component of the circuit can be calculated by multiplying the heat current by the appropriate thermal resistance, such that for the thermal circuit of Fig.\ \ref{Fig2}a):
\begin{eqnarray}
\Delta T_{\mathrm{met}} &= & I_{\mathrm{h,het}} W_{\mathrm{film}} \\
\Delta T_{\mathrm{int}} & = &  I_{\mathrm{h,het}} W_{\mathrm{int}} \\
\Delta T_{\mathrm{sub}} & = &  I_{\mathrm{h,het}} W_{\mathrm{sub}}.
\end{eqnarray}
The somewhat more complicated thermal circuit of Fig.\ \ref{Fig2}b) adds:
\begin{eqnarray}
\Delta T_{\mathrm{int,mo}} & = &  I_{\mathrm{h,het}} W_{\mathrm{int,mo}} \\
\Delta T_{\mathrm{ox}} &= & I_{\mathrm{h,het}} W_{\mathrm{ox}} \\
\Delta T_{\mathrm{int,os}} & = &  I_{\mathrm{h,het}} W_{\mathrm{int,os}}, 
\end{eqnarray}
and of course uses the appropriate definition of $W_{\mathrm{het}}=1/K_{\mathrm{het}}$ from Eq.\ \ref{KhetB}.

Now with chosen values of the various parameters, we can calculate the temperature profile. 
I calculate an example geometry for each type of stack shown in Fig.\ \ref{Fig1}.   For both, I have chosen $w=5\ \mathrm{\mu m}$, and $l=8\ \mathrm{mm}$, to represent a thin slice from the central portion of the type of film/substrate heterostructures that are common in a range of spincaloritronic experiments.\cite{HuangPRL2011}  For the first, simpler, structure, I chose $t_{\mathrm{met}}=10\ \mathrm{nm}$, $t_{\mathrm{sub}}=0.5\ \mathrm{mm}$, and $k_{\mathrm{met}}=10\ \mathrm{W/m\ K}$, and $k_{\mathrm{sub}}=8\ \mathrm{W/m\ K}$.  These values are typical of situations with either a Pt or ferromagnetic metal (such as permalloy, Ni$_{80}$Fe$_{20}$, abbreviated ``Py") deposited on a yttrium iron garnet (YIG) substrate.   The interface thermal conductance for the Pt/YIG interface has very recently been measured, providing the important value $G_{\mathrm{int}}=150\ \mathrm{MW/m^{2}\ K}$.\cite{AngelesPRM2021}  With my choice of $A$, then $K_{\mathrm{int}}=6\ \mathrm{W/K}$.  To model one typical means of making electrical contact to the film, I estimated the thermal conductance from two gold wire bonds with radius $0.001\ \mathrm{inch}=25.4\ \mathrm{\mu m}$, with total length between the sample and the electrical connection to the cryostat (which I assumed was at $T_{\infty}$) of $1\ \mathrm{cm}$.   Drawn wires often have thermal conductivity most comparable to bulk values, for gold near room temperature this is $k_{\mathrm{Au}}=300\ \mathrm{W/m\ K}$. The resulting thermal conductance $K_{\mathrm{wires}}=122\ \mathrm{\mu W/K}$, which we will see below is large enough to significantly affect the thermal profile in the sample.  This choice of material and geometry for the electrical connections lands in a wide range of values most likely achieved in various experiments, with the use of larger manipulator probes or pins most likely much exceeding this value of $K_{\mathrm{wires}}$ used here, and the value for thinner wires formed from Al somewhat smaller.  This estimation should be refined by each reader to match the particular experimental arrangement.

\begin{table*}
\includegraphics[width=\linewidth]{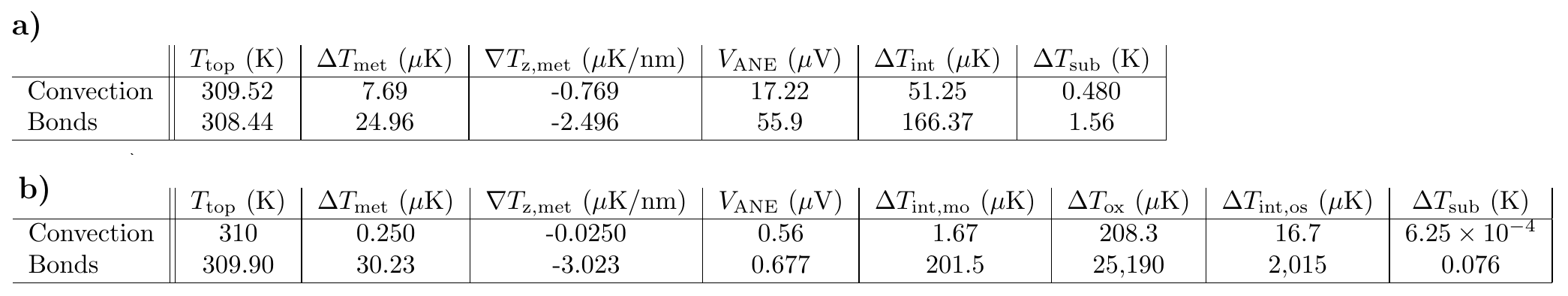}
%
\caption{\textbf{a)} Calculated temperature profile for the metal film on substrate stack of Fig.\ \ref{Fig1}a).  Here $T_{\mathrm{o}}=310\ \mathrm{K}$, $T_{\infty}=300\ \mathrm{K}$.  The thermal parameters and geometry are chosen to match a thin metal film on a YIG substrate (parameters given in main text). \textbf{b)} Calculated temperature profile for the metal film on oxidized substrate stack of Fig.\ \ref{Fig1}b).  Here $T_{\mathrm{o}}=310\ \mathrm{K}$, $T_{\infty}=300\ \mathrm{K}$.  The thermal parameters and geometry are chosen to match a thin metal film on an oxidized Si substrate (parameters given in main text).}
\label{TableStack}
\end{table*}

Table \ref{TableStack}a) presents key results from the simple thermal profile calculation for this film/interface/substrate geometry, where I chose the heated temperature of the bottom of the substrate, $T_{\mathrm{o}}=310\ \mathrm{K}$, with the sample environment held at $T_{\infty}=300\ \mathrm{K}$.  This simple calculation suggests that both pure convection and pure conduction (through electrical connections) would lead to relatively small, but clearly measurable with most thermometry technologies, cooling of the top of the stack.  This is roughly $0.5\ \mathrm{K}$ in the convection case and more than $1.5\ \mathrm{K}$ in the conduction case.  This overall temperature difference demands that the average thermal gradient across this section of the sample points perpendicular to the substrate.  This is a fundamental feature of thermal experiments with thin films on bulk substrates, that has been repeatedly demonstrated in experiments.\cite{HuangPRL2011,MeierNatCom2015,MeierPRB2013}  A second fundamental feature is also obvious in Table\ \ref{TableStack}a): the bulk of the temperature difference, in steady-state, exists across the substrate.  For example, with this particular choice of parameters, the conduction case shows $\Delta T_{\mathrm{met}}\cong25\ \mathrm{\mu K}$, while $\Delta T_{\mathrm{sub}}\cong1.56\ \mathrm{K}$, meaning that the temperature difference on the very thin film would be very challenging to experimentally quantify with any known thermometry technique.  Despite the small size of this temperature difference, the thermal gradient across the film$, \nabla T_{\mathrm{met}}=\Delta T_{\mathrm{z,met}}/t_{\mathrm{met}}$, is not insignificant in either case.  This is particularly important to consider in a geometry such as this, where the electrical contacts are often quite widely spaced.  In such a case even the small thermal gradients I calculate here can drive large voltage contributions if the thin film is a conducting ferromagnet.  I demonstrate this with the column labeled $V_{\mathrm{ANE}}$ using values for the anomalous Nernst effect for Py,\cite{BennetAIPadv2020}, which I will discuss in more detail in section \ref{Zoo}.

The second example geometry models the case of a similar thin metallic film, but now deposited on a silicon substrate with an intervening insulating layer.  My choices for the model parameters start with the same $w$ and $l$, and same $t_{\mathrm{met}}=10\ \mathrm{nm}$, and $k_{\mathrm{met}}=10\ \mathrm{W/m\ K}$.  Measured values for the interfacial thermal conductance between metals and common insulating thin films such as SiO$_{2}$ typically fall near $G_{\mathrm{int,mo}}=150\ \mathrm{MW/m^{2}\ K}$,\cite{ZhuJAP2010} giving a similar $K_{\mathrm{int,mo}}=6\ \mathrm{W/K}$ as seen for the Pt/YIG interface.  SiO$_{2}$ films typically have fairly poor thermal conductivity, $k_{\mathrm{ox}}=1.2\ \mathrm{W/m\ K}$, while bulk single crystalline Si usually has a fairly high thermal conductivity, $k_{si}=200\ \mathrm{W/m\ K}$.  I chose the thickness of these layers to be $t_{\mathrm{ox}}=1\ \mathrm{\mu m}$ and $t_{\mathrm{sub}}=0.5\ \mathrm{mm}$, respectively.  Finally, the interface thermal conductance between the oxide and Si has also been measured, and is roughly a factor of ten smaller than the metal/oxide case, giving $K_{\mathrm{int,os}}=0.6\ \mathrm{W/K}$.     

Table \ref{TableStack}b) presents key results for the film/interface/oxide/interface/substrate geometry.  I keep the same choice for $T_{\mathrm{o}}$, $T_{\infty}$, $h_{\mathrm{c}}$, and $K_{\mathrm{wires}}$.  Though the same main theme emerges here, the details of the temperature profile are modified by the significantly higher substrate thermal conductivity and the insertion of the much lower thermal conductivity oxide film.  For the convection case, the overall increase of the thermal conductance of the sample causes the total temperature difference to drop to most likely undetectable levels, with less than $1\ \mathrm{mK}$ of total change.  However, even here if the metal film is a ferromagnet, a fairly easily measurable ANE voltage component would arise.  In the bond wire conduction case, large $\Delta T$s form on the various elements, and we now see that $\Delta T_{\mathrm{ox}}=25\ \mathrm{mK}$ grows to be a third of the $\Delta T_{\mathrm{sub}}=75\ \mathrm{mK}$. This suggests that in cases where an electrically insulating magnetic film of similar thickness is deposited on a more highly thermally conductive bulk substrate (such as sapphire), appreciable out-of-plane $\nabla T$ could easily appear, and should be carefully considered.  Again, the bond conduction case indicates a significant out-of-plane $\nabla T$ develops on the metal film, which should be expected to drive significant voltage contributions via ANE.  Finally, the relatively poorly conducting oxide/substrate interface also develops significant $\Delta T$ in this model.

\subsection{Simple Analytic Modes: Joule heating from current in metal thin film\label{WithJoule}}

\begin{figure}
\includegraphics[width=3.38in]{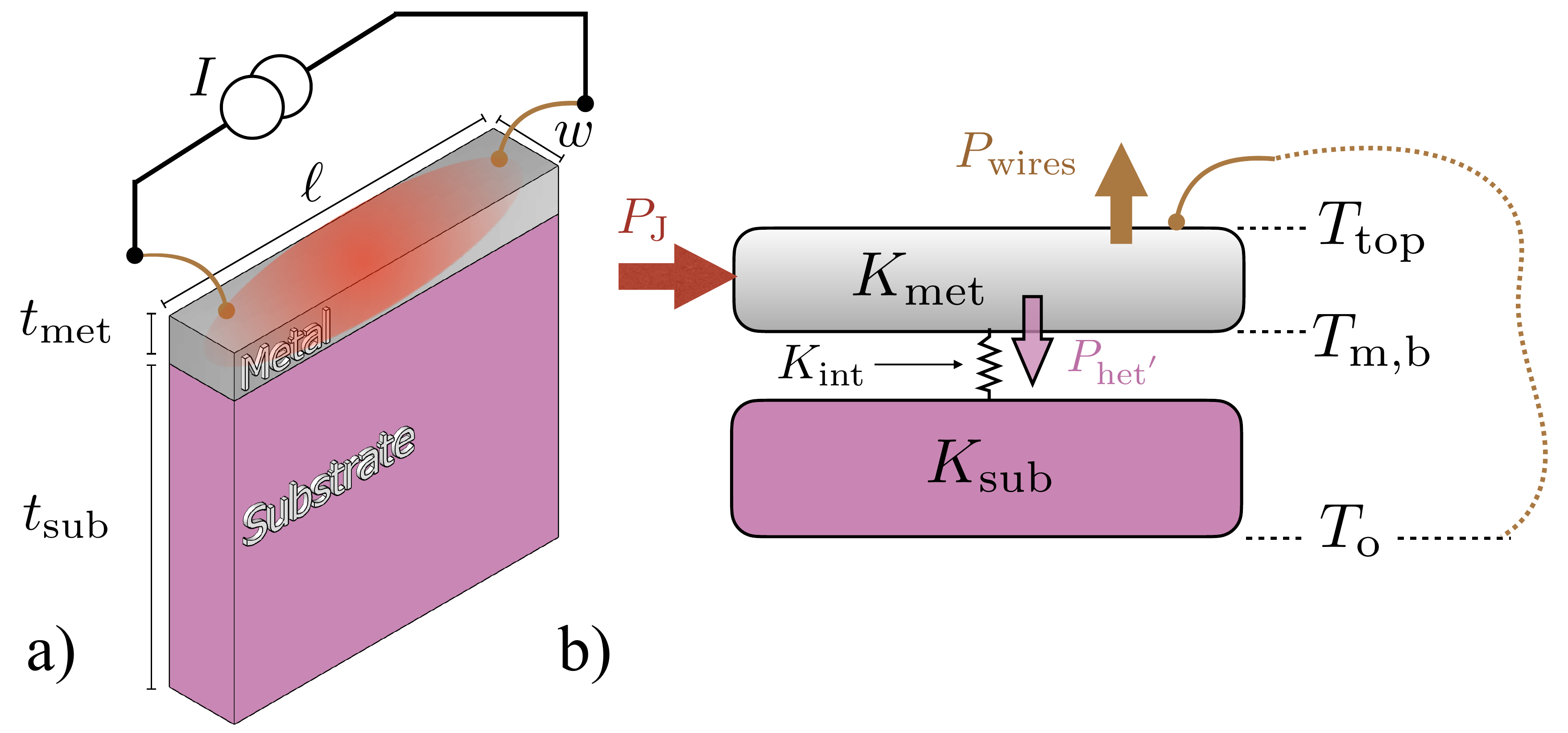}
\caption{\label{Fig3} Schematic view, \textbf{a)},  and simple thermal model \textbf{b)},  for the case where a thin metal film on a bulk substrate is heated by an applied current.   Heat flows into the metal film from Joule heating (red arrow labeled $P_{\mathrm{J}}$).  Heat can leave the top surface of the film via conduction (brown arrow labeled $P_{\mathrm{wires}}$), or via convection (not shown).  Finally heat can leave the bottom surface of the film via conduction to the rest of the sample heterostructure, (pink arrow labeled $P_{\mathrm{het}^{'}}$).  The steady-state thermal profile can be determined by balancing these powers. }
\end{figure}

I now use a similar approach to consider the case when current flows through the metal film while the back of the substate is held at a fixed temperature.  This is another very common experimental situation in spintronic materials and devices.  Fig.\ \ref{Fig3}a) shows a schematic geometry when a current is applied, with Fig.\ \ref{Fig3}b) showing the simple thermal model of the film/interface/substrate stack (I will also perform the calculation for the more complicated stack, but have omitted the schematics).  As indicated in the figure, we now assume both the back of the substrate and the environment are held at $T_{\mathrm{o}}$, and that either conduction through the experimental wiring (as indicated in the figure as $P_{\mathrm{wires}}$) or convection through atmosphere (not shown) can allow heat to flow away from the top surface of the metal thin film.  Heat can flow away from the bottom of the metal film through the rest of the sample heterostructure, indicated as $P_{\mathrm{het^{'}}}$.  Similar contributions were part of the model discussed in Sec.\ \ref{BackHeat}.  The electrical current flowing in the film adds the new possibility that Joule heat is dissipated throughout the metal film, as indicated by $P_{\mathrm{J}}$.    

The first step to determine the resulting steady-state thermal profile is to note that both the experimental wires and the heterostructure below the metal film connect the heated film to thermal ``ground" at $T_{\mathrm{o}}$ (one can easily modify this picture for the case where the environment is held at a different temperature than $T_{\mathrm{o}}$.  We can determine the average temperature of the metal film using the (typically known or measurable) $P_{\mathrm{J}}$ and the sum of these parallel thermal conductances:
\begin{equation}
T_{\mathrm{av,met}}=T_{\mathrm{o}}+\frac{P_{\mathrm{J}}}{K_{\mathrm{het^{'}}}+K_{\mathrm{wires}}}.
\label{Tavmet}
\end{equation}
We can use the definition of the average $T$ to write this as a function of the two unknown temperatures $T_{\mathrm{top}}$ and $T_{\mathrm{m,b}}$: 
\begin{equation}
 T_{\mathrm{av,met}}=\frac{T_{\mathrm{top}}+T_{\mathrm{m,b}}}{2}=T_{\mathrm{o}}+\frac{P_{\mathrm{J}}}{K_{\mathrm{het^{'}}}+K_{\mathrm{wires}}}.
\label{JouleEq1}
\end{equation}
A second relation involving $T_{\mathrm{top}}$ and $T_{\mathrm{m,b}}$ comes from the balance of heat power flows.  Here any net difference between the heat power flowing ``up" away from the top surface of the film, and the heat power flowing ``down" from the bottom surface of the film, in steady-state, will drive a temperature difference across the metal film that is proportional to its thermal conductance, $K_{\mathrm{met}}$.  In vacuum, the power flowing up is $P_{\mathrm{wires}}=K_{\mathrm{wires}} (T_{\mathrm{top}}-T_{\mathrm{o}})$, and the power flowing down is $P_{\mathrm{het^{'}}}=K_{\mathrm{het^{'}}}(T_{\mathrm{m,b}}-T_{\mathrm{o}})$.  $K_{\mathrm{het^{'}}}$ is the thermal conductance of the elements of the heterostructure below the metal film:
\begin{equation}
K_{\mathrm{het^{'}}}= \frac{1}{W_{\mathrm{sub}}+W_{\mathrm{int}}}.
\end{equation}
This then gives:
\begin{equation}
K_{\mathrm{wires}}(T_{\mathrm{top}}-T_{\mathrm{o}}) - K_{\mathrm{het^{'}}}(T_{\mathrm{m,b}}-T_{\mathrm{o}}) = K_{\mathrm{met}}(T_{\mathrm{top}}-T_{\mathrm{m,b}}).
\end{equation}

This system of two equations with two unknowns can now be solved for $T_{\mathrm{top}}$ and $T_{\mathrm{m,b}}$, yielding:
\begin{equation}
T_{\mathrm{top}}=\frac{ 2 T_{\mathrm{av,met}}(K_{\mathrm{het^{'}}}-K_{\mathrm{met}} ) + T_{\mathrm{o}} (K_{\mathrm{wires}}-K_{\mathrm{het^{'}}}  )  }{  K_{\mathrm{wires}} + K_{\mathrm{het^{'}}} -2 K_{\mathrm{met}} },
\label{TtopHeat}
\end{equation} 
and 
\begin{equation}
T_{\mathrm{m,b}}= 2 T_{\mathrm{av,met}}-T_{\mathrm{top}},
\label{TbotHeat}
\end{equation}
where $T_{\mathrm{av,met}}$ is given by Eq.\ \ref{Tavmet}.
As with the case of the heated substrate, determining $T_{\mathrm{top}}$ allows calculation of the total heat current flowing through the heterostructure between $T_{\mathrm{top}}$ and $T_{\mathrm{o}}$ (though this now flows in the opposite direction to that shown in Fig.\ \ref{Fig2}), which gives the temperature drop on the various elements of the stack.  
The same equations, with the substitution $K_{\mathrm{conv}}=h_{\mathrm{c}}A$ for $K_{\mathrm{wires}}$ will model the case for negligible conduction through bonds, but appreciable loss of heat to convection if vacuum is not maintained around the sample.  

\begin{figure*}[t]
\begin{center}
\includegraphics[width=6.0in]{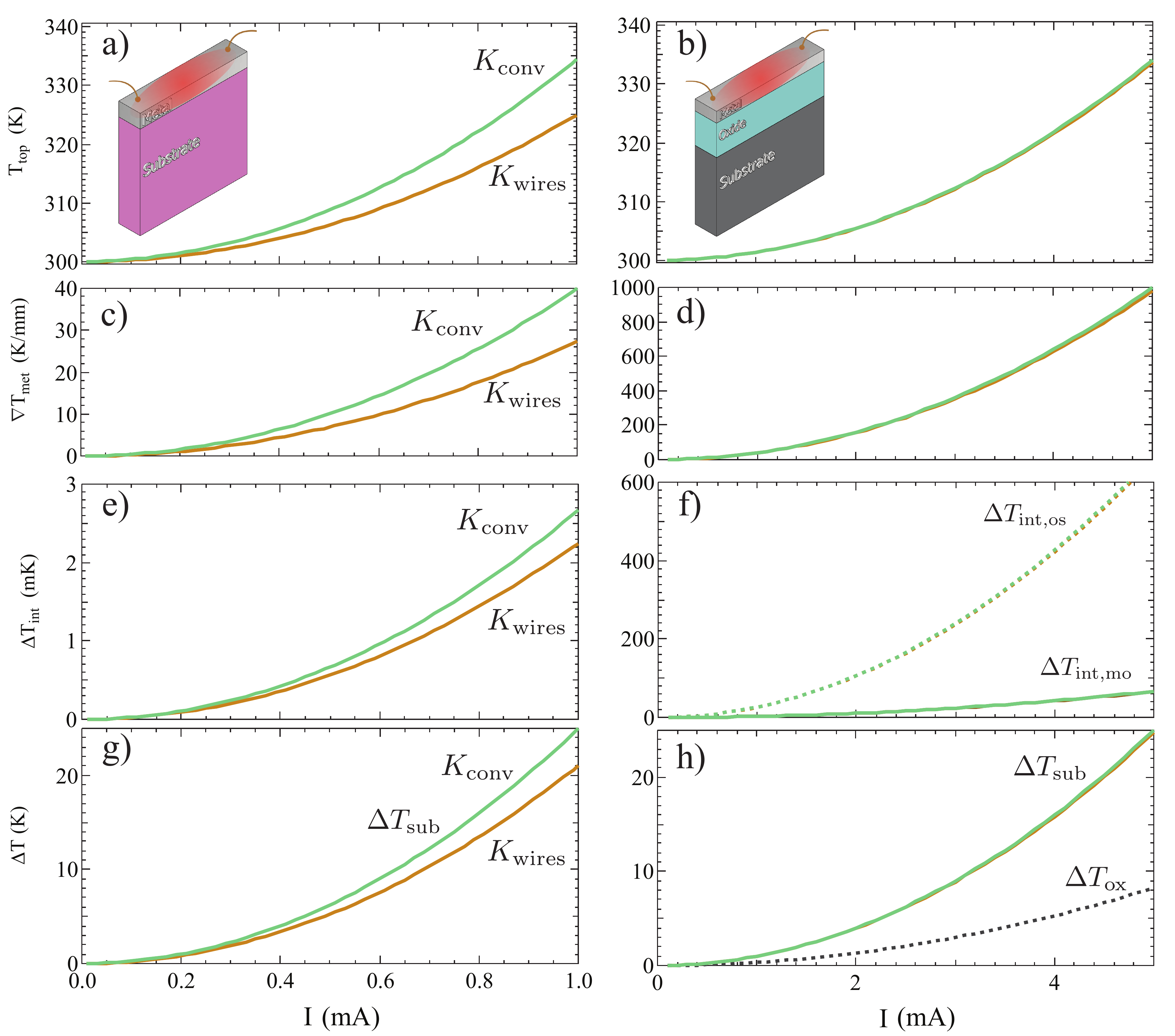}
\end{center}
\caption{\label{Fig5} Example thermal profile for heated thin films on two types of substrates, with various $T$ plotted vs. applied current $I$.   \textbf{a)}$T_{\mathrm{top}}$ vs. $I$ for a film on substrate, for two heat loss conditions from the top surface, convection through air and conduction through experimental wiring.  \textbf{b)} $T_{\mathrm{top}}$ for the same film on an oxide layer, on top of a more highly conductive substrate.  Here both conduction and convection give the same calculated $T_{\mathrm{top}}$.  \textbf{c)} Thermal gradient $\nabla T$ on the metal film. \textbf{d)} $\nabla T$ for the film on oxide case is $>10\times$ larger. \textbf{e)} Temperature difference across the interface, $\Delta T_{\mathrm{int}}$ . \textbf{f)} Temperature differences across the two interfaces that form in the metal/oxide/substrate case. \textbf{g)} $\Delta T$ across the substrate in the film/substrate case for two different surface heat loss modes. \textbf{h)} $\Delta T$ across the substrate and oxide layer for the metal/oxide/substrate case.  Here a significant $\Delta T$ forms on the oxide, despite the very thin layer, due to the poor thermal conductivity.     }
\end{figure*}

Figure \ref{Fig5} shows the resulting temperature profiles as a function of applied current to the metal film, for the same two heterostructures I considered for the case of the heated base of the substrate.  Figs.\ \ref{Fig5}a) and b) plot $T_{\mathrm{top}}$ vs. $I$, for the case of heat loss from the top of the film via pure convection (green line labeled $K_{\mathrm{conv}}$) and pure conduction through electrical connections (copper line labeled $K_{\mathrm{wires}}$).  Here I used the same estimate for the thermal conductance of the electrical connections as described above.  This value will certainly change from experiment to experiment, and should be chosen to match each situation.  For the case of the metal film on YIG substrate, which has a relatively low $k_{\mathrm{sub}}$ the two types of heat loss cause significantly different $T_{\mathrm{top}}$, while for the metal film on oxidized Si substrate, the larger $k_{\mathrm{sub}}$ makes the heat flow into the substrate the dominant loss mechanism, so that the convection and bond conduction cases have nearly identical thermal profiles.  This difference also means that generating a similar $T_{\mathrm{top}}$ for the Si case requires much higher $I_{\mathrm{bias}}$.   However,  in both cases even relatively modest applied dc current drives temperature increase $>10\%$ of $T_{\mathrm{o}}$.  Figs.\ \ref{Fig5}c) and d) plot $\nabla T_{\mathrm{met}}$ for these two cases in $K/mm$.  The positive values indicate $\nabla T_{\mathrm{met}}$ points in the opposite direction as in Tables\ \ref{TableStack}a) and \ref{TableStack}b), as expected.  Though the two types of heterostructure reach very similar $T_{\mathrm{top}}$ with the range of $I$ chosen for each, $\nabla T_{\mathrm{met}}$ is much larger for the film on oxidized Si.  This highlights the importance of understanding the thermal conductivity of the supporting layers beneath a thin film for experiments of this type, as the same metal film on the two different types of supporting structures could show dramatically different thermoelectric voltage contributions driven by this $\nabla T_{\mathrm{met}}$.    

The different $k_{\mathrm{sub}}$ values also lead to significantly different temperature drops across the interfaces in each heterostructure, as shown in Fig.\ \ref{Fig5}e) and f).  While  the $\sim3\ \mathrm{mK}$ values seen at the metal/YIG interface are potentially important, the much larger values on both interfaces in the metal/oxide/Si case almost certainly modify the overall picture of heat flow through the system and cannot be ignored.  Finally, the presence of even the rather thin SiO$_{2}$ layer, because of the quite low $k_{\mathrm{ox}}$ and relatively high $k_{\mathrm{sub}}$, cause a significant $\Delta T_{\mathrm{ox}}$ to appear on this layer even at modest applied $I$, rising to $\sim1/3$ the temperature drop on the entire substrate.  

To conclude this section, I will summarize some of the clear limitations of this very simple framework I have presented to gain understanding of temperature profiles, and emphasize the key lessons learned.  First, one of the main limitations is that, for the sake of much simpler calculations, I did not take the known temperature-dependence of $k$ and $G_{\mathrm{int}}$ into account.  This could easily be done by one wishing to use these equations but needing a more refined picture of the resulting thermal gradients in a given sample.  Second, I have not discussed the trivial addition of the convection and conduction heat loss terms.  Again, if a particular experimental setup demands, this is a simple extension.  Third, as I mentioned earlier, the heat flow away from the thin metal film via experimental wires will be spatially non-uniform in general, though must result in an average out-of-plane thermal gradient that arises in the simple model.  Fourth, there are regimes, especially for very low temperatures, relatively clean materials, and small systems, where additional complications could arise related to nanoscale size effects and the finite mean free path of various heat carriers.\cite{BoonaPRB2014,RegnerNMTE2015,GallJAP2016,BourgeoisCRP2016}  There is a great deal of interesting physics to consider in that area, which I chose to ignore since many experiments and applications in spintronics are at room temperature.   Finally, I chose two idealized sample stacks made from components where reasonable knowledge of $G_{\mathrm{int}}$ already exists.  This is most likely not true for all situations of interest, though this gap in knowledge continues to be filled.   

Despite these obvious limitations, I hope this very simple approach can provide value at least as a guide to intuition, for example when using more complicated numerical modeling.  These much more complicated and computationally intense models, typically now carried out via commercially available software, are in fairly common use, and some approaches now integrate charge, spin, and thermal effects in a single code.\cite{HadamekSSE2022,ShigematsuJAP2022} 
One very important concern with all such models is that they often make assumptions that the user may not realize.  For example, typical engineering codes will often ignore interface effects, which may or may not result in a physically accurate temperature profile.  The finite element approach often assumes temperature can be defined on arbitrarily small length scales, which ignores the possible role of ballistic or quasiballistic phonon transport.\cite{SiemensNatMat2010,MinnichPRL2012,RegnerNatComm2013,MinnichPRL2011,BourgeoisCRP2016,StefanouPRL2021}  Since heat transport in a wide range of materials, including silicon, is now known to rely on surprising long mean free path phonons,\cite{KimPRM2021,LeeNanoLett2015,JohnsonPRL2013,SultanPRB2013,KohPRB2007} the finite-element approach perhaps requires careful examination in a broader range of situations than currently appreciated.  All these codes also all rely on accurate input of thermal properties of constituent materials.  For thin film and nanoscale materials, these are often unknown, and will almost never be accurately represented by bulk values for thermal conductivity that are often the default.  In the absence of measurements of thermal conductivity of films, the Widemann-Franz law can give an estimate of $k$ due to electrons in conducting systems, though significant deviations from that simple estimate have also often been observed for relevant materials such as gold, platinum and tungsten thin films and nanostructures \cite{MasonPRM2020,WangNanotech2019,SawtellePRB2019,EigenfeldNano2015,WangIJHMT2013,WangHMT2011,VolkleinNano2009,ZhangPRB2006,XiaNanotechnology2010}.

The list of important lessons learned in this exercise is shorter and simpler.  First, when a thin film is supported on a bulk substrate, this substrate will typically dominate the heat flow in the structure.  Second, the bulk heat sink of the substrate combined with any of a range of heat loss mechanisms from the top surface of a film will inevitably lead to some component of out-of-plane thermal gradient.  Third, despite comparatively small thermal gradients compared to the substrate, the extreme geometry of a thin film can still lead to important contributions to measured signals in a range of experiments.  In the next section we will explore these in more detail.

\section{The thermoelectric and spin thermoelectric ``Zoo"\label{Zoo}}

\begin{figure}
\includegraphics[width=3.38in]{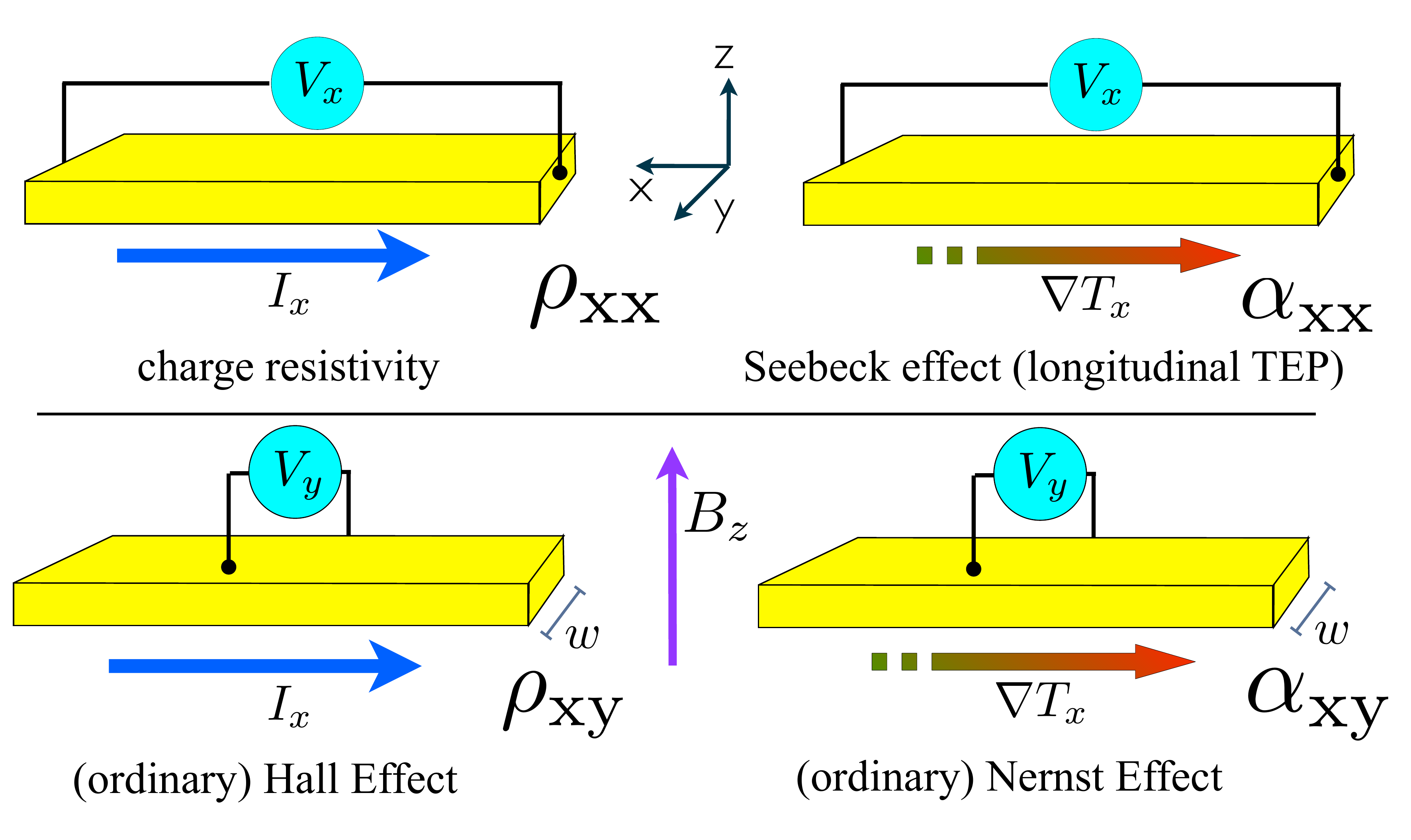}
\caption{\label{Fig4} Schematic views of charge transport and thermoelectric effects in non-magnetic conductors. }
\end{figure}

Consideration of the thermal and transport properties and phenomena relevant to spintronic devices begins with effects in nonmagnetic systems.  As we have already discussed, the heat dissipated by a current flowing through a simple non-magnetic metal is related to the charge resistivity, $\rho$, which can be measured as shown in Fig.\ \ref{Fig4} by applying a known charge current, $I$ and measuring the resulting voltage drop $V$ along the same direction.  It is common to denote the direction of the applied current and measured voltage, respectively, by subscripts such that this common longitudinal charge resistivity is $\rho_{\mathrm{xx}}$.  Even in non-magnetic systems, it is not uncommon to observe that the charge resistivity depends on applied magnetic field, though this magnetoresistance is typically small in the near room temperature regime of most interest in spintronics.\cite{HoffmannPRB1982,RoyPhysB2020} 

\subsection{Seebeck Effect (longitudinal thermopower)}

In the most common thermal analog to this arrangement, we replace the current with a thermal gradient, shown here applied along the $x$ axis, $\nabla T_{\mathrm{x}}$.  The thermal gradient causes mobile charge carriers to flow from hot to cold.  In open circuit conditions, this redistribution of charge leads to an electric field.  In the most general expression, the electric field $\vec{E}$ that results from application of thermal gradient $\vec{\nabla} T$ is 
\begin{equation}
\vec{E}=\tilde{\alpha}\vec{\nabla} T,
\label{matrix}
\end{equation}
where $\tilde{\alpha}$ is the thermoelectric tensor.  For many non-magnetic materials this tensor is diagonal, and these diagonal elements are the Seebeck coefficients of the material.  Then, as shown in Fig.\ \ref{Fig4}, with $\nabla T$ along the $x$ axis, an electric field appears on that axis proportional to $\alpha_{\mathrm{xx}}$.  If we further assume that the thermal gradient is always uniform through the sample in magnitude and direction, then integrating the electric field along the path of the voltage measurement gives the measured voltage $V_{\mathrm{x}}$, and the common formula for the Seebeck coefficient results, $\alpha_{\mathrm{xx}}=V/\Delta T$, where $\Delta T$ is now the temperature difference, rather than the thermal gradient.  There are two important complications that are sometimes missed in this simple textbook descriptions of thermoelectricity and are often important for spintronic systems.  The first is the role of the leads used to complete the measurement circuit.  Except where a superconductor can be used for this material (which has zero Seebeck coefficient well below $T_{\mathrm{c}}$, a temperature difference will drive charge motion in the leads as well, and this always causes the simple ratio of $V$ and $\Delta T$ to include a contribution from these leads, (as described in more detail elsewhere\cite{MasonJAP2020}).   The simple ratio of measured thermovoltage to temperature difference between the ends of the sample is correctly described as the relative thermopower:
\begin{equation}
\alpha_{\mathrm{rel}}=\alpha_{\mathrm{s}}-\alpha_{\mathrm{lead}}=\frac{V}{\Delta T}, 
\label{AlphaRel}
\end{equation}
where $\alpha_{\mathrm{s}}$ is the absolute Seebeck coefficient of the sample, and $\alpha_{\mathrm{lead}}$ the (again absolute) Seebeck coefficient of the leads.  This brings to mind the functional use of a thermocouple to measure temperature.  If two wires of different type are used to connect a voltmeter at one temperature to their junction at a different temperature, Eq.\ \ref{AlphaRel} suggests that the measured voltage signal will be proportional to the difference in temperature and the difference in thermopower between the materials.  This effect can be calibrated to allow simple and quite accurate thermometry, with the additional advantage of not having to apply a current or voltage bias.  This also suggests that if the two wires are formed from the same type of material, there is zero thermovoltage.  While true for bulk systems, in the case of thin films and nanostructures, researchers have commonly observed voltage signals even when the two arms of a thermocouple are nominally the same material.\cite{DuarteNanoLett09,SzakmanyIEEETransNano2014,EvansPNAS2020}  It can very rarely be assumed that a real experimental system should show zero thermopower based on any such argument, and in general one should expect the presence of up to microvolt-size background effects in any thermopower measurement.  This can put an extreme burden on experimentalists wishing to demonstrate novel phenomena.  
An additional consequence of the lead contribution is that using thermopower measurements to quantitatively probe the fundamental physics of a material, when superconducting leads cannot be used, requires careful determination of this lead contribution.  This is typically not exceptionally challenging for bulk samples, but can be extremely difficult for thin films and nanoscale samples, where methods to determine the absolute Seebeck coefficient remain rare.\cite{BohnertPRB2014,ScarioniJPhysD2018,KockertJAP2019,MasonJAP2020}

The nature of the electronic density of states of the material determines the absolute Seebeck coefficient (of sample and leads).  The most straightforward case is that of a non-degenerate semiconductor with a single dominant carrier-type.  In this case, the sign of the resulting thermovoltage will indicate the sign of the carrier.  In more complicated semiconductors, both carrier types can contribute.  The magnitude of $\alpha$ in semiconductors is typically larger than more highly conductive metals, sometimes large enough to ignore the lead contributions.  Even simple non-magnetic metals hold further complications, and a good metal can show any sign of $\alpha$, and the resulting thermovoltage, $V_{\mathrm{x}}$, or zero, depending instead on the slope of the density of states with energy at the Fermi energy.  One can generally also not assume that tabulated bulk values for Seebeck coefficient will be an adequate representation for materials in thin film or nanoscale form, since defects and imperfections always present in thin films affect the Seebeck coefficient.\cite{KondoPTP1965,BermanJPFMP1971,BarnardJPhysE1973,HuebenerPR64,HuebenerPR64b,AngusPhysLettA1970,LinJAP1971,StrunkPRL98,MasonJAP2020}  This means that when the Seebeck coefficient can be important to understand a signal generated in a spintronic device in response to intentional or unintentional heating, the best practice is always to measure $\alpha$ for the thin film used, in as close to the form used in the device as possible.  

In cases where direct measurement of $\alpha$ is not possible, one can gain some insight based on calculations of $\alpha$ from measured charge transport properties, which are typically much more straightforward.   The Mott equation is the simplest theory describing a thermopower that arises from thermal diffusion of charge carriers:
\begin{equation}
\alpha=-\frac{\pi^{2}k_{\mathrm{B}}^{2}T}{3e}\frac{1}{\rho}\left[ \frac{\partial \rho}{\partial E}\right]_{E=E_{F}}, 
\label{Mott}
\end{equation}
which relates the Seebeck coefficient to fundamental constants (the Boltzmann constant, $k_{\mathrm{B}}$, and the electron charge $e$), the charge resistivity $\rho$, and its energy derivative, $\partial \rho/\partial E$, taken here at the Fermi energy, $E_{\mathrm{F}}$.  This expression assumes both conduction through isotropic $s$-like bands, and via carriers that obey the Wiedemann-Franz law.  The energy derivative $\partial \rho/\partial E$ is related to the electron energy-dependence of the dominant charge scattering mechanism(s), which is in some circumstances reasonable to calculate or simulate, but is very difficult to measure.  For extremely thin films, it may be important to take size effects into account, and theories to explain the thickness-dependence of the diffusion thermopower are known. \cite{PichardJPhysF1980,TellierTSF77a,TellierTSF77,TellierTSF78}

An alternate expression that emphasizes materials properties more commonly known for degenerate semiconductors can be helpful:
\begin{equation}
\alpha=\frac{8\pi^{2}k_{\mathrm{B}}^{2}}{3eh^{2}}m^{*}T \left( \frac{\pi}{3n} \right)^{2/3},
\label{SemiMott}
\end{equation}
where $h$ is Planck's constant, $m^{*}$ is the charge carrier effective mass, and $n$ is the carrier concentration.  This form makes clear that as $n$ drops, one generally expects a regime where $\alpha$ becomes large.  Though I have stated that reference to tabulated bulk values is of little use for understanding Seebeck effects in the thin films forming spintronic devices, it is helpful to keep very general ideas about the typical size of $\alpha$ in mind.  For nonmagnetic metals, one generally finds values for $\alpha$ in the range of several to tens of $\mathrm{\mu V/K}$, magnetic metals (to be discussed further below) are often larger, showing $\alpha\sim$ several dozen $\mathrm{\mu V/K}$, and semiconductors can have $\alpha$ ranging from values comparable to metals to well above $1\ \mathrm{mV/K}$.       

In addition to diffusion thermopower, interactions between quasiparticle excitations and charge carriers can also drive important contributions to thermopower.  In non-magnetic systems, phonon drag is such an effect, where momentum transferred between the thermally excited flux of phonons carrying heat and the charge carriers can add to the typical diffusion thermopower.  This contribution is pronounced at low $T$ in relatively high quality semiconductor crystals,\cite{HerringPR58,GeballePR55,FrederiksePR1953} but is often discussed over a broader range of materials and conditions.  However, impurity and defect scattering is usually assumed to limit the phonon mean free path strongly enough to suppress phonon drag effects in thin films near room temperature.  

\subsection{(ordinary) Hall and Nernst effects}

As shown at the bottom of Fig.\ \ref{Fig4}, applying a large magnetic field perpendicular to the film plane introduces transverse voltage components to the resistivity and thermopower experiments.  In the case of the charge transport, this is the well-known (ordinary) Hall effect, which arises from the Lorentz force on a moving charge in a transverse magnetic field.  If the average charge carrier velocity is in the $\hat{x}$ direction with $H$ applied in the $\hat{z}$ direction, the force will be in either the $\pm \hat{y}$ direction.  In both metals and semiconductors the measured transverse voltage can have either sign, indicating the type of charge carrier in a semiconductor, or the details of the density of states at the Fermi level for a metal.   The Hall voltage signal is proportional to $1/n$, so is often large and relatively easy to measure in semiconductors, and often small and challenging to measure in metals.  Again, replacing $I_{\mathrm{x}}$ with $\nabla T$ gives a transverse thermopower, or an off-diagonal term in the thermoelectric tensor.  This is commonly called the (ordinary) Nernst effect.  The word ``ordinary" here and in the Hall effect distinguishes the measurement in a non-magnetic material from the much larger effects in ferromagnets we discuss shortly.  
The ordinary Nernst coefficient in metals is typically small compared to other effects with the same symmetry.  As with the Hall effect, larger Nernst coefficients occur in semiconductors, and also in elemental bismuth, where the unique electronic density of states and exceptionally long electron coherence length contribute to the highest known ordinary Nernst coefficient.\cite{BehniaRPP2016,BehniaJPCM2009}  This has been demonstrated to complicate SOT experiments with Bi-containing films on metallic ferromagnetic films,\cite{RoschewskyPRB2019} and drive spurious signals in other spintronic investigations using Bi films.\cite{YueAPLMats2021}

\subsection{Measurements that require thermometry: Peltier effect and thermal conductivity}

\begin{figure}
\includegraphics[width=3.38in]{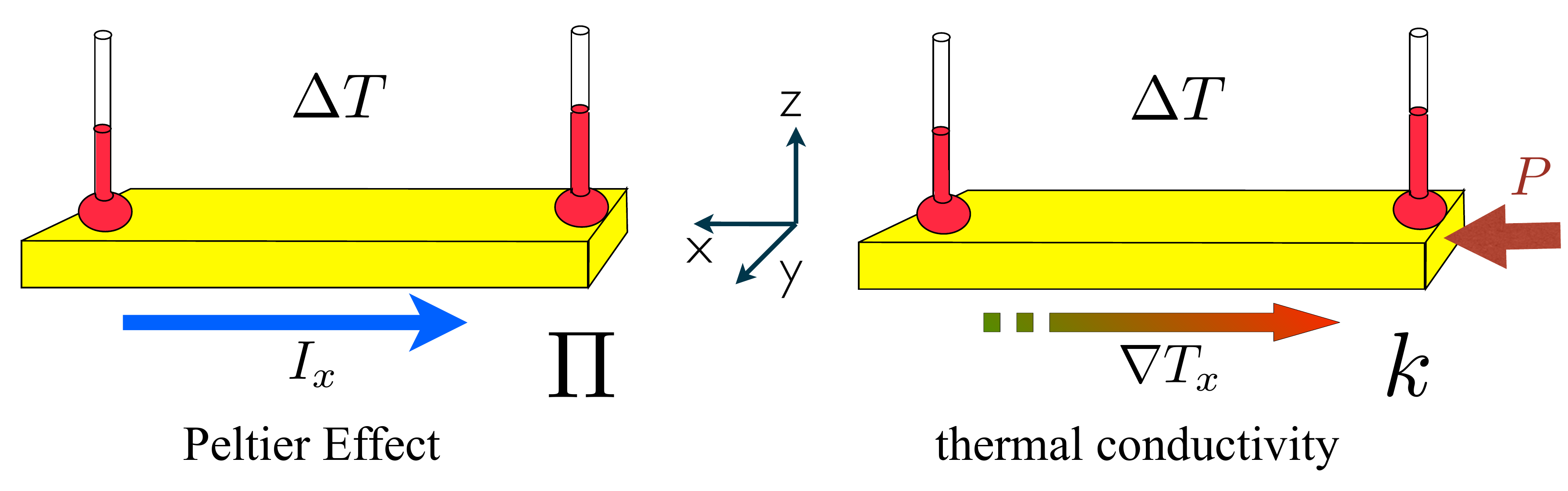}
\caption{\label{Fig4add} Schematic views of Peltier effect and thermal transport experiments in non-magnetic conductors.}
\end{figure}

Measurements of voltage in response to either charge or thermal excitation are among the most accurate and straightforward possible in typical condensed matter systems.  This is not true where one desires a measurement of temperature.  There are many physical reasons for this, including the requirement to identify and calibrate a temperature-dependent measurable transducer, and the introduction of potential interfaces across which thermal energy must flow.  In practical systems achieving accuracy better than $1\%$ in absolute measurements of temperature typically requires extreme care, and reaching 1 part in $10^{5}$ accuracy on measured $T$, a fairly standard accuracy for resistance measurements, is essentially unheard of.    For this reason, for any effect where schematic thermometers appear in the figures below, the reader should assume that only the most heroic measurements can credibly claim even $1\%$ accuracy.  As shown in Fig.\ \ref{Fig4add}, this certainly includes measurements of thermal conductivity, where the experimentalist must measure a temperature difference across a sample in response to known heat power applied.  It also includes the Peltier effect.  This is the time-reversal symmetry conjugate of the Seebeck effect.  By reversing the arrows shown for the Seebeck case in Fig.\ \ref{Fig4}, one can argue that driving a charge current through a conductor should drive a flow of entropy that results in a temperature difference along the current path.  This statement can be made quantifiably correct using the Onsager relation which indicates that, $\Pi=\alpha T$, where $\Pi$ is the Peltier coefficient.   The temperature difference across the sample is proportional to $I\Pi$.  

As already mentioned at the end of the previous section, accurate knowledge of thermal conductivity of thin films and nanostructures is a constant challenge. Isolating the contribution of a thin film or nanostructure from a supporting bulk substrate is challenging, for all the reasons outlined above.  Nevertheless, methods to achieve this continue to be developed and improved.\cite{ZhaoJEP2016,TobererARMR2012,AveryPRB2015} 
Finally, note that in the most general case, the thermal conductivity of a material, like the electrical conductivity, is a tensor that describes possible anisotropy with crystal direction and other effects.  Since anisotropy of thermal conductivity is not an effect that has caused measurable impact on spintronic systems to my knowledge, I have used the simple assumption that $k$ is isotropic in this paper.

\begin{figure}
\includegraphics[width=3.38in]{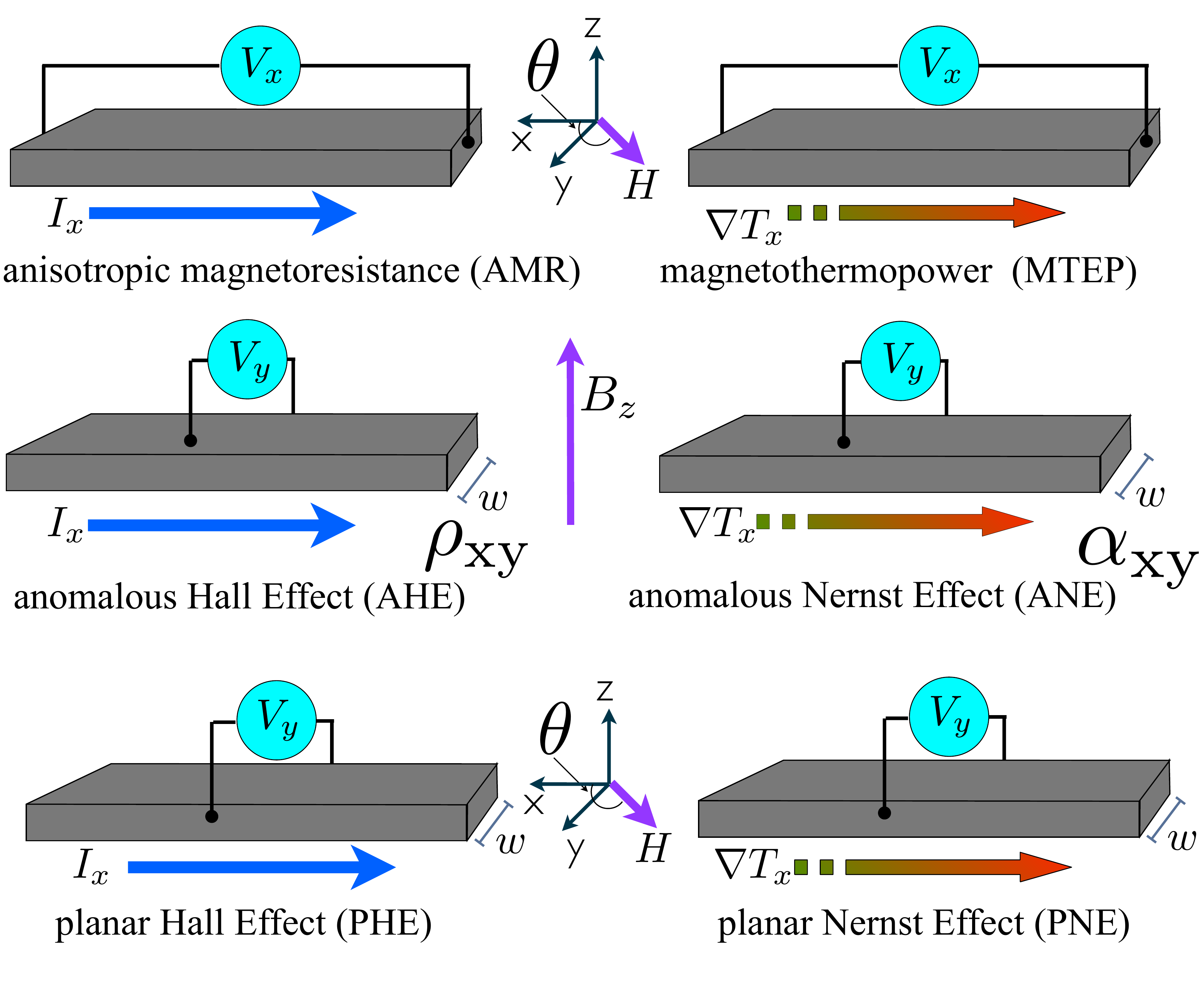}
\caption{\label{FMmet} Schematic views of electric and thermoelectric effects in ferromagnetic conductors.}
\end{figure}

\subsection{Effects in Ferromagnets: magnetization dependent transport (AMR and MTEP, AHE and ANE, PHE and PNE)}

The addition of magnetic order drives important modifications of the picture outlined for non-magnetic conductors.  As shown in Fig.\ \ref{FMmet}, these include both to field-dependent charge transport, and their thermal analogs.  When the magnetization of the FM is controlled in the plane of the film by application of field, where the angle $\theta$ tracks the orientation with respect to current direction, measurements of longitudinal voltage in response to applied current result in the anisotropic magnetoresistance (AMR).\cite{McGuireIEEE75}  Replacing charge current with thermal gradient again gives a Seebeck effect, or longitudinal magnetoresistance, though now with a dependence on $\theta$, often called the magnetothermopower (MTEP).   When the field causes magnetization transverse to the applied current or thermal gradient, a mutually perpendicular electric field (and voltage) appears due to either the anomalous Hall effect (AHE), or anomalous Nernst effect (ANE), respectively.  The term ``anomalous" was used in the earliest measurements of the Hall effect and refers to the unusually large values of Hall effect observed for conducting ferromagnets when compared to non-magnetic metals\cite{HallAmJMath1879}.  In the schematics for AHE and ANE I explicitly labeled the width of the sample, $w$.  With the choice of direction of the current (or thermal gradient) and applied field (and therefore magnetization) indicated, the width lies along the direction of the electric field generated by the Hall or Nernst effect.  As a result, assuming that this electric field is uniform throughout the sample, the measured voltage will be simply proportional to the width.  As stated earlier, it is common for the width of a thin film ferromagnet to be fairly large, and this can easily make significant voltages appear even if thermal gradients or applied currents are small.  The AHE has been studied fairly extensively, and the topic has been reviewed in great detail.\cite{NagaosaRMP2010} Measurements of the ANE are more rare, and studies on a range of thin film ferromagnetic (FM) systems have started to appear only fairly recently.  
To first clarify terminology, the expression for the electric field generated by the ANE, $E_{\mathrm{N}}$, is
\begin{equation}
\label{eq:ANE}
E_{\mathrm{N}=}\nabla V_\mathrm{N} = -S_\mathrm{N}\widehat{m} \times \nabla T,
\end{equation}
where $\widehat{m}$ in the direction of the magnetization of the 
FM, and $\nabla T$ the thermal gradient across the 
contact.  $S_{\mathrm{N}}$ is the transverse Seebeck coefficient, which is often expressed $S_\mathrm{N}=R_\mathrm{N}S_\mathrm{FM}$ where $R_\mathrm{N}$ is termed the anomalous Nernst coefficient (sometimes written $\theta_{\mathrm{ANE}}$ and called the anomalous Nernst angle) and $S_\mathrm{FM}$ is the absolute Seebeck coefficient of the ferromagnet.  This expression makes very clear that, like the resistivity and Seebeck coefficient, the overall size of the ANE will depend on the details of scattering in a given sample.  This provides some context for reports of $S_{\mathrm{N}}$ which may not agree for the nominally same FM material.  The metallic ferromagnet that has the most measurements of the ANE coefficient (still only a handful) is permalloy, the Ni-Fe alloy with $\sim80\%$ Ni.  
Most values for $S_{\mathrm{N}}$ fall near $-2-3\ \mathrm{\mu V/K}$, \cite{SlachterPRB2011,vonBierenAPL13,BrandlAPL2014,ChuangPRB2017,BennetAIPadv2020,BennetPRB2019}, though some authors report much smaller values \cite{YamazakiPRB2022}.  There are two more common approaches to measure the ANE for metals, one a variation of the experiment used to probe the longitudinal spin Seebeck effect (described in more detail below), and the second using nanoscale metallic devices called non local or lateral spin valves. The effect has also been observed in magnetic tunnel junctions, which provided an ANE coefficient value for CoFeB.\cite{MartensCommPhys2018} Chuang, et al. also studied other transition metal ferromagnets,\cite{ChuangPRB2017} and measurements of semiconducting FM ferrites have also been reported.\cite{BougiatiotiPRL2017}
Some authors have shown that combined effects of bulk and surface spin-orbit scattering can both play important roles in the effective ANE voltage.\cite{KannanSciRep2017}  This means that simple assumptions about how a given layer in a magnetic heterostructure should contribute to ANE voltages, or even how a film grown on two different substrates behaves, are most likely not reliable.  This is one of many complications that make separating the ANE from other effects with the same symmetry between the applied magnetic field and measured voltage extremely difficult.

\begin{figure}
\includegraphics[width=3.38in]{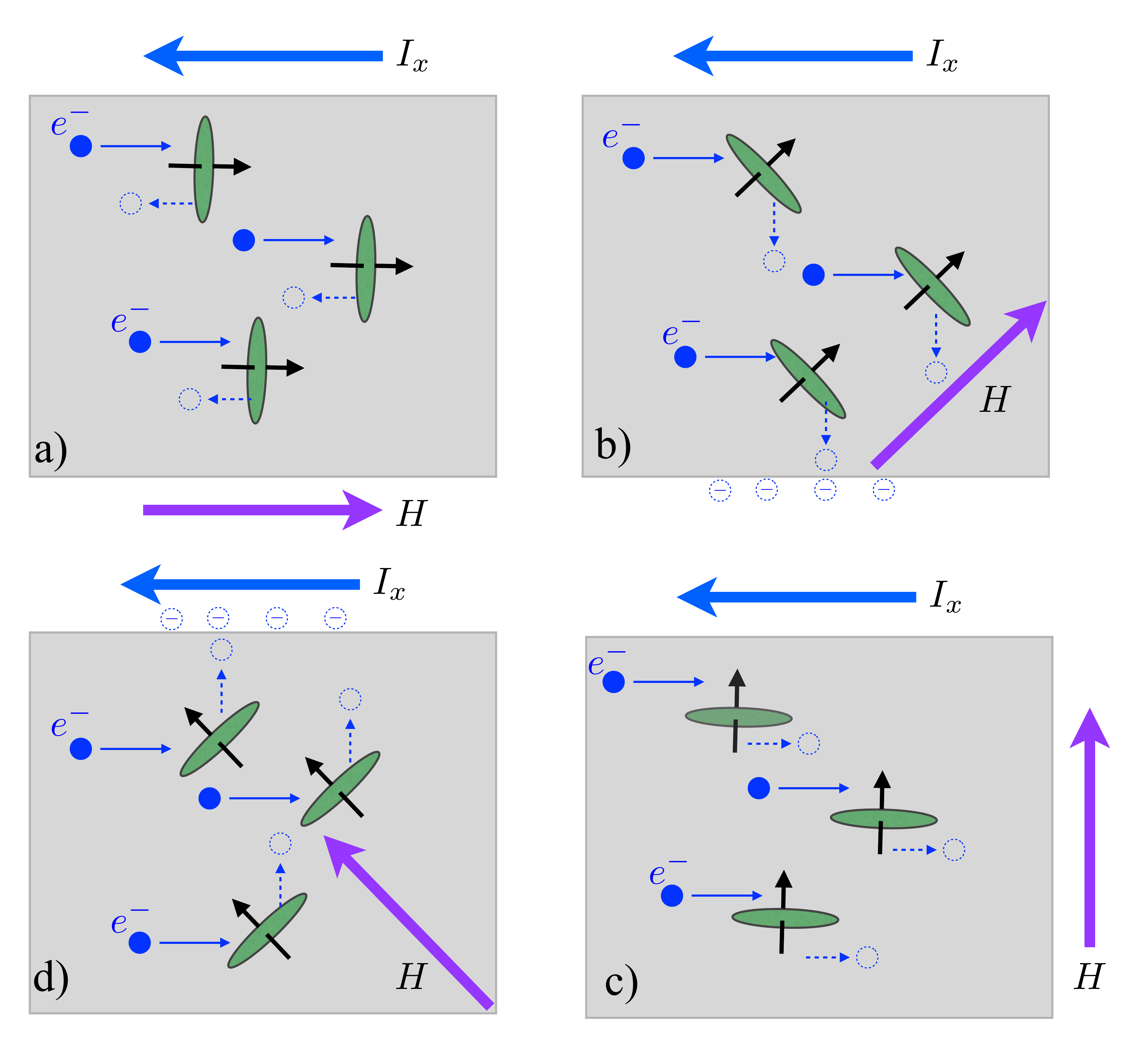}
\caption{\label{FigLSscheme} Simplified physical origin of the $H$ direction dependence of the AMR, MTEP, PHE and PNE in ferromagnetic conductors based on a simple view of scattering of conduction electrons  (blue) from the charge density of orbitals with $L\neq0$ (green disk). \textbf{a)} The highest $R$ state occurs with $I_{\mathrm{x}} \parallel H$. \textbf{b)} with $\theta=45^{o}$ or $225^{o}$, an intermediate $R$ state with a transverse $V$ appears. \textbf{c)} When $H\perp I_{\mathrm{x}}$, the lowest $R$ state occurs. \textbf{d)} for $\theta=135^{o}$ or $315^{o}$, the intermediate $R$ state shows the opposite sign of the transverse $V$.  }
\end{figure}

Ferromagnets also support a third class of effects, where transverse voltage measurements depend on the direction of the in-plane magnetization controlled by a rotating external field.  These are the planar Hall effect (PHE),\cite{ElzwawyJPhysD2021} and planar Nernst effect (PNE)\cite{KyPSSb1967,KyPSSb1966,PuPRL2006,AveryPRL2012,SchmidPRL2013,BrandlAPL2014,ReimerSciRep2017}.  
These effects are driven mainly by spin-orbit coupling that introduces field-direction dependence to the motion of charge carriers.  We can gain some intuitive understanding of the expected symmetry of the AMR, MTEP, and PHE and PNE with an extremely simple view of the spin-orbit scattering as outlined in Fig.\ \ref{FigLSscheme}.  Each panel shows a schematic view of a ferromagnetic conductor, where electrons in an applied current travel on average from left to right (such that the current $I_{\mathrm{x}}$ flows to the left). In each panel, I indicate the direction of applied field, $H$ (assumed large enough to fully magnetize the sample in the indicated direction), and show a number of vectors indicating the alignment of the atomic spin moments (which form the net magnetization), each with a green disk that represents the electron density associated with the orbital quantum number.  One can imagine that scattering of the conduction electrons from this orbital electron distribution will depend on the orientation of this cloud of probability.  In Fig.\ \ref{FigLSscheme}a), the magnetization, $M$, is parallel to $I_{\mathrm{x}}$, forcing the orbital cloud to present a large scattering cross-section to the conduction electrons.  The resulting back scattering would raise the resistance, $R$, of the ferromagnet.  Note that reversing $M$ (by rotating $H$ 180 degrees) leads to the same cross-section, arguing that the resistance is the same in this case.  A change in $R$ occurs when $M$ rotates between these directions, with the lowest scattering cross-section (and lowest $R$) expected when $M\perp I_{\mathrm{x}}$, as shown in panel c).  This simple cartoon also suggests that when $M$ is aligned at $\theta=45^{o}$ (or $225^{o}$), as seen in panel b) that electrons are preferentially scattered to one side of the sample, such that a transverse voltage appears.  The sign of this voltage will reverse when $\theta=135^{o}$ (or $315^{o}$).  Similar patterns of course emerge if the charge motion originates from thermal diffusion.  

This picture also clarifies that the observation of AMR (and MTEP) depends on the orientation of the local magnetization in a given FM film or nanostructure with the overall charge current direction.  Since the magnetization reversal process involves a potentially complicated balance of magnetic anisotropy and domain wall propagation energies, the resulting alignment of magnetic domains with respect to current direction can be difficult to predict.  It is possible, for example, to observe magnetic reversal in narrow wires that occurs entirely by domain wall propagation.  In this case, the large shape anisotropy of the wire prevents the magnetic domains from rotating away from the current direction, and no AMR is observed even in a FM that shows a large effect in other geometries (though the domain wall itself can contribute).\cite{MarrowsAdvPhys2005,KentJPhysCM2001}  This suggests a concomitant lack of MTEP, PHE, and PNE.  However, in a film or device that \emph{does} experimentally demonstrate AMR, the MTEP, PHE, and PNE should also be present.   

It is common to quantify AMR using the AMR ratio defined by
\begin{equation}
\frac{\Delta \rho}{\rho_{\mathrm{av}}}=\frac{\rho_{\mathrm{\parallel}}-\rho_{\mathrm{\perp}}}{\frac{1}{3}\rho_{\mathrm{\parallel}}+\frac{2}{3}\rho_{\mathrm{\perp}}},
\label{AMRratio}
\end{equation}
where $\rho_{\mathrm{\parallel}}$ and $\rho_{\mathrm{\perp}}$ are the charge resistivity measured with magnetization parallel and perpendicular to $I$, respectively.  For transition metal ferromagnets this ratio ranges from $<1\%$ to a few $\%$.\cite{McGuireIEEE75}  The symmetry with applied field direction $\theta$ outlined above is expressed
\begin{equation}
\rho({\theta})=\rho_{\mathrm{\perp}} + \Delta \rho \cos^{2}(\theta).
\label{RhoTheta}
\end{equation} 
The planar Hall effect is then 
\begin{equation}
	\label{PHE}
	{\rho}_{\mathrm{PHE}}(\theta)=\frac{1}{2}[\rho_{\mathrm{\parallel}}-\rho_{\mathrm{\perp}}]\sin{2\theta}.
	\end{equation}
The transverse electric field is $E_{\mathrm{y,PHE}}=\rho_{\mathrm{PHE}}(\theta) I/(t\cdot w)$,  with the sample thickness $t$ and width in the transverse direction $w$ defining the cross-sectional area.  The
transverse PHE voltage is then
\begin{equation}
	\label{VtPHE}
	V_{\mathrm{T,PHE}} = \rho_{\mathrm{PHE}}(\theta)\frac{I}{\ell} w.
	\end{equation}
For the thermal analogs, one can define a MTEP ratio
\begin{equation}
\frac{\Delta \alpha}{\alpha_{\mathrm{av}}}=\frac{\alpha_{\mathrm{\parallel}}-\alpha_{\mathrm{\perp}}}{\frac{1}{3}\alpha_{\mathrm{\parallel}}+\frac{2}{3}\alpha_{\mathrm{\perp}}},
\label{MTEPratio}
\end{equation}
where $\alpha_{\mathrm{\parallel}}$ and $\alpha_{\mathrm{\perp}}$ are the longitudinal Seebeck coefficients measured with magnetization parallel and perpendicular to $\nabla T$, respectively.  
The planar Nernst effect is then 
\begin{equation}
	\label{PNE}
	\alpha_{\mathrm{PNE}}(\theta) = \frac{1}{2} [\alpha_{\mathrm{\parallel}} - \alpha_{\mathrm{\perp}}]  \sin{2\theta}.
	\end{equation}
The transverse electric field generated is then $E_{\mathrm{y,PNE}}=\alpha_{\mathrm{PNE}}(\theta)\partial T/\partial x$ and again if the thermal gradient is uniform the transverse PNE voltage is
\begin{equation}
	\label{VtPNE}
	V_{\mathrm{T,PNE}} = \alpha_{\mathrm{PNE}(\theta)}\frac{\Delta T}{\ell} w,
	\end{equation}
where $w$ is the width of the sample in the transverse direction.  The expected size of the PNE voltage can therefore be determined from longitudinal Seebeck coefficient measurements in the two fully in-plane magnetized directions.  This has been experimentally demonstrated.\cite{AveryPRB2012,Kimling-GoothPRB2013,WesenbergJPhysD2018} 

\begin{figure}
\includegraphics[width=3.38in]{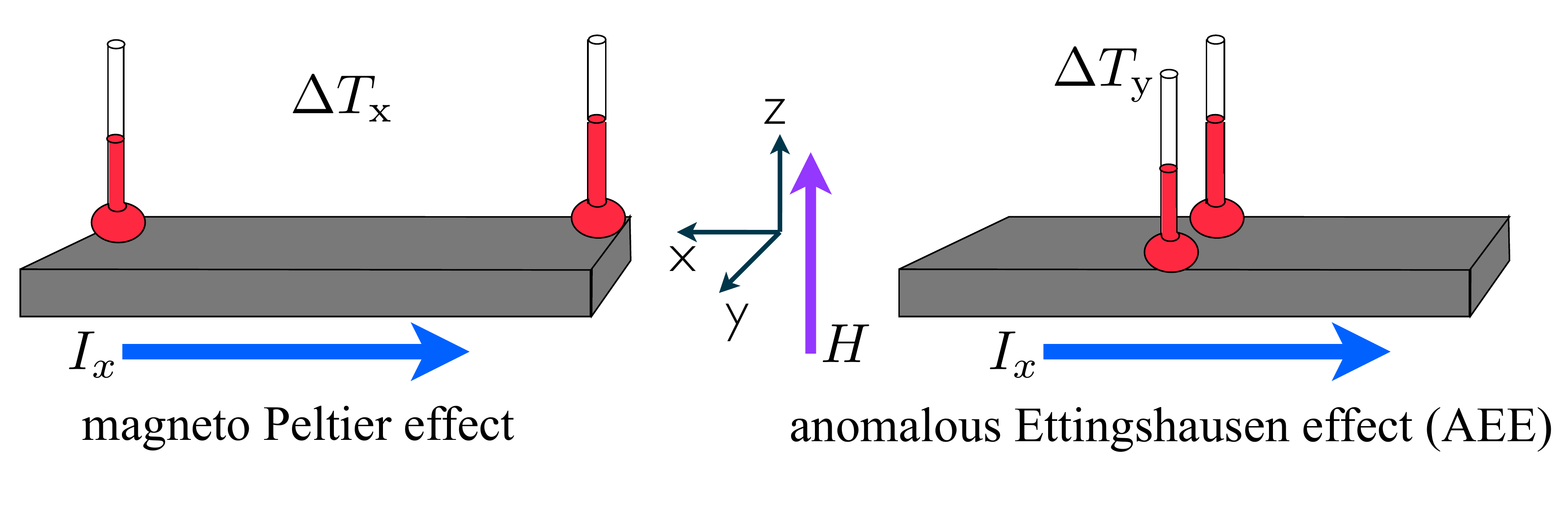}
\caption{\label{FMmetAdd} In FM conductors, \textbf{a)} the Peltier effect becomes field-dependent, as expected based on applying Onsager reciprocity to the MTEP.  \textbf{b)} A transverse thermal gradient is also possible when charge current is applied, termed the anomalous Ettingshausen effect (AEE).  A similar effect with thermal gradient along $\hat{x}$ is the Righi-Leduc, or thermal Hall effect (not shown schematically). }
\end{figure}

Finally, ferromagnetic systems also show effects where charge flow causes heat flow that leads to temperature differences.  This includes a magnetization-dependent Peltier effects (or magneto Peltier effect), and the time reverse symmetry conjugate of the ANE, the anomalous Ettingshausen effect (AEE), both shown in Fig.\ \ref{FMmetAdd}.  As with non-magnetic metals, Onsager reciprocity predicts a strong link between the Seebeck and Peltier coefficients and argues for clearly measurable Peltier cooling, and this has been demonstrated in ferromagnetic films\cite{AveryPRL2013} and nanostructures.\cite{GravierPRB06b,FukushimaJAP06,ShanPRB2015}  The AEE requires even more challenging measurements of transverse temperature differences generated on a sample, which adds more extreme requirements to locally resolve the temperature to the usual difficulties of thermal and transverse measurements.  Nevertheless, impressive optical techniques have recently been developed to demonstrate the AEE in a range of systems.\cite{SekiJPhysD2018,SekiAPL2018}
Though I have not presented a schematic view, it is also possible in some cases to observe a transverse temperature gradient when a longitudinal thermal gradient is applied to a sample.  This is referred to as the Righi-Leduc effect,\cite{MadonPRB2016} or thermal Hall effect.\cite{ChenPRL2016}  Measurements to clearly quantify these are often challenging, as they combine the difficulty of measuring transverse $\Delta T$ with the difficulty in controlling thermal gradients.   

\subsection{Spin Hall and Nernst Effects}
\begin{figure}
\includegraphics[width=3.38in]{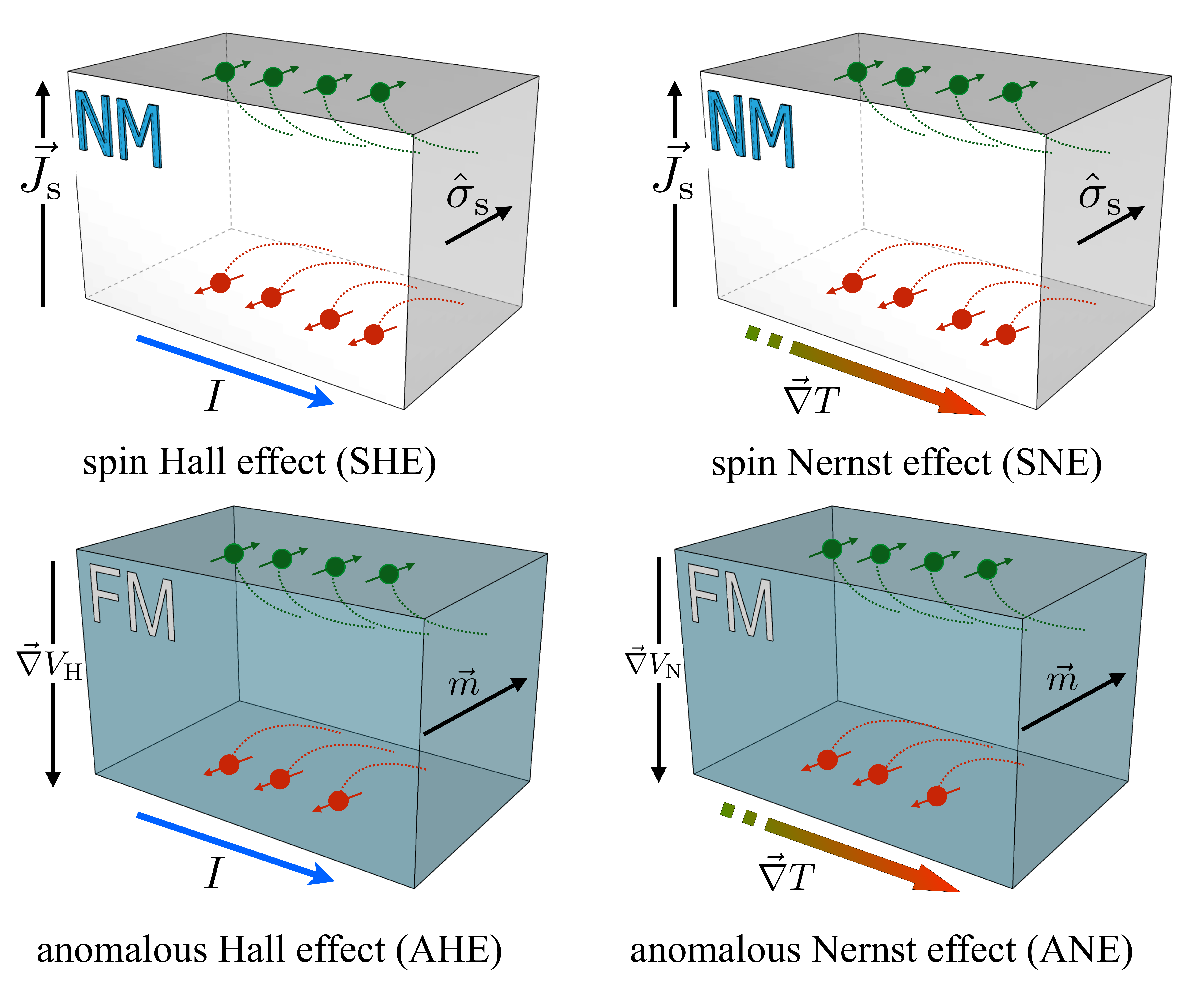}
\caption{\label{SHEetc}Schematic views of spin Hall, spin Nernst, anomalous Hall, and anomalous Nernst effects clarifies the similar physical origin of each. }
\end{figure}

The spin Hall effect (SHE) (and its time-reversed inverse effect the ISHE) play a huge role in current spintronics and spin caloritronics research.\cite{SinovaRMP2015,HoffmannIEEETransMag2013}  As shown schematically in Fig.\ \ref{SHEetc}, when a charge current flows in a non-magnetic conductor, electrons of opposite spin experience opposite transverse forces that drive them toward opposite faces of a narrow wire.  These forces can arise either from intrinsic band structure effects that lead to spin-dependent transverse velocities, or extrinsic effects introduced by spin-dependent (Mott) defect scattering.  The result is a transverse movement of spin angular momentum that is not accompanied by net charge motion, which defines a pure spin current. In the ISHE, applying a spin current to the conductor again causes the two spin species of electron to feel opposite forces, though since these are traveling in opposite directions in the pure spin current, the net result is deflection toward a common face of the wire, generating a charge voltage.  This makes the ISHE a very valuable tool for converting pure spin currents, which are otherwise extremely difficult to detect, to charge voltage that is very easy to measure. The physical origin of both the intrinsic and extrinsic effects typically involves spin orbit coupling, which in the simplest models depends strongly on the mass of the atoms forming the solid.  As a result, the SHE is commonly seen in heavy non-magnetic metals such as platinum and tungsten.  These two materials also happen to have opposite sign of the spin Hall angle, $\theta_{\mathrm{SH}}$ that quantifies the ratio between the transverse spin current and the longitudinal charge current.  Typical magnitudes of $\theta_{\mathrm{SH}}$ realized in experiments are often $\sim 0.1$ or less.  Despite this relatively small number, enough transverse angular momentum can be generated to manipulate nanomagnetic elements without external applied magnetic field, though this often requires high charge current densities to be applied.   As explained at the outset of this paper, this raises understandable concerns for thermal effects.  

In Fig.\ \ref{SHEetc} I show very similar schematic views of transverse motion of charge carriers for FM conductors, which in this scheme differ from the non-magnetic counterpart by possessing a net spin polarization (here there are more green ``right" spins than red ``left" spins.  The same intrinsic and extrinsic sources of transverse electron velocity now generate both a spin current and a net charge voltage.  This is a helpful intuitive picture of how the voltage arises in the anomalous Hall effect, which also clarifies that the AHE should also be expected to generate a spin current. For this and other reasons, ferromagnets are now also explored and used as sources of spin current.\cite{DavidsonPLA2020}  The physics driving these spin-dependent transverse electron velocities is similar when the motion is induced by thermal gradient instead of applied electric field, as in the ANE.  

By the same analogy, one can predict that a thermal gradient applied to a non-magnetic conductor should generate a pure spin current.  This effect, which was theoretically predicted some time ago,\cite{ChengPRB2008,LiuSSC2010,TauberPRL2012,WimmerPRB2013} is much more difficult to quantify than the SHE due to the additional difficulties that arise from controlling thermal gradients on thin films and nanostructures.  Nevertheless, several groups have now reported experiments on the SNE,\cite{MeyerNatMater2017,ShengSciAdv2017,BoseAPL2018}  though some debate about methods continue.  

\subsection{Spin Caloritronic Effects: Interfaces and magnons}

\begin{figure}[t]
\includegraphics[width=3.38in]{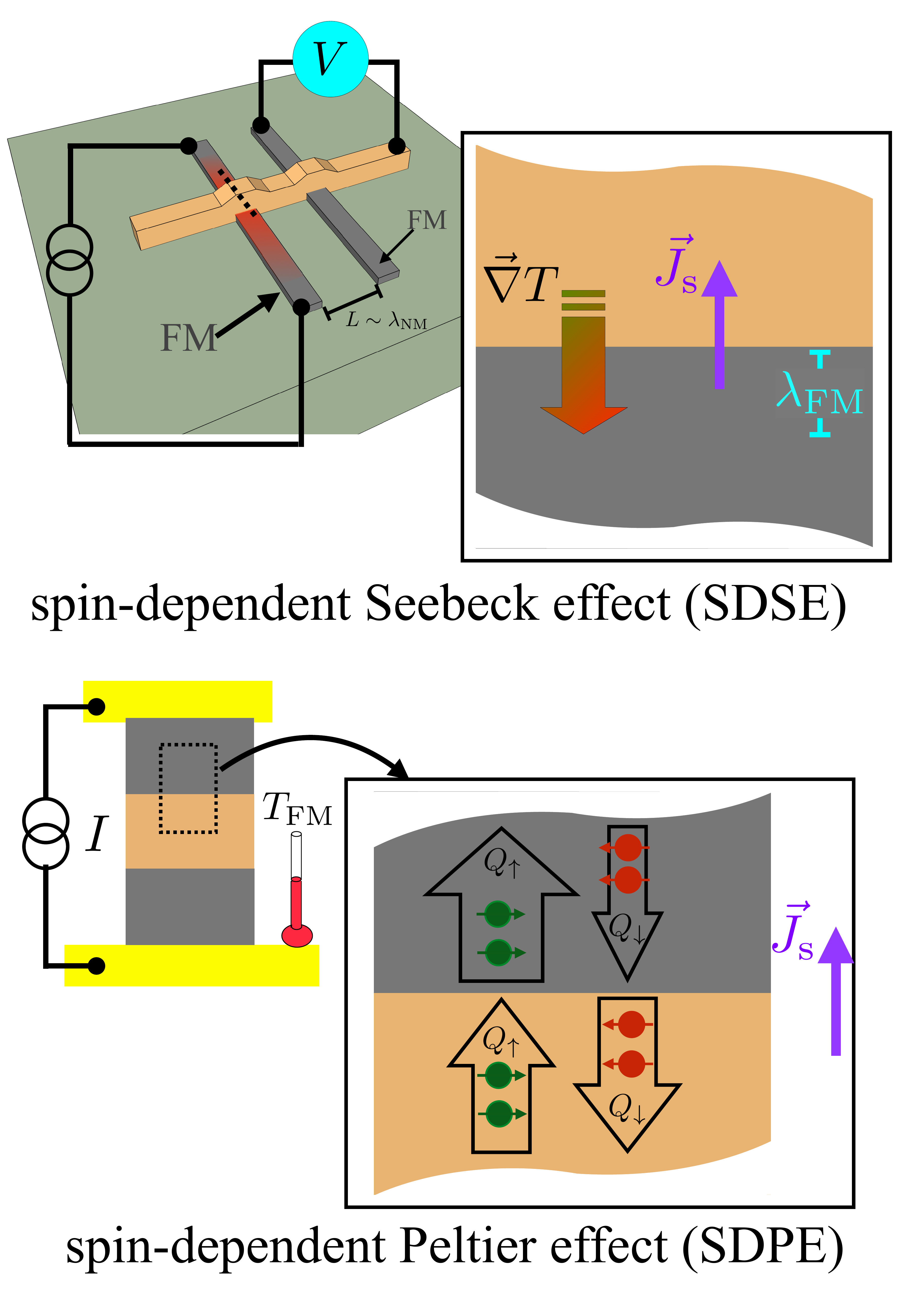}
\caption{\label{ZooFigA} Spin caloritronic effects in metallic systems.  \emph{Top}: In the spin dependent Seebeck effect (SDSE), a thermal gradient applied within the spin diffusion length of a NM/FM interface thermally injects spin into the NM.  \emph{Bottom}: In the spin dependent Peltier effect (SDPE), the spin polarized electron system carries different heat in each branch in the FM, leading to a temperature difference induced across the NM/FM junction. }
\end{figure}

\begin{figure}[t]
\includegraphics[width=3.38in]{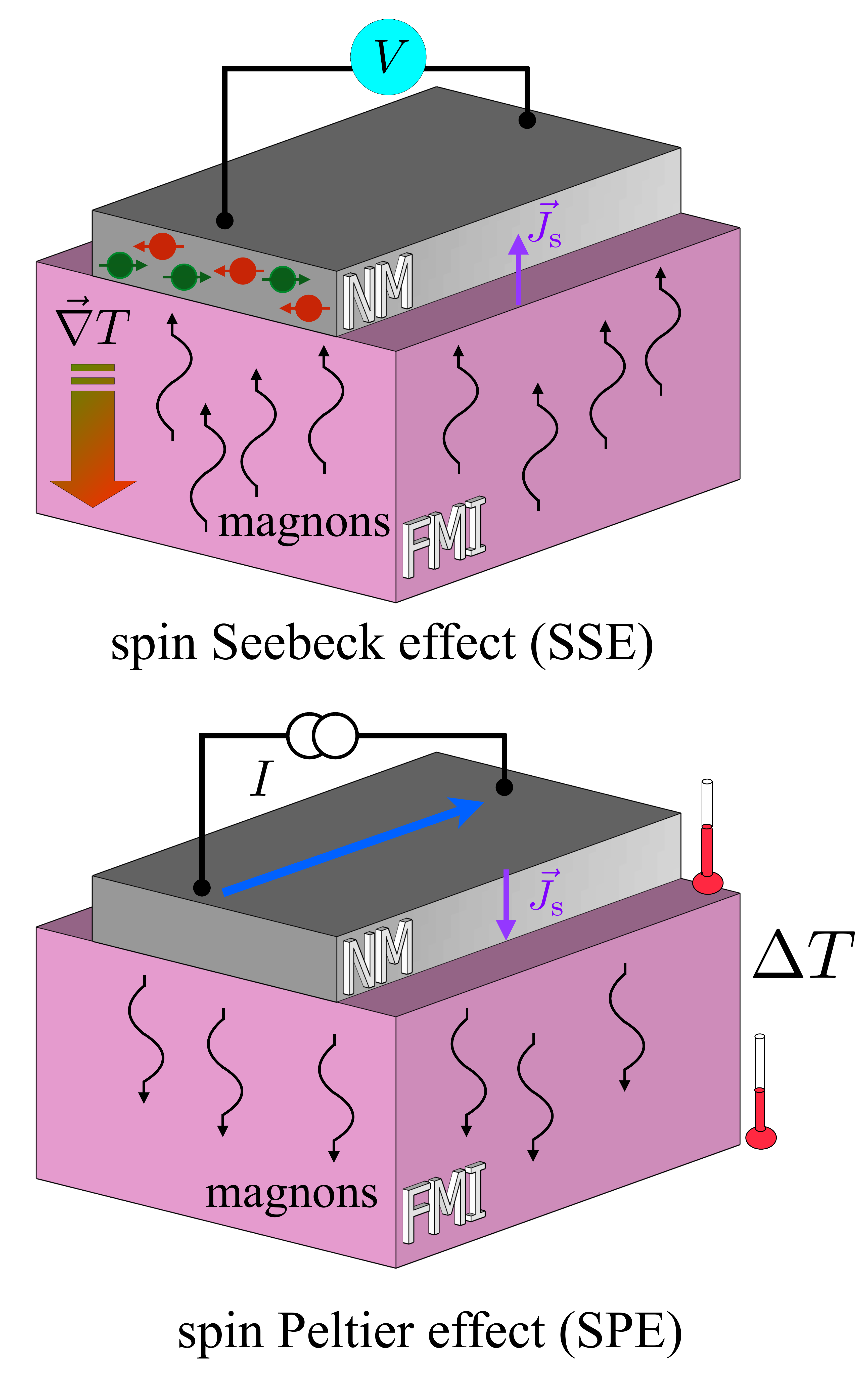}
\caption{\label{ZooFigB} Spin caloritronic effects in hybrid systems where a FM insulator (FMI) is coupled to a SOC metal.  \textbf{a)} the longitudinal spin Seebeck effect, where applied thermal gradient drives an incoherent flow of magnons in the FMI, which flows across the interface as a pure spin current which is converted to a measurable charge voltage in the metal film.  \textbf{b)} In the reciprocal effect, a charge current in the metal drives a transverse spin current into the FMI, causing a measurable temperature difference. }
\end{figure}

In the last $\sim15$ years, studies of the interplay between heat, charge, and spin degrees of freedom in magnetic materials and devices have added several important new effects to this thermoelectric ``zoo."  In Figs.\ \ref{ZooFigA} and \ref{ZooFigB} I graphically summarize the central effects in what has become known as spin caloritronics.\cite{BackJPhysD2019,YuPLA2017,BoonaEES2014,BauerNatMat2012,BauerSSC10}  
One of the original concepts that drove interest in spin caloritronics is the idea that the spin up and spin down electrons in a ferromagnetic metal could have different Seebeck coefficients, such that applying a thermal gradient to the FM metal could generate a difference in spin potential that could provide a source of pure spin currents.  This effect can be realized, but it has now been conclusively shown that the spin separation can only exist on a quite short length scale comparable to the spin diffusion length\cite{BassJPCM2007} in the metallic ferromagnet, $\lambda_{\mathrm{FM}}$.  This effect, which is typically observed in metallic non local spin valves, is called the spin dependent Seebeck effect (SDSE).\cite{SlachterNatPhys2010,ErekhinskyAPL2012,HuNPGAM2014,HuPRB2014,YamasakiAPE2015,PfeifferAPL2015,ChoiNatPhys2015,HojemPRB2016,HuPRB2017} 
As I show at the top of Fig.\ \ref{ZooFigA}, the non local spin valve (NLSV) consists of two FM nanowires bridged by a non magnetic conductor, with separation $L$ on the order of the spin diffusion length, $\lambda_{\mathrm{NM}}$ of the non-magnetic conductor.\cite{JohnsonPRL1993,JedemaNature2001,JiAPL2004,ValenzuelaAPL2004,NiimiPRL2013}  To observe the SDSE, one FM is heated by applying charge current.  As shown in the inset cross-sectional view of the interface between FM and NM, this generates a thermal gradient at this interface, which causes the spin separation due to the different effective $\alpha$ for spin up and spin down electrons.  The result is a spin current, $J_{\mathrm{s}}$, that flows into the NM channel.  As in typical use of the NLSV, one can detect the diffusion of this spin current by measuring the voltage between the channel and the second FM, which will depend on the relative magnetization of the two FM strips, since the role of spin up and down electrons reverse as the magnetizations are changed by in-plane fields.    

As with typical Seebeck effects, one can predict a related Peltier effect via Onsager reciprocity.  This is termed the spin dependent Peltier effect (SDPE) and is shown schematically at the bottom of Fig.\ \ref{ZooFigA}.  Here, a current is driven through a heterostructure containing interfaces between NM and FM layers (again with thickness comparable to appropriate spin diffusion length).  This current becomes spin polarized and contains a component of spin current $J_{\mathrm{s}}$ which flows across the NM/FM interface, the two spin channels carry different amounts of heat in the FM, as shown in the inset cross sectional view.  This generates a temperature difference across the interface.  This can be proven by measuring the temperature of one FM, $T_{\mathrm{FM}}$ relative to the overall base temperature, which again depends on the relative alignment of the FM layers.  Though challenging, this effect has also been experimentally demonstrated,\cite{FlipseNatNano2012}  with a subsequent explicit confirmation of the Onsager relation between the SDSE and SDPE.\cite{DejenePRB2014}

The most commonly and intensively studied of the spin caloritronic effects is now the longitudinal spin Seebeck effect (LSSE).\cite{UchidaAPL2010} In this effect, shown schematically at the top of Fig.\ \ref{ZooFigB}, an out-of-plane thermal gradient, $\nabla T$ is applied to a sample that consists of a ferromagnetic insulator (FMI) and a metallic thin film that supports the SHE (and ISHE).  In the original experiment, and indeed a large fraction of the following work in the field,\cite{UchidaJPCM2014,UchidaIEEE2016,KikkawaarXiv2022} these are yttrium iron garnet (YIG) and platinum (Pt), respectively.  The thermal gradient, which aligns well with the out-of-plane thermal gradients that must always be expected in thin films on bulk substrates as discussed extensively in Secs.\ \ref{BackHeat} and \ref{WithJoule}, drives an incoherent flow of thermal magnons toward the interface.  This magnon flow represents a flow of angular momentum, that can carry across the interface due to spin transfer torque, and flow into the metal, where it is converted into a measurable voltage via the ISHE.\cite{XiaoPRB2010,AdachiPRB2011,RezendePRB2014,RezendeJPhysD2018} YIG is technically neither a simple ferromagnet nor a simple insulator, but rather a ferrimagnet which can be viewed as a simple ferromagnet in many cases, and a semiconductor with band gap on order $2.8$ eV in bulk.\cite{SergaJPhysD2010}  In thin films, this band gap can be reduced.\cite{ThieryPRB2018charge}  Despite these complications, YIG is chosen most often due to its exceptionally low damping of magnetization dynamics, which allows long magnon propagation lengths which help lead to a robust a repeatable signal in LSSE experiments.   

The reader may have noticed that the naming conventions for effects in spin caloritronics can be complicated and seem somewhat arcane, as they are still influenced by the history of the early experiments that have either faded from memory or are not known to new researchers joining the field.  The term "longitudinal" in the LSSE is a prime example, as this does not sensibly align with the use of the term depicted in Figs.\ \ref{Fig4} and \ref{FMmet}.  This term was originally used to distinguish the experiment from the original geometry where the intended thermal gradient was in-plane, which was originally termed simply "spin Seebeck effect" but (as discussed briefly in Sec.\ \ref{noTSSE} below) actually probed ANE and other effects.  It is becoming more common to describe an experiment where a thermally driven magnon spin current is detected via the ISHE simply as the SSE, and this is a reasonable situation.  The main distinction that is important to keep clear is that electronic thermally-driven spin currents must exist inside a spin diffusion length of an interface, as discussed for the SDSE above.  If a metallic ferromagnet is used in the geometry of the LSSE, the signal cannot convincingly be distinguished from the anomalous Nernst effect, which has the same symmetry.  

As the LSSE requires the presence of an interface between the nominally insulating material with magnetic order and the metallic spin-to-charge conversion material, if the thermal gradient can be established parallel to this interface and the magnetization aligned out-of-plane, the mutually perpendicular voltage can be probed for signs of the ANE.  This type of arrangement has been used to put limits on the formation of a metallic layer at the YIG/Pt interface that could be driven ferromagnetic via a proximity effect.\cite{KikkawaPRL2013}  

The LSSE using YIG and Pt has been repeated experimentally by many groups around the world.  Typically the thermal gradient is generated using macroscopic heating or cooling blocks measured with thermometers that are also macroscopic.  This introduces important variations in the location and accuracy of the temperature measurement that makes quantitative comparison of results from one lab to another essentially impossible, since the effective thermal gradient in a given experiment varies widely from one apparatus to another.  This has been explicitly demonstrated by measuring one sample in a ``round robin" comparison in many different labs, and agreement between results was poor.\cite{SolaIEEEtim2019}   It is possible that modifications of the approach to emphasize control of the heat flux rather than thermal gradient could resolve this issue somewhat.\cite{SolaSciRep2017,BoonaJAP2021}
Other groups have demonstrated LSSE experiments where the thermal gradient is established using local heating of a microfabricated thin film structure,\cite{WuPRL2015,WuJAP2015} localized laser heating of a patterned Pt/FMI stack,\cite{WeilerPRL2012} or a microfabricated Pt structure in contact with the FMI.\cite{SchreierAPL2013,WangAPL2014,VlietstraPRB2014}  This approach has the potential advantage of a much more well-controlled and accurately determined temperature profile.  
Despite the challenge of quantifying the thermal gradient in a given LSSE experiment, the technique holds promise for probing a range of magnetic materials, and magnon-related phenomenon in YIG itself.  A key example of this is the shift of the magnon dispersion relation\cite{BarkerPRL2016} revealed by LSSE measurements in high magnetic fields.\cite{ItohPRB2017,KikkawaPRB2015,JinPRB2015,GuoPRX2016}  A second example are sharp features that arise in the SSE response that are driven by coupling between magnons and phonons.  These magnon-polarons can give information on the details of the heterostructures and their magnetization dynamics.\cite{KikkawaPRL2016} 

As with all previous thermoelectric or spin thermoelectric effects, Onsager reciprocity suggests the existence of a spin Peltier effect (SPE), as shown schematically at the bottom of Fig.\ \ref{ZooFigB}.   Here a charge current driven through the non-magnetic metal generates a spin current via the SHE, which excites a magnon spin current in the FMI, which in turn generates a temperature difference in the YIG.  Despite the much more challenging thermometry required, this effect has also been demonstrated by at least two groups,\cite{YagmurJPhysD2018,FlipsePRL2014} and also quantified via thermal imaging.\cite{DaimonNatComms2016} A high field suppression has been observed for the SPE and quantitatively compared to that seen in LSSE.\cite{ItohPRB2017}  As expected, magnon-polarons have also been observed in the SPE.\cite{YahiroPRB2020}

\section{Highlighting important examples and current areas of interest}

\subsection{Challenges in generating in-plane thermal gradients\label{noTSSE}} 

Throughout the development of spintronics and spin caloritronics, experiments or devices that envision in-plane thermal gradients applied to thin film elements supported on bulk substrates have occasionally been promoted. The original geometry proposed for thermal spin current generation is a key example.\cite{UchidaNature08,MyersNatMater2010,XiaoNatMater2010,JaworskiNature2012}  For all the reasons overviewed in the first sections of this paper make clear, ruling out a contribution from an out-of-plane thermal gradient in these experiments is all but impossible.  For systems where FM order coexists with charge conductivity, the out-of-plane gradient leads to contributions from the ANE that overwhelm the intended signals of spin effects, which have the exact same symmetry with field.\cite{HuangPRL2011}  An example of the size of such a background ANE signal is given in Tables \ref{TableStack}a) and b) in the column labeled $V_{\mathrm{ANE}}$, where assuming the metal film at the top of the heterostructure has the ANE coefficient reported by several groups for permalloy.  While this contribution will be shorted somewhat by a Pt strip deposited on top of the permalloy in some experiments, the size of the signal is large enough to explain most claimed spin transport results in these type of experiments.    

Ruling out the out-of-plane thermal gradient is possible when all bulk heat sinks are removed from a thin film structure, which is possible to achieve with micro- and nanofabricated thermal isolation structures.  When such techniques are employed, spin transport over distances greater than the spin diffusion length have been conclusively ruled out.\cite{AveryPRL2012,AveryPRB2012,SchmidPRL2013,WesenbergJPhysD2018,SrichandanJPhysD2018}  By now a large body of evidence has been assembled that clarifies the challenges and contaminating effects in what came to be known as the transverse Spin Seebeck effect.\cite{MeierPRB2013,vonBierenAPL13,YinPRB2013,WegrowePRB2014,SoldatovPRB2014,BrandlAPL2014,MeierNatCom2015,JayathilakaJMMM2015,ShestakovPRB2015,CaoAIPadv2016,ReimerSciRep2017,KimlingPRB2019}  Much of this work has led to the deeper understanding of thermal and thermoelectric effects in nanomagnetic systems that is communicated above.

\subsection{Thermally-generated voltages in nanoscale non local spin valves}

\begin{figure*}
\includegraphics[width=\linewidth]{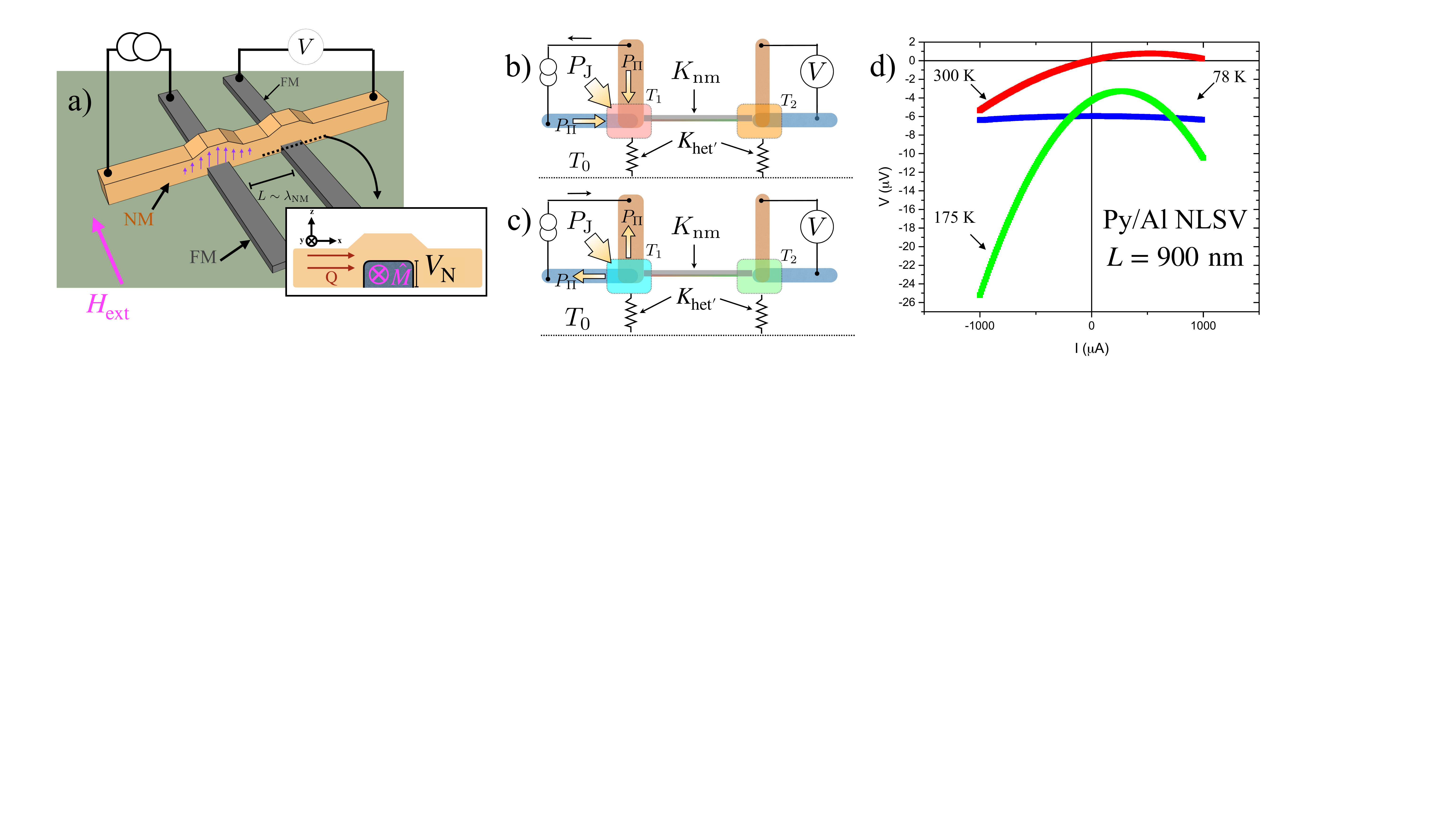}
\caption{\label{NLSVtherm} Overview of a metallic NLSV as an example of thermally driven effects in a spintronic device.  \textbf{a)} Schematic view of a NLSV where charge flow is used to form a spin accumulation, which is detected via a voltage $V$ at the distant contact that switches with external field.   \textbf{b-c)} A simplified thermal model of the NLSV.  The colors at the junctions indicate temperatures $T_{1}$ and $T_{2}$, which change for different polarity of the current.  \textbf{d)} Total measured voltage, $V$ vs. applied current $I$  for a Py/Al NLSV shows linear and quadratic background terms that arise from Seebeck and Peltier effects.  These have strong dependence on base temperature $T_{0}$. }
\end{figure*}

The metallic non local spin valve (NLSV), sometimes called a lateral spin valve or accumulation sensor, is a fundamentally interesting and potentially technologically important spintronic device.  This nanoscale circuit, where two FM nanowires are bridged by a non magnetic metal link, is not only useful for measurements of the SDSE and SDPE as discussed earlier (see Fig.\ \ref{ZooFigA}), but when wired as shown in Fig.\ \ref{NLSVtherm}a) allows a unique separation of spin and charge currents.\cite{JohnsonPRL1993,JedemaNature2001,JiAPL2004,ValenzuelaAPL2004,NiimiPRL2013}  The low resistance and small area with significant magnetoresistance suggests a potential role in magnetic recording, and the NLSV has been very actively explored for applications as read-heads.   \cite{NakataniMRSBull2018,TakahashiAPL2012,YamadaIEEETransMag2013,VedyayevPRApp2018}  The NLSV also has proven to allow a uniquely powerful probe of the electron scattering events that control the spin diffusion length in metals.\cite{WrightPRB2021,WattsPRM2019,WattsPRL2022,OBrienPRB2016,ObrienNatComm2014}  With relatively simple modifications, the NLSV has also been used to measure the spin Hall angle in a technique called spin absorption.\cite{WakamuraNatMater2015,NiimiPRB2014,NiimiPRL2012,KimuraPRL2007,IsshikiPRM2022} These devices also provide an excellent example of thermal  effects contributing new spin effects, the SDSE and SDPE discussed above, and also a range of important thermoelectric background contributions.\cite{BakkerPRL2010,SlachterPRB2011,CasanovaPRB2009,HuPRB2013,KasaiAPL2014}

Fig.\ \ref{NLSVtherm} summarizes the key thermoelectric signals that contribute background effects that often must be carefully removed from the total measured signal to isolate spin effects that are typically of more direct interest.   Panel \textbf{a)} shows the typical connections used for the NLSV, where charge current flows from one FM contact into the NM and is extracted at the end of the NM nanowire most distant from the second FM.  The voltage is measured across this second FM.  The injected current is partially spin polarized by the FM, causing a spin accumulation (shown schematically as purple arrows).  This spin diffuses in the NM and the resulting spin current flow is detected using the magnetic field dependence of the measured voltage, $V$.\cite{JohnsonPRL1993,JedemaNature2001,JiAPL2004,ValenzuelaAPL2004,NiimiPRL2013}  The charge current applied to achieve measurable spin accumulation is often in the range of milliamps, and as stated in the introduction, this leads to large heating and significant thermal gradients.  One major consequence was referenced in our discussion of the ANE and is highlighted in the inset of Fig.\ \ref{NLSVtherm}a.  Here the heat dissipated at the FM/NM junction at left causes heat, $Q$ to flow down the NM nanowire in the plane of this film.  The inset shows that at the junction of the NM nanowire with the detecting FM, the in-plane $Q$ and the in-plane but perpendicularly aligned magnetization $\vec{M}$ generate an ANE voltage in the out-of-plane direction.  This voltage can be detected since the NM and FM cross at the junction.  This effect adds a voltage signal proportional to the hysteresis of the FM (odd in field), and has been demonstrated many times and used to measure the ANE in several FM nanowires.\cite{SlachterPRB2011,HuPRB2013,HuPRB2017,BennetPRB2019,BennetAIPadv2020}

In addition to this field-dependent background effect, there are (in the simplest view) field-independent background contributions to the total measured voltage that arise from interaction of the induced thermal gradients and the Seebeck and Peltier effects. Figs.\ \ref{NLSVtherm}b and c show simple schematics that demonstrate these effects, based on a simple analytic thermal model of the NLSV response discussed in more detail elsewhere.\cite{HojemPRB2016}  The current flowing from FM to NM across the nanowire junction at left dissipates both Joule heat $P_{\mathrm{J}}$, but also Peltier heat $P_{\Pi}$.  The presence of the Peltier component, which would cancel if the nanowires carrying the current to and from the junction are truly identical and formed from identical materials, is proportional to $I$ such that the steady state temperature at this junction, $T_{1}$, is different when the current is positive or negative, as shown schematically in Figs.\ \ref{NLSVtherm}c.  The modification of $T_{1}$ subsequently modifies the temperature at the distant junction, $T_{2}$, since heat flows down the NM nanowire with conductance $K_{\mathrm{nm}}$.  In a real device the heat flow is potentially complicated.  Here I note one can use a similar approach to that introduced in Section\ \ref{WithJoule} above, where the heat flows in parallel through $K_{\mathrm{nm}}$ and the supporting heterostructure, $K_{\mathrm{het\prime}}=1/W_{\mathrm{het\prime}}$.  The resulting difference between the junction at $T_{2}$ and the ends of the measurement wires which must be at equilibrium at $T_{0}$ generates a voltage: $V=S_{\mathrm{rel}}(T_{2}-T_{0})$, where $S_{\mathrm{rel}}$ is the difference of the absolute Seebeck coefficients of the FM and NM wires.  

It is important to understand that $T_{2}$ has purely thermal contributions that are linear with current due to the Peltier effect at the injector junction.  This means that thermal effects \emph{cannot} be removed from spin effects in these devices simply by assuming that all thermal effects should be proportional to $I^{2}$ (or as is often done in practice, by assuming that the $2f$ signal in an ac experiment with a lock-in amplifier at frequency $f$ is the complete thermal response).\cite{IsshikiPRM2022}  Fig.\ \ref{NLSVtherm}d provides an example of the total measured non local voltage $V$ plotted vs applied current $I$ for a NLSV with separation $L=900\ \mathrm{nm}$ formed from Py and Al nanowires (this NLSV is described in more detail elsewhere\cite{HojemPRB2016,HojemThesis}) that clearly demonstrates the importance of the Peltier background components.  At higher temperatures where $S_{\mathrm{rel}}$ remains large, $V$ shows obvious $I^{2}$ contributions added to a large linear background.  The spin signal determined from the magnetic field dependence is more than an order of magnitude smaller than the Peltier-driven linear background.         

In more complicated or unique situations, additional thermal effects have also been discussed in NLSVs, and one can imagine situations where nearly all the thermoelectric and thermomagnetic effects overviewed above could arise in this key spintronic system.  
Though I have focused this discussion on metallic NLSVs, similar physics can arises in systems using semiconducting spin channels.\cite{YamashitaPRApp2018}

\subsection{Thermally-assisted spin transport in tunnel junctions and STT-RAM}

Magnetic tunnel junctions (MTJ) are a very important part of many current and planned spintronic technologies, including spin transfer torque random access memories (STT-RAM) that have been commercialized and provide fast, non-volatile memory for key applications.  Thermal effects in magnetic tunnel junctions are important and have been studied extensively.  I refer interested readers to recent reviews of the tunnel magneto-Seebeck effect (TMS),\cite{KuschelJPhysD2019}, as well as key results demonstrating thermal spin injection,\cite{LeBretonNature2011}, demonstration of the anomalous Nernst\cite{MartensCommPhys2018} and magneto-Peltier effects in MTJs.\cite{ShanPRB2015}   An impressive recent effort led to a very unique measurement of the thermal conductivity of the material forming the oxide tunnel barrier in an MTJ, and elucidated the effect on the thermal profile in the nanoscale device.\cite{JangPRA2020} 
This thermal profile is important for device functionality, as elevated temperatures assist magnetic switching in devices such as MTJ and STT-RAM.\cite{HadamekSSE2022,TaniguchiPRB2011,PrejbeanuJPCM2007}

\subsection{Thermal effects on spin beyond ferromagnets}

Interest in spintronics using materials with forms of magnetic order beyond ferromagnetism is currently growing rapidly.  This includes very active work on antiferromagnets\cite{JungfleischPhysLettA2018,BaltzRMP2018,JungwirthNatNano2016,MartiIEEETransMag2015} (both metals and insulators), ferrimagnets,\cite{KimNatMater2022} and materials with non-trivial topology\cite{HeNatMater2022,SmejkalNatPhys2018} including Weyl semimetals\cite{SakaiNatPhys2018,ChenNatComms2021,LeivaPRM2022}.  Thermal effects play a large role in a wide range of the work on these materials.  As this area of spintronics continues to evolve, I can only point out a few important examples where thermal effects are important.  

Antiferromagnetic systems, despite the lack of a net magnetization in the simplest picture, have shown a range of the phenomenon described here, including AMR\cite{SiddiquiJAP2020}, the spin Seebeck effect (where the antiferromagnetic magnons themselves are heated), \cite{SekiPRL2015,WuPRL2016,RezendePRB2016b,ReitzPRB2020} thermally-driven spin transport (where magnons generated via heating a FM flow through the AFM),\cite{LinPRL2016,CramerJPhysD2018} heat driven spin torque,\cite{BialekJPhysD2018}, and giant ANE.\cite{IkhlasNatPhys2017,LiPRL2017}   The question of if and how an antiferromagnetic element can be switched purely with applied spin torque induced by a charge current is an area of antiferromagnetic spintronics where thermal effects have recently proven to be particularly important.  Following on a large body of work on spin orbit torque (SOT) switching of ferromagnetic nanomagnets,\cite{ShaoIEEETransMag2021,ManchonRMP2019,RamaswamyAPR2018} several research groups performed similar experiments on either metallic antiferromagnets,\cite{WadleyScience2016,OlejnikNatComms2017} or bilayers of heavy metals and antiferromagnetic insulators.\cite{ChenPRL2018,MoriyamaSciRep2018,BaldratiPRL2019}  In these latter experiments, a large charge current was applied to a patterned thin film Pt layer deposited on a NiO thin film supported on a bulk substrate, a situation similar to the two modeled above.  A resulting pattern of transverse voltages was associated with switching of the antiferromagnet.  Chiang and co-workers subsequently made a dramatic observation, that a similar voltage pattern arises even when the Pt thin film was deposited on an entirely non-magnetic glass substrate.\cite{ChiangPRL2019}  After a few years of intense investigation including a range of imaging techniques focused on the AFM domain state,\cite{GrzybowskiPRL2017,WadleyNatNano2018,BodnarPRB2019,MoriyamaSciRep2018,BaldratiPRL2019,GrayPRX2019} and examinations of other AFM materials,\cite{ZhangPRL2019,CoguluPRL2022} the emerging consensus is that AFM domains can be manipulated in this way, but that thermal gradients and coupled magnetostriction play a dominant role.\cite{MeerNanoLetts2021}  This example serves to highlight the critical and enduring role of thermal effects in spintronic materials and devices.

\section{Final Remarks}

The current research landscape in spintronics and magnetism that is impacted by thermal effects is far too large for even the somewhat expansive scope of this review.  Important topics (of which I am aware) that I have not been able to cover include spin Hall magnetoresistance,\cite{MiaoPRL2014,AlthammerPRB2013,NakayamaPRL2013} 2D materials \cite{TianLiuPRB2020,SierraNatNano2017,VeraMarunNatComms2016}, the entire field of heat assisted magnetic recording (HAMR),\cite{KiefMRS2018,WellerIEEETransMag2014,XuJAP2012}, thermally driven neuromorphic computing devices,\cite{DelValleSciRep2020} magnon effects in metallic systems,\cite{LucassenAPL2011,WatzmanPRB2016,TserkovnyakPRB2016,FlebusEPL2016,NatalePRM2021} phase change memory,\cite{LeGalloJPhysD2020} novel use of ANE and SSE for imaging spintronic materials,\cite{GrayPRX2019,ZhangNanoLett2021} and the important role of thermally driven effects in non-local spin transport in magnetic insulators.\cite{AlthammerPSS2021}  I have also limited my consideration to essentially dc effects, and have ignored a large body of fascinating and important research on thermal effects on ultrafast timescales,\cite{ChoiNatComms2014,OstlerNatComms2012,KirilyukRMP2010} including important and illuminating work on the spin Seebeck effect on pico- and femtosecond timescales.\cite{KimlingPRL2017,SeifertNatComms2018}  I am certain I have missed other topics.  Despite these limitations, I hope that this review will provide useful context and tools to gain insight into the important and ongoing impact of thermal physics on spintronic materials and devices.

\section{Conclusion}

In summary, this paper has aimed first to provide a simple framework to understand and calculate the thermal gradients that arise in spintronic systems formed from thin films on bulk substrates.  These arise both when current is applied to a thin film structure and also when heat loss from the top surface of a thin film heterostructure introduces a thermal gradient without large applied currents.  Though an out-of-plane thermal gradient is present in nearly every case imaginable, the details of the thermal profile across a given heterostructure will depend significantly on the thermal conductivity of the constituent thin films and substrate and the thermal conductance of the interfaces between the various layers.  These parameters are often unknown, though they can be measured with modern experimental tools, and rarely match closely with bulk values of thermal conductivity.  Whether intentionally or unintentionally created, these gradients can drive a range of thermoelectric, magnetothermoelectric, and spin-based thermal effects.  The second section of this paper reviewed these in some detail, including anisotropic magnetoresistance, the Seebeck and Peltier effects, and their transverse counterparts.  I provided a physical picture to intuitively link AMR and the transverse effects to spin-orbit coupling.  I also presented the spin Hall effect and the principle spin caloritronic phenomena, including the spin Seebeck effect, the spin dependent Seebeck effect, and their Onsager reciprocal counterparts.  Finally, I provided several key examples that demostrate the importance of thermal gradients and also provide a link between generation of these gradients and the thermoelectric and spin thermoelectric ``zoo."  These included a discussion of the difficulty in generating in-plane gradients on films supported by bulk substrates.  This showed that the size of thermal gradients common in spintronic systems can easily generate artifact signals from, for one example, the anomalous Nernst effect.  The author hopes that this information not only is of use to those joining the field, but also focuses attention on efforts to better understand and thermally characterize the materials and systems used in spintronic systems both now and in coming years.       

\section{Acknowledgements}

I first thank the former and current Ph.D. students who were my partners and collaborators in learning so much of what I have shared here about thermal effects in magnetic systems, thin films, and nanostructures: Rubina Sultan, Azure Avery, Dain Bassett, Sarah Mason, Devin Wesenberg, Alex Hojem, Rachel Bennet, Mike Roos, Matt Natale, Sam Bleser, and Leo Hernandez.  I also thank my colleague Xin Fan for useful discussions, and gratefully acknowledge support from the NSF  (currently via DMR-2004646). 
Finally, I would like to especially thank Prof.\ Chia-Ling Chien, whose contributions this issue celebrates.  Chia-Ling has been both a personal inspiration, and a highly valued colleague and supporter over the years.  I am grateful for his continued leadership in the field. 


\begin{thebibliography}{274}
\expandafter\ifx\csname natexlab\endcsname\relax\def\natexlab#1{#1}\fi
\providecommand{\url}[1]{\texttt{#1}}
\providecommand{\href}[2]{#2}
\providecommand{\path}[1]{#1}
\providecommand{\DOIprefix}{doi:}
\providecommand{\ArXivprefix}{arXiv:}
\providecommand{\URLprefix}{URL: }
\providecommand{\Pubmedprefix}{pmid:}
\providecommand{\doi}[1]{\href{http://dx.doi.org/#1}{\path{#1}}}
\providecommand{\Pubmed}[1]{\href{pmid:#1}{\path{#1}}}
\providecommand{\bibinfo}[2]{#2}
\ifx\xfnm\relax \def\xfnm[#1]{\unskip,\space#1}\fi
\bibitem[{Hirohata et~al.(2020)Hirohata, Yamada, Nakatani, Prejbeanu, Diény,
  Pirro, and Hillebrands}]{HirohataJMMM2020}
\bibinfo{author}{A.~Hirohata}, \bibinfo{author}{K.~Yamada},
  \bibinfo{author}{Y.~Nakatani}, \bibinfo{author}{I.-L. Prejbeanu},
  \bibinfo{author}{B.~Diény}, \bibinfo{author}{P.~Pirro},
  \bibinfo{author}{B.~Hillebrands},
\newblock \bibinfo{title}{Review on spintronics: Principles and device
  applications},
\newblock \bibinfo{journal}{Journal of Magnetism and Magnetic Materials}
  \bibinfo{volume}{509} (\bibinfo{year}{2020}) \bibinfo{pages}{166711}.
  \DOIprefix\doi{https://doi.org/10.1016/j.jmmm.2020.166711}.
\bibitem[{Hoffmann and Bader(2015)}]{HoffmannPRAp2015}
\bibinfo{author}{A.~Hoffmann}, \bibinfo{author}{S.~D. Bader},
\newblock \bibinfo{title}{Opportunities at the frontiers of spintronics},
\newblock \bibinfo{journal}{Phys. Rev. Applied} \bibinfo{volume}{4}
  (\bibinfo{year}{2015}) \bibinfo{pages}{047001}.
  \DOIprefix\doi{10.1103/PhysRevApplied.4.047001}.
\bibitem[{Bader and Parkin(2010)}]{BaderAnnRevCMP2010}
\bibinfo{author}{S.~D. Bader}, \bibinfo{author}{S.~S.~P. Parkin},
\newblock \bibinfo{title}{Spintronics},
\newblock \bibinfo{journal}{Annual Review of Condensed Matter Physics}
  \bibinfo{volume}{1} (\bibinfo{year}{2010}) \bibinfo{pages}{71--88}.
\bibitem[{Zutic et~al.(2004)Zutic, Fabian, and Sarma}]{ZuticRMP04}
\bibinfo{author}{I.~Zutic}, \bibinfo{author}{J.~Fabian}, \bibinfo{author}{S.~D.
  Sarma},
\newblock \bibinfo{title}{Spintronics: Fundamentals and applications},
\newblock \bibinfo{journal}{Reviews of Modern Physics} \bibinfo{volume}{76}
  (\bibinfo{year}{2004}) \bibinfo{pages}{323}.
\bibitem[{Wolf et~al.(2001)Wolf, Awschalom, Buhrman, Daughton, von Molnar,
  Roukes, Chtchelkanova, and Treger}]{WolfScience01}
\bibinfo{author}{S.~A. Wolf}, \bibinfo{author}{D.~D. Awschalom},
  \bibinfo{author}{R.~A. Buhrman}, \bibinfo{author}{J.~M. Daughton},
  \bibinfo{author}{S.~von Molnar}, \bibinfo{author}{M.~L. Roukes},
  \bibinfo{author}{A.~Y. Chtchelkanova}, \bibinfo{author}{D.~M. Treger},
\newblock \bibinfo{title}{Spintronics: A spin-based electronics vision for the
  future},
\newblock \bibinfo{journal}{Science} \bibinfo{volume}{294}
  (\bibinfo{year}{2001}) \bibinfo{pages}{1488}.
\bibitem[{Parkin et~al.(1990)Parkin, More, and Roche}]{ParkinPRL1990}
\bibinfo{author}{S.~S.~P. Parkin}, \bibinfo{author}{N.~More},
  \bibinfo{author}{K.~P. Roche},
\newblock \bibinfo{title}{Oscillations in exchange coupling and
  magnetoresistance in metallic superlattice structures: {C}o/{R}u, {C}o/{C}r,
  and {F}e/{C}r},
\newblock \bibinfo{journal}{Phys. Rev. Lett.} \bibinfo{volume}{64}
  (\bibinfo{year}{1990}) \bibinfo{pages}{2304--2307}.
  \DOIprefix\doi{10.1103/PhysRevLett.64.2304}.
\bibitem[{Unguris et~al.(1991)Unguris, Celotta, and Pierce}]{UngurisPRL1991}
\bibinfo{author}{J.~Unguris}, \bibinfo{author}{R.~J. Celotta},
  \bibinfo{author}{D.~T. Pierce},
\newblock \bibinfo{title}{Observation of two different oscillation periods in
  the exchange coupling of {F}e/{C}r/{F}e(100)},
\newblock \bibinfo{journal}{Phys. Rev. Lett.} \bibinfo{volume}{67}
  (\bibinfo{year}{1991}) \bibinfo{pages}{140--143}.
  \DOIprefix\doi{10.1103/PhysRevLett.67.140}.
\bibitem[{Potter et~al.(1994)Potter, Schad, Beli\"en, Verbanck, Moshchalkov,
  Bruynseraede, Sch\"afer, Sch\"afer, and Gr\"unberg}]{PotterPRB1994}
\bibinfo{author}{C.~D. Potter}, \bibinfo{author}{R.~Schad},
  \bibinfo{author}{P.~Beli\"en}, \bibinfo{author}{G.~Verbanck},
  \bibinfo{author}{V.~V. Moshchalkov}, \bibinfo{author}{Y.~Bruynseraede},
  \bibinfo{author}{M.~Sch\"afer}, \bibinfo{author}{R.~Sch\"afer},
  \bibinfo{author}{P.~Gr\"unberg},
\newblock \bibinfo{title}{Two-monolayer-periodicity oscillations in the
  magnetoresistance of {F}e/{C}r/{F}e trilayers},
\newblock \bibinfo{journal}{Phys. Rev. B} \bibinfo{volume}{49}
  (\bibinfo{year}{1994}) \bibinfo{pages}{16055--16057}.
  \DOIprefix\doi{10.1103/PhysRevB.49.16055}.
\bibitem[{Nakatani et~al.(2018)Nakatani, Gao, and Hono}]{NakataniMRSBull2018}
\bibinfo{author}{T.~Nakatani}, \bibinfo{author}{Z.~Gao},
  \bibinfo{author}{K.~Hono},
\newblock \bibinfo{title}{Read sensor technology for ultrahigh density magnetic
  recording},
\newblock \bibinfo{journal}{{MRS} Bulletin} \bibinfo{volume}{43}
  (\bibinfo{year}{2018}) \bibinfo{pages}{106--111}.
  \DOIprefix\doi{10.1557/mrs.2018.3}.
\bibitem[{Chappert et~al.(2007)Chappert, Fert, and Dau}]{ChappertNatMater2007}
\bibinfo{author}{C.~Chappert}, \bibinfo{author}{A.~Fert},
  \bibinfo{author}{F.~N.~V. Dau},
\newblock \bibinfo{title}{The emergence of spin electronics in data storage},
\newblock \bibinfo{journal}{Nature Materials} \bibinfo{volume}{6}
  (\bibinfo{year}{2007}) \bibinfo{pages}{813--823}.
  \DOIprefix\doi{10.1038/nmat2024}.
\bibitem[{Parkin et~al.(2003)Parkin, Jiang, Kaiser, Panchula, Roche, and
  Samant}]{ParkinIEEE2003}
\bibinfo{author}{S.~Parkin}, \bibinfo{author}{X.~Jiang},
  \bibinfo{author}{C.~Kaiser}, \bibinfo{author}{A.~Panchula},
  \bibinfo{author}{K.~Roche}, \bibinfo{author}{M.~Samant},
\newblock \bibinfo{title}{Magnetically engineered spintronic sensors and
  memory},
\newblock \bibinfo{journal}{Proceedings of the {IEEE}} \bibinfo{volume}{91}
  (\bibinfo{year}{2003}) \bibinfo{pages}{661--680}.
  \DOIprefix\doi{10.1109/jproc.2003.811807}.
\bibitem[{Moser et~al.(2002)Moser, Takano, Margulies, Albrecht, Sonobe, Ikeda,
  Sun, and Fullerton}]{MoserJPhysD2002}
\bibinfo{author}{A.~Moser}, \bibinfo{author}{K.~Takano}, \bibinfo{author}{D.~T.
  Margulies}, \bibinfo{author}{M.~Albrecht}, \bibinfo{author}{Y.~Sonobe},
  \bibinfo{author}{Y.~Ikeda}, \bibinfo{author}{S.~Sun}, \bibinfo{author}{E.~E.
  Fullerton},
\newblock \bibinfo{title}{Magnetic recording: advancing into the future},
\newblock \bibinfo{journal}{Journal of Physics D: Applied Physics}
  \bibinfo{volume}{35} (\bibinfo{year}{2002}) \bibinfo{pages}{R157--R167}.
  \DOIprefix\doi{10.1088/0022-3727/35/19/201}.
\bibitem[{Ralph and Stiles(2008)}]{RalphJMMM2008}
\bibinfo{author}{D.~Ralph}, \bibinfo{author}{M.~Stiles},
\newblock \bibinfo{title}{Spin transfer torques},
\newblock \bibinfo{journal}{Journal of Magnetism and Magnetic Materials}
  \bibinfo{volume}{320} (\bibinfo{year}{2008}) \bibinfo{pages}{1190--1216}.
  \DOIprefix\doi{https://doi.org/10.1016/j.jmmm.2007.12.019}.
\bibitem[{Stiles and Zangwill(2002)}]{StilesPRB2002}
\bibinfo{author}{M.~D. Stiles}, \bibinfo{author}{A.~Zangwill},
\newblock \bibinfo{title}{Anatomy of spin-transfer torque},
\newblock \bibinfo{journal}{Physical Review B} \bibinfo{volume}{66}
  (\bibinfo{year}{2002}). \DOIprefix\doi{10.1103/physrevb.66.014407}.
\bibitem[{Berger(1996)}]{BergerPRB1996}
\bibinfo{author}{L.~Berger},
\newblock \bibinfo{title}{Emission of spin waves by a magnetic multilayer
  traversed by a current},
\newblock \bibinfo{journal}{Phys. Rev. B} \bibinfo{volume}{54}
  (\bibinfo{year}{1996}) \bibinfo{pages}{9353--9358}.
  \DOIprefix\doi{10.1103/PhysRevB.54.9353}.
\bibitem[{Slonczewski(1996)}]{SlonczewskiJMMM96}
\bibinfo{author}{J.~Slonczewski},
\newblock \bibinfo{title}{Current-driven excitation of magnetic multilayers},
\newblock \bibinfo{journal}{Journal of Magnetism and Magnetic Materials}
  \bibinfo{volume}{159} (\bibinfo{year}{1996}) \bibinfo{pages}{L1 -- L7}.
  \DOIprefix\doi{http://dx.doi.org/10.1016/0304-8853(96)00062-5}.
\bibitem[{Back et~al.(2019)Back, Bauer, and Zink}]{BackJPhysD2019}
\bibinfo{author}{C.~H. Back}, \bibinfo{author}{G.~E.~W. Bauer},
  \bibinfo{author}{B.~L. Zink},
\newblock \bibinfo{title}{Special issue on spin caloritronics},
\newblock \bibinfo{journal}{Journal of Physics D: Applied Physics}
  \bibinfo{volume}{52} (\bibinfo{year}{2019}) \bibinfo{pages}{230301}.
  \DOIprefix\doi{10.1088/1361-6463/ab070a}.
\bibitem[{Boona et~al.(2014)Boona, Myers, and Heremans}]{BoonaEES2014}
\bibinfo{author}{S.~R. Boona}, \bibinfo{author}{R.~C. Myers},
  \bibinfo{author}{J.~P. Heremans},
\newblock \bibinfo{title}{Spin caloritronics},
\newblock \bibinfo{journal}{Energy Environ. Sci.} \bibinfo{volume}{7}
  (\bibinfo{year}{2014}) \bibinfo{pages}{885--910}.
  \DOIprefix\doi{10.1039/C3EE43299H}.
\bibitem[{Bauer et~al.(2012)Bauer, Saitoh, and van Wees}]{BauerNatMat2012}
\bibinfo{author}{G.~E.~W. Bauer}, \bibinfo{author}{E.~Saitoh},
  \bibinfo{author}{B.~J. van Wees},
\newblock \bibinfo{title}{Spin caloritronics},
\newblock \bibinfo{journal}{Nature Materials} \bibinfo{volume}{11}
  (\bibinfo{year}{2012}) \bibinfo{pages}{391--399}.
\bibitem[{Bauer et~al.(2010)Bauer, MacDonald, and Maekawa}]{BauerSSC10}
\bibinfo{author}{G.~E.~W. Bauer}, \bibinfo{author}{A.~H. MacDonald},
  \bibinfo{author}{S.~Maekawa},
\newblock \bibinfo{title}{Spin caloritronics},
\newblock \bibinfo{journal}{Solid State Communications} \bibinfo{volume}{150}
  (\bibinfo{year}{2010}) \bibinfo{pages}{459--460}. \DOIprefix\doi{DOI:
  10.1016/j.ssc.2009.11.003}.
\bibitem[{Rea(mple)}]{ReactorNote}
\bibinfo{year}{I will give one example}. \bibinfo{note}{The RMBK-1000 nuclear
  reactor (the type involved in the 1986 Chernobyl accident) is designed to
  produce $1000$ MW of electricity, while generating $3200$ MW of thermal
  energy. Even if we make the wild underestimation that the active volume of
  the reactor core is 1 cubic meter, this gives a power density of only $\sim
  3\times10^{6}$ W/m$^{3}$. Reactor details available at https://inis.iaea.org/
  article reference number 48080457}.
\bibitem[{Daimon et~al.(2017)Daimon, Uchida, Iguchi, Hioki, and
  Saitoh}]{DaimonPRB2017}
\bibinfo{author}{S.~Daimon}, \bibinfo{author}{K.-i. Uchida},
  \bibinfo{author}{R.~Iguchi}, \bibinfo{author}{T.~Hioki},
  \bibinfo{author}{E.~Saitoh},
\newblock \bibinfo{title}{Thermographic measurements of the spin {P}eltier
  effect in metal/yttrium-iron-garnet junction systems},
\newblock \bibinfo{journal}{Phys. Rev. B} \bibinfo{volume}{96}
  (\bibinfo{year}{2017}) \bibinfo{pages}{024424}.
  \DOIprefix\doi{10.1103/PhysRevB.96.024424}.
\bibitem[{Cahill et~al.(2003)Cahill, Ford, Goodson, Mahan, Majumdar, Maris,
  Merlin, and Phillpot}]{CahillJAP2003}
\bibinfo{author}{D.~G. Cahill}, \bibinfo{author}{W.~K. Ford},
  \bibinfo{author}{K.~E. Goodson}, \bibinfo{author}{G.~D. Mahan},
  \bibinfo{author}{A.~Majumdar}, \bibinfo{author}{H.~J. Maris},
  \bibinfo{author}{R.~Merlin}, \bibinfo{author}{S.~R. Phillpot},
\newblock \bibinfo{title}{Nanoscale thermal transport},
\newblock \bibinfo{journal}{Journal of Applied Physics} \bibinfo{volume}{93}
  (\bibinfo{year}{2003}) \bibinfo{pages}{793}.
\bibitem[{Cahill et~al.(2014)Cahill, Braun, Chen, Clarke, Fan, Goodson,
  Keblinski, King, Mahan, Majumdar, Maris, Phillpot, Pop, and
  Shi}]{CahillAPR2014}
\bibinfo{author}{D.~G. Cahill}, \bibinfo{author}{P.~V. Braun},
  \bibinfo{author}{G.~Chen}, \bibinfo{author}{D.~R. Clarke},
  \bibinfo{author}{S.~Fan}, \bibinfo{author}{K.~E. Goodson},
  \bibinfo{author}{P.~Keblinski}, \bibinfo{author}{W.~P. King},
  \bibinfo{author}{G.~D. Mahan}, \bibinfo{author}{A.~Majumdar},
  \bibinfo{author}{H.~J. Maris}, \bibinfo{author}{S.~R. Phillpot},
  \bibinfo{author}{E.~Pop}, \bibinfo{author}{L.~Shi},
\newblock \bibinfo{title}{Nanoscale thermal transport {II}. 2003-2012},
\newblock \bibinfo{journal}{Applied Physics Reviews} \bibinfo{volume}{1}
  (\bibinfo{year}{2014}) \bibinfo{pages}{--}.
  \DOIprefix\doi{http://dx.doi.org/10.1063/1.4832615}.
\bibitem[{Chen et~al.(2022)Chen, Xu, Zhou, and Li}]{ChenRMP2022}
\bibinfo{author}{J.~Chen}, \bibinfo{author}{X.~Xu}, \bibinfo{author}{J.~Zhou},
  \bibinfo{author}{B.~Li},
\newblock \bibinfo{title}{Interfacial thermal resistance: Past, present, and
  future},
\newblock \bibinfo{journal}{Rev. Mod. Phys.} \bibinfo{volume}{94}
  (\bibinfo{year}{2022}) \bibinfo{pages}{025002}.
  \DOIprefix\doi{10.1103/RevModPhys.94.025002}.
\bibitem[{Monachon et~al.(2016)Monachon, Weber, and Dames}]{MonachonARMR2016}
\bibinfo{author}{C.~Monachon}, \bibinfo{author}{L.~Weber},
  \bibinfo{author}{C.~Dames},
\newblock \bibinfo{title}{Thermal boundary conductance: A materials science
  perspective},
\newblock \bibinfo{journal}{Annual Review of Materials Research}
  \bibinfo{volume}{46} (\bibinfo{year}{2016}) \bibinfo{pages}{433--463}.
  \DOIprefix\doi{10.1146/annurev-matsci-070115-031719}.
\bibitem[{Angeles et~al.(2021)Angeles, Sun, Ortiz, Shi, Li, and
  Wilson}]{AngelesPRM2021}
\bibinfo{author}{F.~Angeles}, \bibinfo{author}{Q.~Sun}, \bibinfo{author}{V.~H.
  Ortiz}, \bibinfo{author}{J.~Shi}, \bibinfo{author}{C.~Li},
  \bibinfo{author}{R.~B. Wilson},
\newblock \bibinfo{title}{Interfacial thermal transport in spin caloritronic
  material systems},
\newblock \bibinfo{journal}{Phys. Rev. Materials} \bibinfo{volume}{5}
  (\bibinfo{year}{2021}) \bibinfo{pages}{114403}.
  \DOIprefix\doi{10.1103/PhysRevMaterials.5.114403}.
\bibitem[{Pollack(1969)}]{PollackRMP1969}
\bibinfo{author}{G.~L. Pollack},
\newblock \bibinfo{title}{Kapitza resistance},
\newblock \bibinfo{journal}{Rev. Mod. Phys.} \bibinfo{volume}{41}
  (\bibinfo{year}{1969}) \bibinfo{pages}{48--81}.
  \DOIprefix\doi{10.1103/RevModPhys.41.48}.
\bibitem[{Huang et~al.(2011)Huang, Wang, Lee, Kwo, and Chien}]{HuangPRL2011}
\bibinfo{author}{S.~Y. Huang}, \bibinfo{author}{W.~G. Wang},
  \bibinfo{author}{S.~F. Lee}, \bibinfo{author}{J.~Kwo}, \bibinfo{author}{C.~L.
  Chien},
\newblock \bibinfo{title}{Intrinsic spin-dependent thermal transport},
\newblock \bibinfo{journal}{Phys. Rev. Lett.} \bibinfo{volume}{107}
  (\bibinfo{year}{2011}) \bibinfo{pages}{216604}.
  \DOIprefix\doi{10.1103/PhysRevLett.107.216604}.
\bibitem[{Meier et~al.(2013)Meier, Reinhardt, Schmid, Back, Schmalhorst,
  Kuschel, and Reiss}]{MeierPRB2013}
\bibinfo{author}{D.~Meier}, \bibinfo{author}{D.~Reinhardt},
  \bibinfo{author}{M.~Schmid}, \bibinfo{author}{C.~H. Back},
  \bibinfo{author}{J.-M. Schmalhorst}, \bibinfo{author}{T.~Kuschel},
  \bibinfo{author}{G.~Reiss},
\newblock \bibinfo{title}{Influence of heat flow directions on {N}ernst effects
  in {P}y/{P}t bilayers},
\newblock \bibinfo{journal}{Phys. Rev. B} \bibinfo{volume}{88}
  (\bibinfo{year}{2013}) \bibinfo{pages}{184425}.
  \DOIprefix\doi{10.1103/PhysRevB.88.184425}.
\bibitem[{Schmid et~al.(2013)Schmid, Srichandan, Meier, Kuschel, Schmalhorst,
  Vogel, Reiss, Strunk, and Back}]{SchmidPRL2013}
\bibinfo{author}{M.~Schmid}, \bibinfo{author}{S.~Srichandan},
  \bibinfo{author}{D.~Meier}, \bibinfo{author}{T.~Kuschel},
  \bibinfo{author}{J.-M. Schmalhorst}, \bibinfo{author}{M.~Vogel},
  \bibinfo{author}{G.~Reiss}, \bibinfo{author}{C.~Strunk},
  \bibinfo{author}{C.~H. Back},
\newblock \bibinfo{title}{Transverse spin {S}eebeck effect versus anomalous and
  planar {N}ernst effects in permalloy thin films},
\newblock \bibinfo{journal}{Phys. Rev. Lett.} \bibinfo{volume}{111}
  (\bibinfo{year}{2013}) \bibinfo{pages}{187201}.
  \DOIprefix\doi{10.1103/PhysRevLett.111.187201}.
\bibitem[{Shestakov et~al.(2015)Shestakov, Schmid, Meier, Kuschel, and
  Back}]{ShestakovPRB2015}
\bibinfo{author}{A.~S. Shestakov}, \bibinfo{author}{M.~Schmid},
  \bibinfo{author}{D.~Meier}, \bibinfo{author}{T.~Kuschel},
  \bibinfo{author}{C.~H. Back},
\newblock \bibinfo{title}{Dependence of transverse magnetothermoelectric
  effects on inhomogeneous magnetic fields},
\newblock \bibinfo{journal}{Phys. Rev. B} \bibinfo{volume}{92}
  (\bibinfo{year}{2015}) \bibinfo{pages}{224425}.
  \DOIprefix\doi{10.1103/PhysRevB.92.224425}.
\bibitem[{Meier et~al.(2015)Meier, Reinhardt, van Straaten, Klewe, Althammer,
  Schreier, Goennenwein, Gupta, Schmid, Back, Schmalhorst, Kuschel, and
  Reiss}]{MeierNatCom2015}
\bibinfo{author}{D.~Meier}, \bibinfo{author}{D.~Reinhardt},
  \bibinfo{author}{M.~van Straaten}, \bibinfo{author}{C.~Klewe},
  \bibinfo{author}{M.~Althammer}, \bibinfo{author}{M.~Schreier},
  \bibinfo{author}{S.~T.~B. Goennenwein}, \bibinfo{author}{A.~Gupta},
  \bibinfo{author}{M.~Schmid}, \bibinfo{author}{C.~H. Back},
  \bibinfo{author}{J.-M. Schmalhorst}, \bibinfo{author}{T.~Kuschel},
  \bibinfo{author}{G.~Reiss},
\newblock \bibinfo{title}{Longitudinal spin {S}eebeck effect contribution in
  transverse spin {S}eebeck effect experiments in {P}t/{YIG} and {P}t/{NFO}},
\newblock \bibinfo{journal}{Nature Communications} \bibinfo{volume}{6}
  (\bibinfo{year}{2015}) \bibinfo{pages}{8211}.
  \DOIprefix\doi{10.1038/ncomms9211}.
\bibitem[{Bennet et~al.(2020)Bennet, Hojem, and Zink}]{BennetAIPadv2020}
\bibinfo{author}{R.~K. Bennet}, \bibinfo{author}{A.~Hojem},
  \bibinfo{author}{B.~L. Zink},
\newblock \bibinfo{title}{Temperature dependence of the anomalous {N}ernst
  coefficient for {N}i$_{80}${F}e$_{20}$ determined with metallic nonlocal spin
  valves},
\newblock \bibinfo{journal}{{AIP} Advances} \bibinfo{volume}{10}
  (\bibinfo{year}{2020}) \bibinfo{pages}{065127}.
  \DOIprefix\doi{10.1063/5.0006599}.
\bibitem[{Zhu et~al.(2010)Zhu, Tang, Wang, Liu, Holub, and Yang}]{ZhuJAP2010}
\bibinfo{author}{J.~Zhu}, \bibinfo{author}{D.~Tang}, \bibinfo{author}{W.~Wang},
  \bibinfo{author}{J.~Liu}, \bibinfo{author}{K.~W. Holub},
  \bibinfo{author}{R.~Yang},
\newblock \bibinfo{title}{Ultrafast thermoreflectance techniques for measuring
  thermal conductivity and interface thermal conductance of thin films},
\newblock \bibinfo{journal}{Journal of Applied Physics} \bibinfo{volume}{108}
  (\bibinfo{year}{2010}) \bibinfo{pages}{094315}.
  \DOIprefix\doi{10.1063/1.3504213}.
\bibitem[{Boona and Heremans(2014)}]{BoonaPRB2014}
\bibinfo{author}{S.~R. Boona}, \bibinfo{author}{J.~P. Heremans},
\newblock \bibinfo{title}{Magnon thermal mean free path in yttrium iron
  garnet},
\newblock \bibinfo{journal}{Phys. Rev. B} \bibinfo{volume}{90}
  (\bibinfo{year}{2014}) \bibinfo{pages}{064421}.
  \DOIprefix\doi{10.1103/PhysRevB.90.064421}.
\bibitem[{Regner et~al.(2015)Regner, Freedman, and Malen}]{RegnerNMTE2015}
\bibinfo{author}{K.~T. Regner}, \bibinfo{author}{J.~P. Freedman},
  \bibinfo{author}{J.~A. Malen},
\newblock \bibinfo{title}{Advances in studying phonon mean free path dependent
  contributions to thermal conductivity},
\newblock \bibinfo{journal}{Nanoscale and Microscale Thermophysical
  Engineering} \bibinfo{volume}{19} (\bibinfo{year}{2015})
  \bibinfo{pages}{183--205}. \DOIprefix\doi{10.1080/15567265.2015.1045640}.
\bibitem[{Gall(2016)}]{GallJAP2016}
\bibinfo{author}{D.~Gall},
\newblock \bibinfo{title}{Electron mean free path in elemental metals},
\newblock \bibinfo{journal}{Journal of Applied Physics} \bibinfo{volume}{119}
  (\bibinfo{year}{2016}) \bibinfo{pages}{085101}.
  \DOIprefix\doi{10.1063/1.4942216}.
\bibitem[{Bourgeois et~al.(2016)Bourgeois, Tainoff, Tavakoli, Liu, Blanc,
  Boukhari, Barski, and Hadji}]{BourgeoisCRP2016}
\bibinfo{author}{O.~Bourgeois}, \bibinfo{author}{D.~Tainoff},
  \bibinfo{author}{A.~Tavakoli}, \bibinfo{author}{Y.~Liu},
  \bibinfo{author}{C.~Blanc}, \bibinfo{author}{M.~Boukhari},
  \bibinfo{author}{A.~Barski}, \bibinfo{author}{E.~Hadji},
\newblock \bibinfo{title}{Reduction of phonon mean free path: From
  low-temperature physics to room temperature applications in
  thermoelectricity},
\newblock \bibinfo{journal}{Comptes Rendus Physique} \bibinfo{volume}{17}
  (\bibinfo{year}{2016}) \bibinfo{pages}{1154--1160}.
  \DOIprefix\doi{10.1016/j.crhy.2016.08.008}.
\bibitem[{Had\'{a}mek et~al.(2022)Had\'{a}mek, Fiorentini, Bendra, Ender, {de
  Orio}, Goes, Selberherr, and Sverdlov}]{HadamekSSE2022}
\bibinfo{author}{T.~Had\'{a}mek}, \bibinfo{author}{S.~Fiorentini},
  \bibinfo{author}{M.~Bendra}, \bibinfo{author}{J.~Ender},
  \bibinfo{author}{R.~{de Orio}}, \bibinfo{author}{W.~Goes},
  \bibinfo{author}{S.~Selberherr}, \bibinfo{author}{V.~Sverdlov},
\newblock \bibinfo{title}{Temperature increase in {STT-MRAM} at writing: A
  fully three-dimensional finite element approach},
\newblock \bibinfo{journal}{Solid-State Electronics} \bibinfo{volume}{193}
  (\bibinfo{year}{2022}) \bibinfo{pages}{108269}.
  \DOIprefix\doi{https://doi.org/10.1016/j.sse.2022.108269}.
\bibitem[{Shigematsu et~al.(2022)Shigematsu, Tamura, Ohshima, Ando, and
  Shiraishi}]{ShigematsuJAP2022}
\bibinfo{author}{E.~Shigematsu}, \bibinfo{author}{E.~Tamura},
  \bibinfo{author}{R.~Ohshima}, \bibinfo{author}{Y.~Ando},
  \bibinfo{author}{M.~Shiraishi},
\newblock \bibinfo{title}{Full calculation of inter-conversion between charge,
  spin, and heat current using a common partial differential equation
  platform},
\newblock \bibinfo{journal}{Journal of Applied Physics} \bibinfo{volume}{131}
  (\bibinfo{year}{2022}) \bibinfo{pages}{243903}.
  \DOIprefix\doi{10.1063/5.0088343}.
\bibitem[{Siemens et~al.(2010)Siemens, Li, Yang, Nelson, Anderson, Murnane, and
  Kapteyn}]{SiemensNatMat2010}
\bibinfo{author}{M.~E. Siemens}, \bibinfo{author}{Q.~Li},
  \bibinfo{author}{R.~Yang}, \bibinfo{author}{K.~A. Nelson},
  \bibinfo{author}{E.~H. Anderson}, \bibinfo{author}{M.~M. Murnane},
  \bibinfo{author}{H.~C. Kapteyn},
\newblock \bibinfo{title}{Quasi-ballistic thermal transport from nanoscale
  interfaces observed using ultrafast coherent soft x-ray beams},
\newblock \bibinfo{journal}{Nature Materials} \bibinfo{volume}{9}
  (\bibinfo{year}{2010}) \bibinfo{pages}{26--30}.
\bibitem[{Minnich(2012)}]{MinnichPRL2012}
\bibinfo{author}{A.~J. Minnich},
\newblock \bibinfo{title}{Determining phonon mean free paths from observations
  of quasiballistic thermal transport},
\newblock \bibinfo{journal}{Phys. Rev. Lett.} \bibinfo{volume}{109}
  (\bibinfo{year}{2012}) \bibinfo{pages}{205901}.
  \DOIprefix\doi{10.1103/PhysRevLett.109.205901}.
\bibitem[{Regner et~al.(2013)Regner, Sellan, Su, Amon, McGaughey, and
  Malen}]{RegnerNatComm2013}
\bibinfo{author}{K.~T. Regner}, \bibinfo{author}{D.~P. Sellan},
  \bibinfo{author}{Z.~Su}, \bibinfo{author}{C.~H. Amon},
  \bibinfo{author}{A.~J.~H. McGaughey}, \bibinfo{author}{J.~A. Malen},
\newblock \bibinfo{title}{Broadband phonon mean free path contributions to
  thermal conductivity measured using frequency domain thermoreflectance},
\newblock \bibinfo{journal}{Nature Communications} \bibinfo{volume}{4}
  (\bibinfo{year}{2013}) \bibinfo{pages}{1640}.
\bibitem[{Minnich et~al.(2011)Minnich, Johnson, Schmidt, Esfarjani,
  Dresselhaus, Nelson, and Chen}]{MinnichPRL2011}
\bibinfo{author}{A.~J. Minnich}, \bibinfo{author}{J.~A. Johnson},
  \bibinfo{author}{A.~J. Schmidt}, \bibinfo{author}{K.~Esfarjani},
  \bibinfo{author}{M.~S. Dresselhaus}, \bibinfo{author}{K.~A. Nelson},
  \bibinfo{author}{G.~Chen},
\newblock \bibinfo{title}{Thermal conductivity spectroscopy technique to
  measure phonon mean free paths},
\newblock \bibinfo{journal}{Phys. Rev. Lett.} \bibinfo{volume}{107}
  (\bibinfo{year}{2011}) \bibinfo{pages}{095901}.
  \DOIprefix\doi{10.1103/PhysRevLett.107.095901}.
\bibitem[{Stefanou et~al.(2021)Stefanou, Menges, Boehm, Moran, Adams, Ali,
  Rosamond, Gotsmann, Allenspach, Burnell, and Hickey}]{StefanouPRL2021}
\bibinfo{author}{G.~Stefanou}, \bibinfo{author}{F.~Menges},
  \bibinfo{author}{B.~Boehm}, \bibinfo{author}{K.~A. Moran},
  \bibinfo{author}{J.~Adams}, \bibinfo{author}{M.~Ali}, \bibinfo{author}{M.~C.
  Rosamond}, \bibinfo{author}{B.~Gotsmann}, \bibinfo{author}{R.~Allenspach},
  \bibinfo{author}{G.~Burnell}, \bibinfo{author}{B.~J. Hickey},
\newblock \bibinfo{title}{Scanning thermal microscopy and ballistic phonon
  transport in lateral spin valves},
\newblock \bibinfo{journal}{Phys. Rev. Lett.} \bibinfo{volume}{127}
  (\bibinfo{year}{2021}) \bibinfo{pages}{035901}.
  \DOIprefix\doi{10.1103/PhysRevLett.127.035901}.
\bibitem[{Kim et~al.(2021)Kim, Moon, and Minnich}]{KimPRM2021}
\bibinfo{author}{T.~Kim}, \bibinfo{author}{J.~Moon}, \bibinfo{author}{A.~J.
  Minnich},
\newblock \bibinfo{title}{Origin of micrometer-scale propagation lengths of
  heat-carrying acoustic excitations in amorphous silicon},
\newblock \bibinfo{journal}{Phys. Rev. Materials} \bibinfo{volume}{5}
  (\bibinfo{year}{2021}) \bibinfo{pages}{065602}.
  \DOIprefix\doi{10.1103/PhysRevMaterials.5.065602}.
\bibitem[{Lee et~al.(2015)Lee, Lim, and Yang}]{LeeNanoLett2015}
\bibinfo{author}{J.~Lee}, \bibinfo{author}{J.~Lim}, \bibinfo{author}{P.~Yang},
\newblock \bibinfo{title}{Ballistic phonon transport in holey silicon},
\newblock \bibinfo{journal}{Nano Letters} \bibinfo{volume}{15}
  (\bibinfo{year}{2015}) \bibinfo{pages}{3273--3279}.
  \DOIprefix\doi{10.1021/acs.nanolett.5b00495}, \bibinfo{note}{pMID: 25861026}.
\bibitem[{Johnson et~al.(2013)Johnson, Maznev, Cuffe, Eliason, Minnich, Kehoe,
  Torres, Chen, and Nelson}]{JohnsonPRL2013}
\bibinfo{author}{J.~A. Johnson}, \bibinfo{author}{A.~A. Maznev},
  \bibinfo{author}{J.~Cuffe}, \bibinfo{author}{J.~K. Eliason},
  \bibinfo{author}{A.~J. Minnich}, \bibinfo{author}{T.~Kehoe},
  \bibinfo{author}{C.~M.~S. Torres}, \bibinfo{author}{G.~Chen},
  \bibinfo{author}{K.~A. Nelson},
\newblock \bibinfo{title}{Direct measurement of room-temperature nondiffusive
  thermal transport over micron distances in a silicon membrane},
\newblock \bibinfo{journal}{Phys. Rev. Lett.} \bibinfo{volume}{110}
  (\bibinfo{year}{2013}) \bibinfo{pages}{025901}.
  \DOIprefix\doi{10.1103/PhysRevLett.110.025901}.
\bibitem[{Sultan et~al.(2013)Sultan, Avery, Underwood, Mason, Bassett, and
  Zink}]{SultanPRB2013}
\bibinfo{author}{R.~Sultan}, \bibinfo{author}{A.~D. Avery},
  \bibinfo{author}{J.~M. Underwood}, \bibinfo{author}{S.~J. Mason},
  \bibinfo{author}{D.~Bassett}, \bibinfo{author}{B.~L. Zink},
\newblock \bibinfo{title}{Heat transport by long mean free path vibrations in
  amorphous silicon nitride near room temperature},
\newblock \bibinfo{journal}{Phys. Rev. B} \bibinfo{volume}{87}
  (\bibinfo{year}{2013}) \bibinfo{pages}{214305}.
  \DOIprefix\doi{10.1103/PhysRevB.87.214305}.
\bibitem[{Koh and Cahill(2007)}]{KohPRB2007}
\bibinfo{author}{Y.~K. Koh}, \bibinfo{author}{D.~G. Cahill},
\newblock \bibinfo{title}{Frequency dependence of the thermal conductivity of
  semiconductor alloys},
\newblock \bibinfo{journal}{Phys. Rev. B} \bibinfo{volume}{76}
  (\bibinfo{year}{2007}) \bibinfo{pages}{075207}.
  \DOIprefix\doi{10.1103/PhysRevB.76.075207}.
\bibitem[{Mason et~al.(2020)Mason, Wesenberg, Hojem, Manno, Leighton, and
  Zink}]{MasonPRM2020}
\bibinfo{author}{S.~J. Mason}, \bibinfo{author}{D.~J. Wesenberg},
  \bibinfo{author}{A.~Hojem}, \bibinfo{author}{M.~Manno},
  \bibinfo{author}{C.~Leighton}, \bibinfo{author}{B.~L. Zink},
\newblock \bibinfo{title}{Violation of the {W}iedemann-{F}ranz law through
  reduction of thermal conductivity in gold thin films},
\newblock \bibinfo{journal}{Phys. Rev. Materials} \bibinfo{volume}{4}
  (\bibinfo{year}{2020}) \bibinfo{pages}{065003}.
  \DOIprefix\doi{10.1103/PhysRevMaterials.4.065003}.
\bibitem[{Wang et~al.(2019)Wang, Yu, Walbert, Antoni, Wang, Xi, Muench, Yang,
  Chen, and Ensinger}]{WangNanotech2019}
\bibinfo{author}{J.~Wang}, \bibinfo{author}{H.~Yu},
  \bibinfo{author}{T.~Walbert}, \bibinfo{author}{M.~Antoni},
  \bibinfo{author}{C.~Wang}, \bibinfo{author}{W.~Xi},
  \bibinfo{author}{F.~Muench}, \bibinfo{author}{J.~Yang},
  \bibinfo{author}{Y.~Chen}, \bibinfo{author}{W.~Ensinger},
\newblock \bibinfo{title}{Electrical and thermal conductivities of
  polycrystalline platinum nanowires},
\newblock \bibinfo{journal}{Nanotechnology} \bibinfo{volume}{30}
  (\bibinfo{year}{2019}) \bibinfo{pages}{455706}.
  \DOIprefix\doi{10.1088/1361-6528/ab37a7}.
\bibitem[{Sawtelle and Reed(2019)}]{SawtellePRB2019}
\bibinfo{author}{S.~D. Sawtelle}, \bibinfo{author}{M.~A. Reed},
\newblock \bibinfo{title}{Temperature-dependent thermal conductivity and
  suppressed {L}orenz number in ultrathin gold nanowires},
\newblock \bibinfo{journal}{Phys. Rev. B} \bibinfo{volume}{99}
  (\bibinfo{year}{2019}) \bibinfo{pages}{054304}.
  \DOIprefix\doi{10.1103/PhysRevB.99.054304}.
\bibitem[{Eigenfeld et~al.(2015)Eigenfeld, Gertsch, Skidmore, George, and
  Bright}]{EigenfeldNano2015}
\bibinfo{author}{N.~T. Eigenfeld}, \bibinfo{author}{J.~C. Gertsch},
  \bibinfo{author}{G.~D. Skidmore}, \bibinfo{author}{S.~M. George},
  \bibinfo{author}{V.~M. Bright},
\newblock \bibinfo{title}{Electrical and thermal conduction in ultra-thin
  freestanding atomic layer deposited {W} nanobridges},
\newblock \bibinfo{journal}{Nanoscale} \bibinfo{volume}{7}
  (\bibinfo{year}{2015}) \bibinfo{pages}{17923--17928}.
  \DOIprefix\doi{10.1039/C5NR04885K}.
\bibitem[{Wang et~al.(2013)Wang, Liu, Zhang, and Takahashi}]{WangIJHMT2013}
\bibinfo{author}{H.~Wang}, \bibinfo{author}{J.~Liu},
  \bibinfo{author}{X.~Zhang}, \bibinfo{author}{K.~Takahashi},
\newblock \bibinfo{title}{Breakdown of {W}iedemann-{F}ranz law in individual
  suspended polycrystalline gold nanofilms down to 3 {K}},
\newblock \bibinfo{journal}{International Journal of Heat and Mass Transfer}
  \bibinfo{volume}{66} (\bibinfo{year}{2013}) \bibinfo{pages}{585 -- 591}.
  \DOIprefix\doi{http://dx.doi.org/10.1016/j.ijheatmasstransfer.2013.07.066}.
\bibitem[{Wang et~al.(2011)Wang, Liu, Zhang, Guo, and Takahashi}]{WangHMT2011}
\bibinfo{author}{H.-D. Wang}, \bibinfo{author}{J.-H. Liu},
  \bibinfo{author}{X.~Zhang}, \bibinfo{author}{Z.-Y. Guo},
  \bibinfo{author}{K.~Takahashi},
\newblock \bibinfo{title}{Experimental study on the influences of grain
  boundary scattering on the charge and heat transport in gold and platinum
  nanofilms},
\newblock \bibinfo{journal}{Heat and Mass Transfer} \bibinfo{volume}{47}
  (\bibinfo{year}{2011}) \bibinfo{pages}{893--898}.
\bibitem[{Volklein et~al.(2009)Volklein, Reith, Cornelius, Rauber, and
  Neumann}]{VolkleinNano2009}
\bibinfo{author}{F.~Volklein}, \bibinfo{author}{H.~Reith},
  \bibinfo{author}{T.~W. Cornelius}, \bibinfo{author}{M.~Rauber},
  \bibinfo{author}{R.~Neumann},
\newblock \bibinfo{title}{The experimental investigation of thermal
  conductivity and the {W}iedemann-{F}ranz law for single metallic nanowires},
\newblock \bibinfo{journal}{Nanotechnology} \bibinfo{volume}{20}
  (\bibinfo{year}{2009}) \bibinfo{pages}{325706}.
\bibitem[{Zhang et~al.(2006)Zhang, Cao, Zhang, Fujii, and
  Takahashi}]{ZhangPRB2006}
\bibinfo{author}{Q.~G. Zhang}, \bibinfo{author}{B.~Y. Cao},
  \bibinfo{author}{X.~Zhang}, \bibinfo{author}{M.~Fujii},
  \bibinfo{author}{K.~Takahashi},
\newblock \bibinfo{title}{Influence of grain boundary scattering on the
  electrical and thermal conductivities of polycrystalline gold nanofilms},
\newblock \bibinfo{journal}{Phys. Rev. B} \bibinfo{volume}{74}
  (\bibinfo{year}{2006}) \bibinfo{pages}{134109}.
  \DOIprefix\doi{10.1103/PhysRevB.74.134109}.
\bibitem[{Xia et~al.(2010)Xia, Wang, Wang, Li, Zhang, Feng, and
  Ding}]{XiaNanotechnology2010}
\bibinfo{author}{R.~Xia}, \bibinfo{author}{J.~L. Wang},
  \bibinfo{author}{R.~Wang}, \bibinfo{author}{X.~Li},
  \bibinfo{author}{X.~Zhang}, \bibinfo{author}{X.-Q. Feng},
  \bibinfo{author}{Y.~Ding},
\newblock \bibinfo{title}{Correlation of the thermal and electrical
  conductivities of nanoporous gold},
\newblock \bibinfo{journal}{Nanotechnology} \bibinfo{volume}{21}
  (\bibinfo{year}{2010}) \bibinfo{pages}{085703}.
\bibitem[{Hoffmann et~al.(1982)Hoffmann, Hofmann, and
  Schoepe}]{HoffmannPRB1982}
\bibinfo{author}{H.~Hoffmann}, \bibinfo{author}{F.~Hofmann},
  \bibinfo{author}{W.~Schoepe},
\newblock \bibinfo{title}{Magnetoresistance and non-ohmic conductivity of thin
  platinum films at low temperatures},
\newblock \bibinfo{journal}{Phys. Rev. B} \bibinfo{volume}{25}
  (\bibinfo{year}{1982}) \bibinfo{pages}{5563--5565}.
  \DOIprefix\doi{10.1103/PhysRevB.25.5563}.
\bibitem[{Roy et~al.(2020)Roy, {Akshaya Shiva}, M\'{a}t\'{e}fi-Tempfli,
  M\'{a}t\'{e}fi-Tempfli, Kiebooms, and Menon}]{RoyPhysB2020}
\bibinfo{author}{A.~Roy}, \bibinfo{author}{V.~{Akshaya Shiva}},
  \bibinfo{author}{S.~M\'{a}t\'{e}fi-Tempfli},
  \bibinfo{author}{M.~M\'{a}t\'{e}fi-Tempfli}, \bibinfo{author}{R.~Kiebooms},
  \bibinfo{author}{R.~Menon},
\newblock \bibinfo{title}{Resistance and magnetoresistance in platinum
  nanowire: Implications to {K}ohler's rule},
\newblock \bibinfo{journal}{Physica B: Condensed Matter} \bibinfo{volume}{591}
  (\bibinfo{year}{2020}) \bibinfo{pages}{412239}.
  \DOIprefix\doi{https://doi.org/10.1016/j.physb.2020.412239}.
\bibitem[{Mason et~al.(2020)Mason, Hojem, Wesenberg, Avery, and
  Zink}]{MasonJAP2020}
\bibinfo{author}{S.~J. Mason}, \bibinfo{author}{A.~Hojem},
  \bibinfo{author}{D.~J. Wesenberg}, \bibinfo{author}{A.~D. Avery},
  \bibinfo{author}{B.~L. Zink},
\newblock \bibinfo{title}{Determining absolute {S}eebeck coefficients from
  relative thermopower measurements of thin films and nanostructures},
\newblock \bibinfo{journal}{Journal of Applied Physics} \bibinfo{volume}{127}
  (\bibinfo{year}{2020}) \bibinfo{pages}{085101}.
  \DOIprefix\doi{10.1063/1.5143447}.
\bibitem[{Duarte et~al.(2009)Duarte, Mahan, and Tadigadapa}]{DuarteNanoLett09}
\bibinfo{author}{N.~B. Duarte}, \bibinfo{author}{G.~D. Mahan},
  \bibinfo{author}{S.~Tadigadapa},
\newblock \bibinfo{title}{Thermopower enhancement in nanowires via junction
  effects},
\newblock \bibinfo{journal}{Nano Letters} \bibinfo{volume}{9}
  (\bibinfo{year}{2009}) \bibinfo{pages}{617}.
\bibitem[{Szakmany et~al.(2014)Szakmany, Orlov, Bernstein, and
  Porod}]{SzakmanyIEEETransNano2014}
\bibinfo{author}{G.~Szakmany}, \bibinfo{author}{A.~Orlov},
  \bibinfo{author}{G.~Bernstein}, \bibinfo{author}{W.~Porod},
\newblock \bibinfo{title}{Single-metal nanoscale thermocouples},
\newblock \bibinfo{journal}{Nanotechnology, IEEE Transactions on}
  \bibinfo{volume}{13} (\bibinfo{year}{2014}) \bibinfo{pages}{1234--1239}.
  \DOIprefix\doi{10.1109/TNANO.2014.2358532}.
\bibitem[{Evans et~al.(2020)Evans, Yang, Gan, Abbasi, Wang, Traylor, Fan, and
  Natelson}]{EvansPNAS2020}
\bibinfo{author}{C.~I. Evans}, \bibinfo{author}{R.~Yang},
  \bibinfo{author}{L.~T. Gan}, \bibinfo{author}{M.~Abbasi},
  \bibinfo{author}{X.~Wang}, \bibinfo{author}{R.~Traylor},
  \bibinfo{author}{J.~A. Fan}, \bibinfo{author}{D.~Natelson},
\newblock \bibinfo{title}{Thermoelectric response from grain boundaries and
  lattice distortions in crystalline gold devices},
\newblock \bibinfo{journal}{Proceedings of the National Academy of Sciences}
  \bibinfo{volume}{117} (\bibinfo{year}{2020}) \bibinfo{pages}{23350--23355}.
  \DOIprefix\doi{10.1073/pnas.2002284117}.
\bibitem[{B\"{o}hnert et~al.(2014)B\"{o}hnert, Niemann, Michel, B\"{a}{\ss}ler,
  Gooth, T{\'{o}}th, Neur{\'{o}}hr, P{\'{e}}ter, Bakonyi, Vega, Prida, and
  Nielsch}]{BohnertPRB2014}
\bibinfo{author}{T.~B\"{o}hnert}, \bibinfo{author}{A.~C. Niemann},
  \bibinfo{author}{A.-K. Michel}, \bibinfo{author}{S.~B\"{a}{\ss}ler},
  \bibinfo{author}{J.~Gooth}, \bibinfo{author}{B.~G. T{\'{o}}th},
  \bibinfo{author}{K.~Neur{\'{o}}hr}, \bibinfo{author}{L.~P{\'{e}}ter},
  \bibinfo{author}{I.~Bakonyi}, \bibinfo{author}{V.~Vega},
  \bibinfo{author}{V.~M. Prida}, \bibinfo{author}{K.~Nielsch},
\newblock \bibinfo{title}{Magnetothermopower and magnetoresistance of single
  {C}o-{N}i/{C}u multilayered nanowires},
\newblock \bibinfo{journal}{Physical Review B} \bibinfo{volume}{90}
  (\bibinfo{year}{2014}). \DOIprefix\doi{10.1103/physrevb.90.165416}.
\bibitem[{Scarioni et~al.(2018)Scarioni, Krzysteczko, Sievers, Hu, and
  Schumacher}]{ScarioniJPhysD2018}
\bibinfo{author}{A.~F. Scarioni}, \bibinfo{author}{P.~Krzysteczko},
  \bibinfo{author}{S.~Sievers}, \bibinfo{author}{X.~Hu}, \bibinfo{author}{H.~W.
  Schumacher},
\newblock \bibinfo{title}{Temperature dependence of the domain wall
  magneto-{S}eebeck effect: avoiding artifacts of lead contributions},
\newblock \bibinfo{journal}{Journal of Physics D: Applied Physics}
  \bibinfo{volume}{51} (\bibinfo{year}{2018}) \bibinfo{pages}{234004}.
  \DOIprefix\doi{10.1088/1361-6463/aac0a7}.
\bibitem[{Kockert et~al.(2019)Kockert, Mitdank, Zykov, Kowarik, and
  Fischer}]{KockertJAP2019}
\bibinfo{author}{M.~Kockert}, \bibinfo{author}{R.~Mitdank},
  \bibinfo{author}{A.~Zykov}, \bibinfo{author}{S.~Kowarik},
  \bibinfo{author}{S.~F. Fischer},
\newblock \bibinfo{title}{Absolute {S}eebeck coefficient of thin platinum
  films},
\newblock \bibinfo{journal}{Journal of Applied Physics} \bibinfo{volume}{126}
  (\bibinfo{year}{2019}) \bibinfo{pages}{105106}.
  \DOIprefix\doi{10.1063/1.5101028}.
\bibitem[{Kondo(1965)}]{KondoPTP1965}
\bibinfo{author}{J.~Kondo},
\newblock \bibinfo{title}{Giant thermo-electric power of dilute magnetic
  alloys},
\newblock \bibinfo{journal}{Progress of Theoretical Physics}
  \bibinfo{volume}{34} (\bibinfo{year}{1965}) \bibinfo{pages}{372--382}.
  \DOIprefix\doi{10.1143/ptp.34.372}.
\bibitem[{Berman and Kopp(1971)}]{BermanJPFMP1971}
\bibinfo{author}{R.~Berman}, \bibinfo{author}{J.~Kopp},
\newblock \bibinfo{title}{The thermoelectric power of dilute gold-iron alloys},
\newblock \bibinfo{journal}{Journal of Physics F: Metal Physics}
  \bibinfo{volume}{1} (\bibinfo{year}{1971}) \bibinfo{pages}{457}.
\bibitem[{Barnard(1973)}]{BarnardJPhysE1973}
\bibinfo{author}{R.~D. Barnard},
\newblock \bibinfo{title}{Magnetic impurities and the thermopower of gold at
  low temperatures},
\newblock \bibinfo{journal}{Journal of Physics E: Scientific Instruments}
  \bibinfo{volume}{6} (\bibinfo{year}{1973}) \bibinfo{pages}{508--511}.
  \DOIprefix\doi{10.1088/0022-3735/6/6/002}.
\bibitem[{Huebener(1964{\natexlab{a}})}]{HuebenerPR64}
\bibinfo{author}{R.~P. Huebener},
\newblock \bibinfo{title}{Thermoelectric size effect in pure gold},
\newblock \bibinfo{journal}{Phys. Rev.} \bibinfo{volume}{136}
  (\bibinfo{year}{1964}{\natexlab{a}}) \bibinfo{pages}{A1740--A1744}.
  \DOIprefix\doi{10.1103/PhysRev.136.A1740}.
\bibitem[{Huebener(1964{\natexlab{b}})}]{HuebenerPR64b}
\bibinfo{author}{R.~P. Huebener},
\newblock \bibinfo{title}{Thermoelectric power of lattice vacancies in gold},
\newblock \bibinfo{journal}{Phys. Rev.} \bibinfo{volume}{135}
  (\bibinfo{year}{1964}{\natexlab{b}}) \bibinfo{pages}{A1281--A1291}.
  \DOIprefix\doi{10.1103/PhysRev.135.A1281}.
\bibitem[{Angus and Dalgliesh(1970)}]{AngusPhysLettA1970}
\bibinfo{author}{R.~Angus}, \bibinfo{author}{I.~Dalgliesh},
\newblock \bibinfo{title}{Thermopower and resistivity of thin metal films},
\newblock \bibinfo{journal}{Physics Letters A} \bibinfo{volume}{31}
  (\bibinfo{year}{1970}) \bibinfo{pages}{280--281}.
  \DOIprefix\doi{10.1016/0375-9601(70)90973-4}.
\bibitem[{Lin and Leonard(1971)}]{LinJAP1971}
\bibinfo{author}{S.~F. Lin}, \bibinfo{author}{W.~F. Leonard},
\newblock \bibinfo{title}{Thermoelectric power of thin gold films},
\newblock \bibinfo{journal}{Journal of Applied Physics} \bibinfo{volume}{42}
  (\bibinfo{year}{1971}) \bibinfo{pages}{3634--3639}.
  \DOIprefix\doi{10.1063/1.1660781}.
\bibitem[{Strunk et~al.(1998)Strunk, Henny, Sch\"onenberger, Neuttiens, and
  Van~Haesendonck}]{StrunkPRL98}
\bibinfo{author}{C.~Strunk}, \bibinfo{author}{M.~Henny},
  \bibinfo{author}{C.~Sch\"onenberger}, \bibinfo{author}{G.~Neuttiens},
  \bibinfo{author}{C.~Van~Haesendonck},
\newblock \bibinfo{title}{Size dependent thermopower in mesoscopic {A}u{F}e
  wires},
\newblock \bibinfo{journal}{Phys. Rev. Lett.} \bibinfo{volume}{81}
  (\bibinfo{year}{1998}) \bibinfo{pages}{2982--2985}.
  \DOIprefix\doi{10.1103/PhysRevLett.81.2982}.
\bibitem[{Pichard et~al.(1980)Pichard, Tellier, and Tosser}]{PichardJPhysF1980}
\bibinfo{author}{C.~R. Pichard}, \bibinfo{author}{C.~R. Tellier},
  \bibinfo{author}{A.~J. Tosser},
\newblock \bibinfo{title}{Thermoelectric power of thin polycrystalline metal
  films in an effective mean free path model},
\newblock \bibinfo{journal}{Journal of Physics F: Metal Physics}
  \bibinfo{volume}{10} (\bibinfo{year}{1980}) \bibinfo{pages}{2009}.
\bibitem[{Tellier and Tosser(1977)}]{TellierTSF77a}
\bibinfo{author}{C.~Tellier}, \bibinfo{author}{A.~Tosser},
\newblock \bibinfo{title}{Thermoelectric power of metallic films in the
  {M}ayadas-{S}hatzkes model},
\newblock \bibinfo{journal}{Thin Solid Films} \bibinfo{volume}{41}
  (\bibinfo{year}{1977}) \bibinfo{pages}{161--166}.
  \DOIprefix\doi{10.1016/0040-6090(77)90399-6}.
\bibitem[{Tellier et~al.(1977)Tellier, Tosser, and Boutrit}]{TellierTSF77}
\bibinfo{author}{C.~Tellier}, \bibinfo{author}{A.~Tosser},
  \bibinfo{author}{C.~Boutrit},
\newblock \bibinfo{title}{The {M}ayadas-{S}hatzkes conduction model treated as
  a {F}uchs-{S}ondheimer model},
\newblock \bibinfo{journal}{Thin Solid Films} \bibinfo{volume}{44}
  (\bibinfo{year}{1977}) \bibinfo{pages}{201 -- 208}.
  \DOIprefix\doi{http://dx.doi.org/10.1016/0040-6090(77)90455-2}.
\bibitem[{Tellier(1978)}]{TellierTSF78}
\bibinfo{author}{C.~Tellier},
\newblock \bibinfo{title}{A theoretical description of grain boundary electron
  scattering by an effective mean free path},
\newblock \bibinfo{journal}{Thin Solid Films} \bibinfo{volume}{51}
  (\bibinfo{year}{1978}) \bibinfo{pages}{311 -- 317}.
  \DOIprefix\doi{http://dx.doi.org/10.1016/0040-6090(78)90293-6}.
\bibitem[{Herring et~al.(1958)Herring, Geballe, and Kunzler}]{HerringPR58}
\bibinfo{author}{C.~Herring}, \bibinfo{author}{T.~H. Geballe},
  \bibinfo{author}{J.~E. Kunzler},
\newblock \bibinfo{title}{Phonon-drag thermomagnetic effects in $n$-type
  germanium. i. general survey},
\newblock \bibinfo{journal}{Phys. Rev.} \bibinfo{volume}{111}
  (\bibinfo{year}{1958}) \bibinfo{pages}{36--57}.
  \DOIprefix\doi{10.1103/PhysRev.111.36}.
\bibitem[{Geballe and Hull(1955)}]{GeballePR55}
\bibinfo{author}{T.~H. Geballe}, \bibinfo{author}{G.~W. Hull},
\newblock \bibinfo{title}{Seebeck effect in silicon},
\newblock \bibinfo{journal}{Phys. Rev.} \bibinfo{volume}{98}
  (\bibinfo{year}{1955}) \bibinfo{pages}{940--947}.
  \DOIprefix\doi{10.1103/PhysRev.98.940}.
\bibitem[{Frederikse(1953)}]{FrederiksePR1953}
\bibinfo{author}{H.~P.~R. Frederikse},
\newblock \bibinfo{title}{Thermoelectric power of germanium below room
  temperature},
\newblock \bibinfo{journal}{Phys. Rev.} \bibinfo{volume}{92}
  (\bibinfo{year}{1953}) \bibinfo{pages}{248--252}.
  \DOIprefix\doi{10.1103/PhysRev.92.248}.
\bibitem[{Behnia and Aubin(2016)}]{BehniaRPP2016}
\bibinfo{author}{K.~Behnia}, \bibinfo{author}{H.~Aubin},
\newblock \bibinfo{title}{Nernst effect in metals and superconductors: a review
  of concepts and experiments},
\newblock \bibinfo{journal}{Reports on Progress in Physics}
  \bibinfo{volume}{79} (\bibinfo{year}{2016}) \bibinfo{pages}{046502}.
  \DOIprefix\doi{10.1088/0034-4885/79/4/046502}.
\bibitem[{Behnia(2009)}]{BehniaJPCM2009}
\bibinfo{author}{K.~Behnia},
\newblock \bibinfo{title}{The {N}ernst effect and the boundaries of the {F}ermi
  liquid picture},
\newblock \bibinfo{journal}{Journal of Physics: Condensed Matter}
  \bibinfo{volume}{21} (\bibinfo{year}{2009}) \bibinfo{pages}{113101}.
  \DOIprefix\doi{10.1088/0953-8984/21/11/113101}.
\bibitem[{Roschewsky et~al.(2019)Roschewsky, Walker, Gowtham, Muschinske,
  Hellman, Bank, and Salahuddin}]{RoschewskyPRB2019}
\bibinfo{author}{N.~Roschewsky}, \bibinfo{author}{E.~S. Walker},
  \bibinfo{author}{P.~Gowtham}, \bibinfo{author}{S.~Muschinske},
  \bibinfo{author}{F.~Hellman}, \bibinfo{author}{S.~R. Bank},
  \bibinfo{author}{S.~Salahuddin},
\newblock \bibinfo{title}{Spin-orbit torque and {N}ernst effect in
  {B}i-{S}b/{C}o heterostructures},
\newblock \bibinfo{journal}{Physical Review B} \bibinfo{volume}{99}
  (\bibinfo{year}{2019}) \bibinfo{pages}{195103}.
  \DOIprefix\doi{10.1103/physrevb.99.195103}.
\bibitem[{Yue et~al.(2021)Yue, Lin, and Chien}]{YueAPLMats2021}
\bibinfo{author}{D.~Yue}, \bibinfo{author}{W.~Lin}, \bibinfo{author}{C.~L.
  Chien},
\newblock \bibinfo{title}{Negligible spin-charge conversion in {B}i films and
  {B}i/{A}g({C}u) bilayers},
\newblock \bibinfo{journal}{APL Materials} \bibinfo{volume}{9}
  (\bibinfo{year}{2021}) \bibinfo{pages}{050904}.
  \DOIprefix\doi{10.1063/5.0048042}.
\bibitem[{Zhao et~al.(2016)Zhao, Qian, Gu, Jajja, and Yang}]{ZhaoJEP2016}
\bibinfo{author}{D.~Zhao}, \bibinfo{author}{X.~Qian}, \bibinfo{author}{X.~Gu},
  \bibinfo{author}{S.~A. Jajja}, \bibinfo{author}{R.~Yang},
\newblock \bibinfo{title}{{Measurement Techniques for Thermal Conductivity and
  Interfacial Thermal Conductance of Bulk and Thin Film Materials}},
\newblock \bibinfo{journal}{Journal of Electronic Packaging}
  \bibinfo{volume}{138} (\bibinfo{year}{2016}).
  \DOIprefix\doi{10.1115/1.4034605}, \bibinfo{note}{040802}.
\bibitem[{Toberer et~al.(2012)Toberer, Baranowski, and Dames}]{TobererARMR2012}
\bibinfo{author}{E.~S. Toberer}, \bibinfo{author}{L.~L. Baranowski},
  \bibinfo{author}{C.~Dames},
\newblock \bibinfo{title}{Advances in thermal conductivity},
\newblock \bibinfo{journal}{Annual Review of Materials Research}
  \bibinfo{volume}{42} (\bibinfo{year}{2012}) \bibinfo{pages}{179--209}.
\bibitem[{Avery et~al.(2015)Avery, Mason, Bassett, Wesenberg, and
  Zink}]{AveryPRB2015}
\bibinfo{author}{A.~D. Avery}, \bibinfo{author}{S.~J. Mason},
  \bibinfo{author}{D.~Bassett}, \bibinfo{author}{D.~Wesenberg},
  \bibinfo{author}{B.~L. Zink},
\newblock \bibinfo{title}{Thermal and electrical conductivity of approximately
  100-nm permalloy, {N}i, {C}o, {A}l, and {C}u films and examination of the
  {W}iedemann-{F}ranz law},
\newblock \bibinfo{journal}{Phys. Rev. B} \bibinfo{volume}{92}
  (\bibinfo{year}{2015}) \bibinfo{pages}{214410}.
  \DOIprefix\doi{10.1103/PhysRevB.92.214410}.
\bibitem[{McGuire and Potter(1975)}]{McGuireIEEE75}
\bibinfo{author}{T.~R. McGuire}, \bibinfo{author}{R.~I. Potter},
\newblock \bibinfo{title}{Anisotropic magnetoresistance in ferromagnetic 3d
  alloys},
\newblock \bibinfo{journal}{IEEE Transactions on Magnetics}
  \bibinfo{volume}{MAG-11} (\bibinfo{year}{1975}) \bibinfo{pages}{1018}.
\bibitem[{Hall(1879)}]{HallAmJMath1879}
\bibinfo{author}{E.~Hall},
\newblock \bibinfo{title}{On a new action of magnetic on electric currents},
\newblock \bibinfo{journal}{Am. J. Math} \bibinfo{volume}{2}
  (\bibinfo{year}{1879}) \bibinfo{pages}{287--292}.
\bibitem[{Nagaosa et~al.(2010)Nagaosa, Sinova, Onoda, MacDonald, and
  Ong}]{NagaosaRMP2010}
\bibinfo{author}{N.~Nagaosa}, \bibinfo{author}{J.~Sinova},
  \bibinfo{author}{S.~Onoda}, \bibinfo{author}{A.~H. MacDonald},
  \bibinfo{author}{N.~P. Ong},
\newblock \bibinfo{title}{Anomalous {H}all effect},
\newblock \bibinfo{journal}{Rev. Mod. Phys.} \bibinfo{volume}{82}
  (\bibinfo{year}{2010}) \bibinfo{pages}{1539--1592}.
  \DOIprefix\doi{10.1103/RevModPhys.82.1539}.
\bibitem[{Slachter et~al.(2011)Slachter, Bakker, and van
  Wees}]{SlachterPRB2011}
\bibinfo{author}{A.~Slachter}, \bibinfo{author}{F.~L. Bakker},
  \bibinfo{author}{B.~J. van Wees},
\newblock \bibinfo{title}{Anomalous {N}ernst and anisotropic magnetoresistive
  heating in a lateral spin valve},
\newblock \bibinfo{journal}{Phys. Rev. B} \bibinfo{volume}{84}
  (\bibinfo{year}{2011}) \bibinfo{pages}{020412}.
  \DOIprefix\doi{10.1103/PhysRevB.84.020412}.
\bibitem[{von Bieren et~al.(2013)von Bieren, Brandl, Grundler, and
  Ansermet}]{vonBierenAPL13}
\bibinfo{author}{A.~von Bieren}, \bibinfo{author}{F.~Brandl},
  \bibinfo{author}{D.~Grundler}, \bibinfo{author}{J.-P. Ansermet},
\newblock \bibinfo{title}{Space- and time-resolved {S}eebeck and {N}ernst
  voltages in laser-heated permalloy/gold microstructures},
\newblock \bibinfo{journal}{Applied Physics Letters} \bibinfo{volume}{102}
  (\bibinfo{year}{2013}) \bibinfo{pages}{052408}.
  \DOIprefix\doi{10.1063/1.4789974}.
\bibitem[{Brandl and Grundler(2014)}]{BrandlAPL2014}
\bibinfo{author}{F.~Brandl}, \bibinfo{author}{D.~Grundler},
\newblock \bibinfo{title}{Fabrication and local laser heating of freestanding
  {N}i$_{80}${F}e$_{20}$ bridges with pt contacts displaying anisotropic
  magnetoresistance and anomalous {N}ernst effect},
\newblock \bibinfo{journal}{Applied Physics Letters} \bibinfo{volume}{104}
  (\bibinfo{year}{2014}) \bibinfo{pages}{172401}.
  \DOIprefix\doi{10.1063/1.4874302}.
\bibitem[{Chuang et~al.(2017)Chuang, Su, Wu, and Huang}]{ChuangPRB2017}
\bibinfo{author}{T.~C. Chuang}, \bibinfo{author}{P.~L. Su},
  \bibinfo{author}{P.~H. Wu}, \bibinfo{author}{S.~Y. Huang},
\newblock \bibinfo{title}{Enhancement of the anomalous {N}ernst effect in
  ferromagnetic thin films},
\newblock \bibinfo{journal}{Phys. Rev. B} \bibinfo{volume}{96}
  (\bibinfo{year}{2017}) \bibinfo{pages}{174406}.
  \DOIprefix\doi{10.1103/PhysRevB.96.174406}.
\bibitem[{Bennet et~al.(2019)Bennet, Hojem, and Zink}]{BennetPRB2019}
\bibinfo{author}{R.~K. Bennet}, \bibinfo{author}{A.~Hojem},
  \bibinfo{author}{B.~L. Zink},
\newblock \bibinfo{title}{Thermal gradients and anomalous {N}ernst effects in
  membrane-supported nonlocal spin valves},
\newblock \bibinfo{journal}{Phys. Rev. B} \bibinfo{volume}{100}
  (\bibinfo{year}{2019}) \bibinfo{pages}{104404}.
  \DOIprefix\doi{10.1103/PhysRevB.100.104404}.
\bibitem[{Yamazaki et~al.(2022)Yamazaki, Seki, Modak, Nakagawara, Hirai, Ito,
  Uchida, and Takanashi}]{YamazakiPRB2022}
\bibinfo{author}{T.~Yamazaki}, \bibinfo{author}{T.~Seki},
  \bibinfo{author}{R.~Modak}, \bibinfo{author}{K.~Nakagawara},
  \bibinfo{author}{T.~Hirai}, \bibinfo{author}{K.~Ito}, \bibinfo{author}{K.-i.
  Uchida}, \bibinfo{author}{K.~Takanashi},
\newblock \bibinfo{title}{Thickness dependence of anomalous hall and nernst
  effects in {N}i-{F}e thin films},
\newblock \bibinfo{journal}{Phys. Rev. B} \bibinfo{volume}{105}
  (\bibinfo{year}{2022}) \bibinfo{pages}{214416}.
  \DOIprefix\doi{10.1103/PhysRevB.105.214416}.
\bibitem[{Martens et~al.(2018)Martens, Huebner, Ulrichs, Reimer, Kuschel,
  Tamming, Chang, Tobey, Thomas, M\"{u}nzenberg, and
  Walowski}]{MartensCommPhys2018}
\bibinfo{author}{U.~Martens}, \bibinfo{author}{T.~Huebner},
  \bibinfo{author}{H.~Ulrichs}, \bibinfo{author}{O.~Reimer},
  \bibinfo{author}{T.~Kuschel}, \bibinfo{author}{R.~R. Tamming},
  \bibinfo{author}{C.-L. Chang}, \bibinfo{author}{R.~I. Tobey},
  \bibinfo{author}{A.~Thomas}, \bibinfo{author}{M.~M\"{u}nzenberg},
  \bibinfo{author}{J.~Walowski},
\newblock \bibinfo{title}{Anomalous {N}ernst effect and three-dimensional
  temperature gradients in magnetic tunnel junctions},
\newblock \bibinfo{journal}{Communications Physics} \bibinfo{volume}{1}
  (\bibinfo{year}{2018}) \bibinfo{pages}{65}.
  \DOIprefix\doi{10.1038/s42005-018-0063-y}.
\bibitem[{Bougiatioti et~al.(2017)Bougiatioti, Klewe, Meier, Manos, Kuschel,
  Wollschl\"ager, Bouchenoire, Brown, Schmalhorst, Reiss, and
  Kuschel}]{BougiatiotiPRL2017}
\bibinfo{author}{P.~Bougiatioti}, \bibinfo{author}{C.~Klewe},
  \bibinfo{author}{D.~Meier}, \bibinfo{author}{O.~Manos},
  \bibinfo{author}{O.~Kuschel}, \bibinfo{author}{J.~Wollschl\"ager},
  \bibinfo{author}{L.~Bouchenoire}, \bibinfo{author}{S.~D. Brown},
  \bibinfo{author}{J.-M. Schmalhorst}, \bibinfo{author}{G.~Reiss},
  \bibinfo{author}{T.~Kuschel},
\newblock \bibinfo{title}{Quantitative disentanglement of the spin {S}eebeck,
  proximity-induced, and ferromagnetic-induced anomalous {N}ernst effect in
  normal-metal/ferromagnet bilayers},
\newblock \bibinfo{journal}{Phys. Rev. Lett.} \bibinfo{volume}{119}
  (\bibinfo{year}{2017}) \bibinfo{pages}{227205}.
  \DOIprefix\doi{10.1103/PhysRevLett.119.227205}.
\bibitem[{Kannan et~al.(2017)Kannan, Fan, Celik, Han, and
  Xiao}]{KannanSciRep2017}
\bibinfo{author}{H.~Kannan}, \bibinfo{author}{X.~Fan},
  \bibinfo{author}{H.~Celik}, \bibinfo{author}{X.~Han}, \bibinfo{author}{J.~Q.
  Xiao},
\newblock \bibinfo{title}{Thickness dependence of anomalous {N}ernst
  coefficient and longitudinal spin {S}eebeck effect in ferromagnetic
  {N}i$_{x}${F}e$_{100-x}$ films},
\newblock \bibinfo{journal}{Scientific Reports} \bibinfo{volume}{7}
  (\bibinfo{year}{2017}). \DOIprefix\doi{10.1038/s41598-017-05946-1}.
\bibitem[{Elzwawy et~al.(2021)Elzwawy, Piskin, Akdo{\u{g}}an, Volmer, Reiss,
  Marnitz, Moskaltsova, Gurel, and Schmalhorst}]{ElzwawyJPhysD2021}
\bibinfo{author}{A.~Elzwawy}, \bibinfo{author}{H.~Piskin},
  \bibinfo{author}{N.~Akdo{\u{g}}an}, \bibinfo{author}{M.~Volmer},
  \bibinfo{author}{G.~Reiss}, \bibinfo{author}{L.~Marnitz},
  \bibinfo{author}{A.~Moskaltsova}, \bibinfo{author}{O.~Gurel},
  \bibinfo{author}{J.~Schmalhorst},
\newblock \bibinfo{title}{Current trends in planar {H}all effect sensors:
  evolution, optimization, and applications},
\newblock \bibinfo{journal}{Journal of Physics D: Applied Physics}
  (\bibinfo{year}{2021}). \DOIprefix\doi{10.1088/1361-6463/abfbfb}.
\bibitem[{Ky(1967)}]{KyPSSb1967}
\bibinfo{author}{V.~D. Ky},
\newblock \bibinfo{title}{Planar {H}all and {N}ernst effect in ferromagnetic
  metals},
\newblock \bibinfo{journal}{physica status solidi (b)} \bibinfo{volume}{22}
  (\bibinfo{year}{1967}) \bibinfo{pages}{729--736}.
  \DOIprefix\doi{10.1002/pssb.19670220242}.
\bibitem[{Ky(1966)}]{KyPSSb1966}
\bibinfo{author}{V.~D. Ky},
\newblock \bibinfo{title}{The planar {N}ernst effect in permalloy films},
\newblock \bibinfo{journal}{physica status solidi (b)} \bibinfo{volume}{17}
  (\bibinfo{year}{1966}) \bibinfo{pages}{K207--K209}.
  \DOIprefix\doi{10.1002/pssb.19660170262}.
\bibitem[{Pu et~al.(2006)Pu, Johnston-Halperin, Awschalom, and Shi}]{PuPRL2006}
\bibinfo{author}{Y.~Pu}, \bibinfo{author}{E.~Johnston-Halperin},
  \bibinfo{author}{D.~D. Awschalom}, \bibinfo{author}{J.~Shi},
\newblock \bibinfo{title}{Anisotropic thermopower and planar {N}ernst effect in
  {G}a$_{1-x}${M}n$_{x}${A}s ferromagnetic semiconductors},
\newblock \bibinfo{journal}{Phys. Rev. Lett.} \bibinfo{volume}{97}
  (\bibinfo{year}{2006}) \bibinfo{pages}{036601}.
  \DOIprefix\doi{10.1103/PhysRevLett.97.036601}.
\bibitem[{Avery et~al.(2012)Avery, Pufall, and Zink}]{AveryPRL2012}
\bibinfo{author}{A.~D. Avery}, \bibinfo{author}{M.~R. Pufall},
  \bibinfo{author}{B.~L. Zink},
\newblock \bibinfo{title}{Observation of the planar {N}ernst effect in
  permalloy and nickel thin films with in-plane thermal gradients},
\newblock \bibinfo{journal}{Physical Review Letters} \bibinfo{volume}{109}
  (\bibinfo{year}{2012}) \bibinfo{pages}{196602}.
\bibitem[{Reimer et~al.(2017)Reimer, Meier, Bovender, Helmich, Dreessen,
  Krieft, Shestakov, Back, Schmalhorst, H\"{u}tten, Reiss, and
  Kuschel}]{ReimerSciRep2017}
\bibinfo{author}{O.~Reimer}, \bibinfo{author}{D.~Meier},
  \bibinfo{author}{M.~Bovender}, \bibinfo{author}{L.~Helmich},
  \bibinfo{author}{J.-O. Dreessen}, \bibinfo{author}{J.~Krieft},
  \bibinfo{author}{A.~S. Shestakov}, \bibinfo{author}{C.~H. Back},
  \bibinfo{author}{J.-M. Schmalhorst}, \bibinfo{author}{A.~H\"{u}tten},
  \bibinfo{author}{G.~Reiss}, \bibinfo{author}{T.~Kuschel},
\newblock \bibinfo{title}{Quantitative separation of the anisotropic
  magnetothermopower and planar {N}ernst effect by the rotation of an in-plane
  thermal gradient},
\newblock \bibinfo{journal}{Scientific Reports} \bibinfo{volume}{7}
  (\bibinfo{year}{2017}) \bibinfo{pages}{40586}.
  \DOIprefix\doi{10.1038/srep40586}.
\bibitem[{Marrows(2005)}]{MarrowsAdvPhys2005}
\bibinfo{author}{C.~H. Marrows},
\newblock \bibinfo{title}{Spin-polarised currents and magnetic domain walls},
\newblock \bibinfo{journal}{Advances in Physics} \bibinfo{volume}{54}
  (\bibinfo{year}{2005}) \bibinfo{pages}{585--713}.
  \DOIprefix\doi{10.1080/00018730500442209}.
\bibitem[{Kent et~al.(2001)Kent, Yu, Rüdiger, and Parkin}]{KentJPhysCM2001}
\bibinfo{author}{A.~D. Kent}, \bibinfo{author}{J.~Yu},
  \bibinfo{author}{U.~Rüdiger}, \bibinfo{author}{S.~S.~P. Parkin},
\newblock \bibinfo{title}{Domain wall resistivity in epitaxial thin film
  microstructures},
\newblock \bibinfo{journal}{Journal of Physics: Condensed Matter}
  \bibinfo{volume}{13} (\bibinfo{year}{2001}) \bibinfo{pages}{R461--R488}.
  \DOIprefix\doi{10.1088/0953-8984/13/25/202}.
\bibitem[{Avery et~al.(2012)Avery, Pufall, and Zink}]{AveryPRB2012}
\bibinfo{author}{A.~D. Avery}, \bibinfo{author}{M.~R. Pufall},
  \bibinfo{author}{B.~L. Zink},
\newblock \bibinfo{title}{Predicting the planar {N}ernst effect from
  magnetic-field-dependent thermopower and resistance in nickel and permalloy
  thin films},
\newblock \bibinfo{journal}{Physical Review B} \bibinfo{volume}{86}
  (\bibinfo{year}{2012}) \bibinfo{pages}{184408}.
\bibitem[{Kimling et~al.(2013)Kimling, Gooth, and
  Nielsch}]{Kimling-GoothPRB2013}
\bibinfo{author}{J.~Kimling}, \bibinfo{author}{J.~Gooth},
  \bibinfo{author}{K.~Nielsch},
\newblock \bibinfo{title}{Anisotropic magnetothermal resistance in {N}i
  nanowires},
\newblock \bibinfo{journal}{Phys. Rev. B} \bibinfo{volume}{87}
  (\bibinfo{year}{2013}) \bibinfo{pages}{094409}.
  \DOIprefix\doi{10.1103/PhysRevB.87.094409}.
\bibitem[{Wesenberg et~al.(2018)Wesenberg, Hojem, Bennet, and
  Zink}]{WesenbergJPhysD2018}
\bibinfo{author}{D.~Wesenberg}, \bibinfo{author}{A.~Hojem},
  \bibinfo{author}{R.~K. Bennet}, \bibinfo{author}{B.~L. Zink},
\newblock \bibinfo{title}{Relation of planar {H}all and planar {N}ernst effects
  in thin film permalloy},
\newblock \bibinfo{journal}{Journal of Physics D: Applied Physics}
  \bibinfo{volume}{51} (\bibinfo{year}{2018}) \bibinfo{pages}{244005}.
  \DOIprefix\doi{10.1088/1361-6463/aac2b3}.
\bibitem[{Avery and Zink(2013)}]{AveryPRL2013}
\bibinfo{author}{A.~D. Avery}, \bibinfo{author}{B.~L. Zink},
\newblock \bibinfo{title}{Peltier cooling and {O}nsager reciprocity in
  ferromagnetic thin films},
\newblock \bibinfo{journal}{Phys. Rev. Lett.} \bibinfo{volume}{111}
  (\bibinfo{year}{2013}) \bibinfo{pages}{126602}.
  \DOIprefix\doi{10.1103/PhysRevLett.111.126602}.
\bibitem[{Gravier et~al.(2006)Gravier, Serrano-Guisan, Reuse, and
  Ansermet}]{GravierPRB06b}
\bibinfo{author}{L.~Gravier}, \bibinfo{author}{S.~Serrano-Guisan},
  \bibinfo{author}{F.~Reuse}, \bibinfo{author}{J.-P. Ansermet},
\newblock \bibinfo{title}{Spin-dependent {P}eltier effect of perpendicular
  currents in multilayered nanowires},
\newblock \bibinfo{journal}{Physical Review B (Condensed Matter and Materials
  Physics)} \bibinfo{volume}{73} (\bibinfo{year}{2006})
  \bibinfo{pages}{052410}. \DOIprefix\doi{10.1103/PhysRevB.73.052410}.
\bibitem[{Fukushima et~al.(2006)Fukushima, Kubota, Yamamoto, Suzuki, and
  Yuasa}]{FukushimaJAP06}
\bibinfo{author}{A.~Fukushima}, \bibinfo{author}{H.~Kubota},
  \bibinfo{author}{A.~Yamamoto}, \bibinfo{author}{Y.~Suzuki},
  \bibinfo{author}{S.~Yuasa},
\newblock \bibinfo{title}{Peltier cooling in current-perpendicular-to-plane
  metallic junctions},
\newblock \bibinfo{journal}{Journal of Applied Physics} \bibinfo{volume}{99}
  (\bibinfo{year}{2006}) \bibinfo{pages}{08H706}.
  \DOIprefix\doi{10.1063/1.2172211}.
\bibitem[{Shan et~al.(2015)Shan, Dejene, Leutenantsmeyer, Flipse, M\"unzenberg,
  and van Wees}]{ShanPRB2015}
\bibinfo{author}{J.~Shan}, \bibinfo{author}{F.~K. Dejene},
  \bibinfo{author}{J.~C. Leutenantsmeyer}, \bibinfo{author}{J.~Flipse},
  \bibinfo{author}{M.~M\"unzenberg}, \bibinfo{author}{B.~J. van Wees},
\newblock \bibinfo{title}{Comparison of the magneto-{P}eltier and
  magneto-{S}eebeck effects in magnetic tunnel junctions},
\newblock \bibinfo{journal}{Phys. Rev. B} \bibinfo{volume}{92}
  (\bibinfo{year}{2015}) \bibinfo{pages}{020414}.
  \DOIprefix\doi{10.1103/PhysRevB.92.020414}.
\bibitem[{Seki et~al.(2018{\natexlab{a}})Seki, Iguchi, Takanashi, and
  Uchida}]{SekiJPhysD2018}
\bibinfo{author}{T.~Seki}, \bibinfo{author}{R.~Iguchi},
  \bibinfo{author}{K.~Takanashi}, \bibinfo{author}{K.~Uchida},
\newblock \bibinfo{title}{Relationship between anomalous {E}ttingshausen effect
  and anomalous {N}ernst effect in an {FePt} thin film},
\newblock \bibinfo{journal}{Journal of Physics D: Applied Physics}
  \bibinfo{volume}{51} (\bibinfo{year}{2018}{\natexlab{a}})
  \bibinfo{pages}{254001}. \DOIprefix\doi{10.1088/1361-6463/aac481}.
\bibitem[{Seki et~al.(2018{\natexlab{b}})Seki, Iguchi, Takanashi, and
  Uchida}]{SekiAPL2018}
\bibinfo{author}{T.~Seki}, \bibinfo{author}{R.~Iguchi},
  \bibinfo{author}{K.~Takanashi}, \bibinfo{author}{K.~Uchida},
\newblock \bibinfo{title}{Visualization of anomalous {E}ttingshausen effect in
  a ferromagnetic film: Direct evidence of different symmetry from spin
  {P}eltier effect},
\newblock \bibinfo{journal}{Applied Physics Letters} \bibinfo{volume}{112}
  (\bibinfo{year}{2018}{\natexlab{b}}) \bibinfo{pages}{152403}.
  \DOIprefix\doi{10.1063/1.5022759}.
\bibitem[{Madon et~al.(2016)Madon, Pham, Wegrowe, Lacour, Hehn, Polewczyk,
  Anane, and Cros}]{MadonPRB2016}
\bibinfo{author}{B.~Madon}, \bibinfo{author}{D.~C. Pham},
  \bibinfo{author}{J.-E. Wegrowe}, \bibinfo{author}{D.~Lacour},
  \bibinfo{author}{M.~Hehn}, \bibinfo{author}{V.~Polewczyk},
  \bibinfo{author}{A.~Anane}, \bibinfo{author}{V.~Cros},
\newblock \bibinfo{title}{Anomalous and planar {R}ighi-{L}educ effects in
  {N}i$_{80}${F}e$_{20}$ ferromagnets},
\newblock \bibinfo{journal}{Phys. Rev. B} \bibinfo{volume}{94}
  (\bibinfo{year}{2016}) \bibinfo{pages}{144423}.
  \DOIprefix\doi{10.1103/PhysRevB.94.144423}.
\bibitem[{Chen and Huang(2016)}]{ChenPRL2016}
\bibinfo{author}{Y.-J. Chen}, \bibinfo{author}{S.-Y. Huang},
\newblock \bibinfo{title}{Absence of the thermal {H}all effect in anomalous
  {N}ernst and spin {S}eebeck effects},
\newblock \bibinfo{journal}{Phys. Rev. Lett.} \bibinfo{volume}{117}
  (\bibinfo{year}{2016}) \bibinfo{pages}{247201}.
  \DOIprefix\doi{10.1103/PhysRevLett.117.247201}.
\bibitem[{Sinova et~al.(2015)Sinova, Valenzuela, Wunderlich, Back, and
  Jungwirth}]{SinovaRMP2015}
\bibinfo{author}{J.~Sinova}, \bibinfo{author}{S.~O. Valenzuela},
  \bibinfo{author}{J.~Wunderlich}, \bibinfo{author}{C.~H. Back},
  \bibinfo{author}{T.~Jungwirth},
\newblock \bibinfo{title}{Spin {H}all effects},
\newblock \bibinfo{journal}{Rev. Mod. Phys.} \bibinfo{volume}{87}
  (\bibinfo{year}{2015}) \bibinfo{pages}{1213--1260}.
  \DOIprefix\doi{10.1103/RevModPhys.87.1213}.
\bibitem[{Hoffmann(2013)}]{HoffmannIEEETransMag2013}
\bibinfo{author}{A.~Hoffmann},
\newblock \bibinfo{title}{Spin {H}all effects in metals},
\newblock \bibinfo{journal}{Magnetics, IEEE Transactions on}
  \bibinfo{volume}{49} (\bibinfo{year}{2013}) \bibinfo{pages}{5172--5193}.
  \DOIprefix\doi{10.1109/TMAG.2013.2262947}.
\bibitem[{Davidson et~al.(2020)Davidson, Amin, Aljuaid, Haney, and
  Fan}]{DavidsonPLA2020}
\bibinfo{author}{A.~Davidson}, \bibinfo{author}{V.~P. Amin},
  \bibinfo{author}{W.~S. Aljuaid}, \bibinfo{author}{P.~M. Haney},
  \bibinfo{author}{X.~Fan},
\newblock \bibinfo{title}{Perspectives of electrically generated spin currents
  in ferromagnetic materials},
\newblock \bibinfo{journal}{Physics Letters A} \bibinfo{volume}{384}
  (\bibinfo{year}{2020}) \bibinfo{pages}{126228}.
  \DOIprefix\doi{10.1016/j.physleta.2019.126228}.
\bibitem[{Cheng et~al.(2008)Cheng, Xing, Sun, and Xie}]{ChengPRB2008}
\bibinfo{author}{S.-g. Cheng}, \bibinfo{author}{Y.~Xing},
  \bibinfo{author}{Q.-f. Sun}, \bibinfo{author}{X.~C. Xie},
\newblock \bibinfo{title}{Spin {N}ernst effect and {N}ernst effect in
  two-dimensional electron systems},
\newblock \bibinfo{journal}{Phys. Rev. B} \bibinfo{volume}{78}
  (\bibinfo{year}{2008}) \bibinfo{pages}{045302}.
  \DOIprefix\doi{10.1103/PhysRevB.78.045302}.
\bibitem[{Liu and Xie(2010)}]{LiuSSC2010}
\bibinfo{author}{X.~Liu}, \bibinfo{author}{X.~Xie},
\newblock \bibinfo{title}{Spin {N}ernst effect in the absence of a magnetic
  field},
\newblock \bibinfo{journal}{Solid State Communications} \bibinfo{volume}{150}
  (\bibinfo{year}{2010}) \bibinfo{pages}{471--474}.
  \DOIprefix\doi{https://doi.org/10.1016/j.ssc.2009.12.017},
  \bibinfo{note}{spin Caloritronics}.
\bibitem[{Tauber et~al.(2012)Tauber, Gradhand, Fedorov, and
  Mertig}]{TauberPRL2012}
\bibinfo{author}{K.~Tauber}, \bibinfo{author}{M.~Gradhand},
  \bibinfo{author}{D.~V. Fedorov}, \bibinfo{author}{I.~Mertig},
\newblock \bibinfo{title}{Extrinsic spin {N}ernst effect from first
  principles},
\newblock \bibinfo{journal}{Phys. Rev. Lett.} \bibinfo{volume}{109}
  (\bibinfo{year}{2012}) \bibinfo{pages}{026601}.
  \DOIprefix\doi{10.1103/PhysRevLett.109.026601}.
\bibitem[{Wimmer et~al.(2013)Wimmer, K\"odderitzsch, Chadova, and
  Ebert}]{WimmerPRB2013}
\bibinfo{author}{S.~Wimmer}, \bibinfo{author}{D.~K\"odderitzsch},
  \bibinfo{author}{K.~Chadova}, \bibinfo{author}{H.~Ebert},
\newblock \bibinfo{title}{First-principles linear response description of the
  spin {N}ernst effect},
\newblock \bibinfo{journal}{Phys. Rev. B} \bibinfo{volume}{88}
  (\bibinfo{year}{2013}) \bibinfo{pages}{201108}.
  \DOIprefix\doi{10.1103/PhysRevB.88.201108}.
\bibitem[{Meyer et~al.(2017)Meyer, Chen, Wimmer, Althammer, Wimmer, Schlitz,
  Gepr\"{a}gs, Huebl, K\"{o}dderitzsch, Ebert, Bauer, Gross, and
  Goennenwein}]{MeyerNatMater2017}
\bibinfo{author}{S.~Meyer}, \bibinfo{author}{Y.-T. Chen},
  \bibinfo{author}{S.~Wimmer}, \bibinfo{author}{M.~Althammer},
  \bibinfo{author}{T.~Wimmer}, \bibinfo{author}{R.~Schlitz},
  \bibinfo{author}{S.~Gepr\"{a}gs}, \bibinfo{author}{H.~Huebl},
  \bibinfo{author}{D.~K\"{o}dderitzsch}, \bibinfo{author}{H.~Ebert},
  \bibinfo{author}{G.~E.~W. Bauer}, \bibinfo{author}{R.~Gross},
  \bibinfo{author}{S.~T.~B. Goennenwein},
\newblock \bibinfo{title}{Observation of the spin {N}ernst effect},
\newblock \bibinfo{journal}{Nature Materials} \bibinfo{volume}{16}
  (\bibinfo{year}{2017}) \bibinfo{pages}{977--981}.
  \DOIprefix\doi{10.1038/nmat4964}.
\bibitem[{Sheng et~al.(2017)Sheng, Sakuraba, Lau, Takahashi, Mitani, and
  Hayashi}]{ShengSciAdv2017}
\bibinfo{author}{P.~Sheng}, \bibinfo{author}{Y.~Sakuraba},
  \bibinfo{author}{Y.-C. Lau}, \bibinfo{author}{S.~Takahashi},
  \bibinfo{author}{S.~Mitani}, \bibinfo{author}{M.~Hayashi},
\newblock \bibinfo{title}{The spin {N}ernst effect in tungsten},
\newblock \bibinfo{journal}{Science Advances} \bibinfo{volume}{3}
  (\bibinfo{year}{2017}) \bibinfo{pages}{e1701503}.
  \DOIprefix\doi{10.1126/sciadv.1701503}.
\bibitem[{Bose et~al.(2018)Bose, Bhuktare, Singh, Dutta, Achanta, and
  Tulapurkar}]{BoseAPL2018}
\bibinfo{author}{A.~Bose}, \bibinfo{author}{S.~Bhuktare},
  \bibinfo{author}{H.~Singh}, \bibinfo{author}{S.~Dutta},
  \bibinfo{author}{V.~G. Achanta}, \bibinfo{author}{A.~A. Tulapurkar},
\newblock \bibinfo{title}{Direct detection of spin {N}ernst effect in
  platinum},
\newblock \bibinfo{journal}{Applied Physics Letters} \bibinfo{volume}{112}
  (\bibinfo{year}{2018}) \bibinfo{pages}{162401}.
  \DOIprefix\doi{10.1063/1.5021731}.
\bibitem[{Yu et~al.(2017)Yu, Brechet, and Ansermet}]{YuPLA2017}
\bibinfo{author}{H.~Yu}, \bibinfo{author}{S.~D. Brechet},
  \bibinfo{author}{J.-P. Ansermet},
\newblock \bibinfo{title}{Spin caloritronics, origin and outlook},
\newblock \bibinfo{journal}{Physics Letters A} \bibinfo{volume}{381}
  (\bibinfo{year}{2017}) \bibinfo{pages}{825--837}.
  \DOIprefix\doi{https://doi.org/10.1016/j.physleta.2016.12.038}.
\bibitem[{Bass and Pratt(2007)}]{BassJPCM2007}
\bibinfo{author}{J.~Bass}, \bibinfo{author}{W.~P. Pratt},
\newblock \bibinfo{title}{Spin-diffusion lengths in metals and alloys, and
  spin-flipping at metal/metal interfaces: an experimentalist's critical
  review},
\newblock \bibinfo{journal}{Journal of Physics: Condensed Matter}
  \bibinfo{volume}{19} (\bibinfo{year}{2007}) \bibinfo{pages}{183201}.
\bibitem[{Slachter et~al.(2010)Slachter, Bakker, Adam, and van
  Wees}]{SlachterNatPhys2010}
\bibinfo{author}{A.~Slachter}, \bibinfo{author}{F.~L. Bakker},
  \bibinfo{author}{J.-P. Adam}, \bibinfo{author}{B.~J. van Wees},
\newblock \bibinfo{title}{Thermally driven spin injection from a ferromagnet
  into a non-magnetic metal},
\newblock \bibinfo{journal}{Nature Physics} \bibinfo{volume}{6}
  (\bibinfo{year}{2010}) \bibinfo{pages}{879}.
\bibitem[{Erekhinsky et~al.(2012)Erekhinsky, Casanova, Schuller, and
  Sharoni}]{ErekhinskyAPL2012}
\bibinfo{author}{M.~Erekhinsky}, \bibinfo{author}{F.~Casanova},
  \bibinfo{author}{I.~K. Schuller}, \bibinfo{author}{A.~Sharoni},
\newblock \bibinfo{title}{Spin-dependent {S}eebeck effect in non-local spin
  valve devices},
\newblock \bibinfo{journal}{Applied Physics Letters} \bibinfo{volume}{100}
  (\bibinfo{year}{2012}) \bibinfo{pages}{--}.
  \DOIprefix\doi{http://dx.doi.org/10.1063/1.4717752}.
\bibitem[{Hu et~al.(2014)Hu, Itoh, and Kimura}]{HuNPGAM2014}
\bibinfo{author}{S.~Hu}, \bibinfo{author}{H.~Itoh},
  \bibinfo{author}{T.~Kimura},
\newblock \bibinfo{title}{Efficient thermal spin injection using {C}o{F}e{A}l
  nanowire},
\newblock \bibinfo{journal}{NPG Asia Materials} \bibinfo{volume}{6}
  (\bibinfo{year}{2014}) \bibinfo{pages}{e127}.
  \DOIprefix\doi{10.1038/am.2014.74}.
\bibitem[{Hu and Kimura(2014)}]{HuPRB2014}
\bibinfo{author}{S.~Hu}, \bibinfo{author}{T.~Kimura},
\newblock \bibinfo{title}{Significant modulation of electrical spin
  accumulation by efficient thermal spin injection},
\newblock \bibinfo{journal}{Phys. Rev. B} \bibinfo{volume}{90}
  (\bibinfo{year}{2014}) \bibinfo{pages}{134412}.
  \DOIprefix\doi{10.1103/PhysRevB.90.134412}.
\bibitem[{Yamasaki et~al.(2015)Yamasaki, Oki, Yamada, Kanashima, and
  Hamaya}]{YamasakiAPE2015}
\bibinfo{author}{K.~Yamasaki}, \bibinfo{author}{S.~Oki},
  \bibinfo{author}{S.~Yamada}, \bibinfo{author}{T.~Kanashima},
  \bibinfo{author}{K.~Hamaya},
\newblock \bibinfo{title}{Spin-related thermoelectric conversion in lateral
  spin-valve devices with single-crystalline {C}o$_{2}${F}e{S}i electrodes},
\newblock \bibinfo{journal}{Applied Physics Express} \bibinfo{volume}{8}
  (\bibinfo{year}{2015}) \bibinfo{pages}{043003}.
\bibitem[{Pfeiffer et~al.(2015)Pfeiffer, Hu, Reeve, Kronenberg, Jourdan,
  Kimura, and Klaui}]{PfeifferAPL2015}
\bibinfo{author}{A.~Pfeiffer}, \bibinfo{author}{S.~Hu}, \bibinfo{author}{R.~M.
  Reeve}, \bibinfo{author}{A.~Kronenberg}, \bibinfo{author}{M.~Jourdan},
  \bibinfo{author}{T.~Kimura}, \bibinfo{author}{M.~Klaui},
\newblock \bibinfo{title}{Spin currents injected electrically and thermally
  from highly spin polarized {C}o$_{2}${M}n{S}i},
\newblock \bibinfo{journal}{Applied Physics Letters} \bibinfo{volume}{107}
  (\bibinfo{year}{2015}) \bibinfo{pages}{--}.
  \DOIprefix\doi{http://dx.doi.org/10.1063/1.4929423}.
\bibitem[{Choi et~al.(2015)Choi, Moon, Min, Lee, and Cahill}]{ChoiNatPhys2015}
\bibinfo{author}{G.-M. Choi}, \bibinfo{author}{C.-H. Moon},
  \bibinfo{author}{B.-C. Min}, \bibinfo{author}{K.-J. Lee},
  \bibinfo{author}{D.~G. Cahill},
\newblock \bibinfo{title}{Thermal spin-transfer torque driven by the
  spin-dependent {S}eebeck effect in metallic spin-valves},
\newblock \bibinfo{journal}{Nature Physics} \bibinfo{volume}{11}
  (\bibinfo{year}{2015}) \bibinfo{pages}{576}.
  \DOIprefix\doi{10.1038/nphys3355}.
\bibitem[{Hojem et~al.(2016)Hojem, Wesenberg, and Zink}]{HojemPRB2016}
\bibinfo{author}{A.~Hojem}, \bibinfo{author}{D.~Wesenberg},
  \bibinfo{author}{B.~L. Zink},
\newblock \bibinfo{title}{Thermal spin injection and interface insensitivity in
  permalloy/aluminum metallic nonlocal spin valves},
\newblock \bibinfo{journal}{Phys. Rev. B} \bibinfo{volume}{94}
  (\bibinfo{year}{2016}) \bibinfo{pages}{024426}.
  \DOIprefix\doi{10.1103/PhysRevB.94.024426}.
\bibitem[{Hu et~al.(2017)Hu, Cui, Nomura, Min, and Kimura}]{HuPRB2017}
\bibinfo{author}{S.~Hu}, \bibinfo{author}{X.~Cui}, \bibinfo{author}{T.~Nomura},
  \bibinfo{author}{T.~Min}, \bibinfo{author}{T.~Kimura},
\newblock \bibinfo{title}{Nonreciprocity of electrically excited thermal spin
  signals in {C}o{F}e{A}l-{C}u-{P}y lateral spin valves},
\newblock \bibinfo{journal}{Phys. Rev. B} \bibinfo{volume}{95}
  (\bibinfo{year}{2017}) \bibinfo{pages}{100403}.
  \DOIprefix\doi{10.1103/PhysRevB.95.100403}.
\bibitem[{Johnson(1993)}]{JohnsonPRL1993}
\bibinfo{author}{M.~Johnson},
\newblock \bibinfo{title}{Spin accumulation in gold films},
\newblock \bibinfo{journal}{Phys. Rev. Lett.} \bibinfo{volume}{70}
  (\bibinfo{year}{1993}) \bibinfo{pages}{2142--2145}.
\bibitem[{Jedema et~al.(2001)Jedema, Filip, and van Wees}]{JedemaNature2001}
\bibinfo{author}{F.~J. Jedema}, \bibinfo{author}{A.~T. Filip},
  \bibinfo{author}{B.~J. van Wees},
\newblock \bibinfo{title}{Electrical spin injection and accumulation at room
  temperature in an all-metal mesoscopic spin valve},
\newblock \bibinfo{journal}{Nature} \bibinfo{volume}{410}
  (\bibinfo{year}{2001}) \bibinfo{pages}{345--348}.
\bibitem[{Ji et~al.(2004)Ji, Hoffmann, Jiang, and Bader}]{JiAPL2004}
\bibinfo{author}{Y.~Ji}, \bibinfo{author}{A.~Hoffmann}, \bibinfo{author}{J.~S.
  Jiang}, \bibinfo{author}{S.~D. Bader},
\newblock \bibinfo{title}{Spin injection, diffusion, and detection in lateral
  spin-valves},
\newblock \bibinfo{journal}{Applied Physics Letters} \bibinfo{volume}{85}
  (\bibinfo{year}{2004}) \bibinfo{pages}{6218--6220}.
\bibitem[{Valenzuela and Tinkham(2004)}]{ValenzuelaAPL2004}
\bibinfo{author}{S.~O. Valenzuela}, \bibinfo{author}{M.~Tinkham},
\newblock \bibinfo{title}{Spin-polarized tunneling in room-temperature
  mesoscopic spin valves},
\newblock \bibinfo{journal}{Appl. Phys. Lett.} \bibinfo{volume}{85}
  (\bibinfo{year}{2004}) \bibinfo{pages}{5914}.
  \DOIprefix\doi{10.1063/1.1830685}.
\bibitem[{Niimi et~al.(2013)Niimi, Wei, Idzuchi, Wakamura, Kato, and
  Otani}]{NiimiPRL2013}
\bibinfo{author}{Y.~Niimi}, \bibinfo{author}{D.~Wei},
  \bibinfo{author}{H.~Idzuchi}, \bibinfo{author}{T.~Wakamura},
  \bibinfo{author}{T.~Kato}, \bibinfo{author}{Y.~Otani},
\newblock \bibinfo{title}{Experimental verification of comparability between
  spin-orbit and spin-diffusion lengths},
\newblock \bibinfo{journal}{Phys. Rev. Lett.} \bibinfo{volume}{110}
  (\bibinfo{year}{2013}) \bibinfo{pages}{016805}.
  \DOIprefix\doi{10.1103/PhysRevLett.110.016805}.
\bibitem[{Flipse et~al.(2012)Flipse, Bakker, Slachter, Dejene, and van
  Wees}]{FlipseNatNano2012}
\bibinfo{author}{J.~Flipse}, \bibinfo{author}{F.~L. Bakker},
  \bibinfo{author}{A.~Slachter}, \bibinfo{author}{F.~K. Dejene},
  \bibinfo{author}{B.~J. van Wees},
\newblock \bibinfo{title}{Direct observation of the spin-dependent {P}eltier
  effect},
\newblock \bibinfo{journal}{Nature Nanotechnology} \bibinfo{volume}{7}
  (\bibinfo{year}{2012}) \bibinfo{pages}{166--168}.
\bibitem[{Dejene et~al.(2014)Dejene, Flipse, and van Wees}]{DejenePRB2014}
\bibinfo{author}{F.~K. Dejene}, \bibinfo{author}{J.~Flipse},
  \bibinfo{author}{B.~J. van Wees},
\newblock \bibinfo{title}{Verification of the {T}homson-{O}nsager reciprocity
  relation for spin caloritronics},
\newblock \bibinfo{journal}{Phys. Rev. B} \bibinfo{volume}{90}
  (\bibinfo{year}{2014}) \bibinfo{pages}{180402}.
  \DOIprefix\doi{10.1103/PhysRevB.90.180402}.
\bibitem[{Uchida et~al.(2010)Uchida, Adachi, Ota, Nakayama, Maekawa, and
  Saitoh}]{UchidaAPL2010}
\bibinfo{author}{K.-i. Uchida}, \bibinfo{author}{H.~Adachi},
  \bibinfo{author}{T.~Ota}, \bibinfo{author}{H.~Nakayama},
  \bibinfo{author}{S.~Maekawa}, \bibinfo{author}{E.~Saitoh},
\newblock \bibinfo{title}{Observation of longitudinal spin-{S}eebeck effect in
  magnetic insulators},
\newblock \bibinfo{journal}{Applied Physics Letters} \bibinfo{volume}{97}
  (\bibinfo{year}{2010}). \DOIprefix\doi{http://dx.doi.org/10.1063/1.3507386}.
\bibitem[{Uchida et~al.(2014)Uchida, Ishida, Kikkawa, Kirihara, Murakami, and
  Saitoh}]{UchidaJPCM2014}
\bibinfo{author}{K.~Uchida}, \bibinfo{author}{M.~Ishida},
  \bibinfo{author}{T.~Kikkawa}, \bibinfo{author}{A.~Kirihara},
  \bibinfo{author}{T.~Murakami}, \bibinfo{author}{E.~Saitoh},
\newblock \bibinfo{title}{Longitudinal spin {S}eebeck effect: from fundamentals
  to applications},
\newblock \bibinfo{journal}{Journal of Physics: Condensed Matter}
  \bibinfo{volume}{26} (\bibinfo{year}{2014}) \bibinfo{pages}{343202}.
\bibitem[{i.~Uchida et~al.(2016)i.~Uchida, Adachi, Kikkawa, Kirihara, Ishida,
  Yorozu, Maekawa, and Saitoh}]{UchidaIEEE2016}
\bibinfo{author}{K.~i.~Uchida}, \bibinfo{author}{H.~Adachi},
  \bibinfo{author}{T.~Kikkawa}, \bibinfo{author}{A.~Kirihara},
  \bibinfo{author}{M.~Ishida}, \bibinfo{author}{S.~Yorozu},
  \bibinfo{author}{S.~Maekawa}, \bibinfo{author}{E.~Saitoh},
\newblock \bibinfo{title}{Thermoelectric generation based on spin {S}eebeck
  effects},
\newblock \bibinfo{journal}{Proceedings of the IEEE} \bibinfo{volume}{104}
  (\bibinfo{year}{2016}) \bibinfo{pages}{1946--1973}.
  \DOIprefix\doi{10.1109/JPROC.2016.2535167}.
\bibitem[{Kikkawa and Saitoh(2022)}]{KikkawaarXiv2022}
\bibinfo{author}{T.~Kikkawa}, \bibinfo{author}{E.~Saitoh}, \bibinfo{title}{Spin
  {S}eebeck effect: Sensitive probe for elementary excitation, spin
  correlation, transport, magnetic order, and domains in solids},
  \bibinfo{year}{2022}. \URLprefix \url{https://arxiv.org/abs/2205.10509}.
  \DOIprefix\doi{10.48550/ARXIV.2205.10509}.
\bibitem[{Xiao et~al.(2010)Xiao, Bauer, Uchida, Saitoh, and
  Maekawa}]{XiaoPRB2010}
\bibinfo{author}{J.~Xiao}, \bibinfo{author}{G.~E.~W. Bauer},
  \bibinfo{author}{K.-c. Uchida}, \bibinfo{author}{E.~Saitoh},
  \bibinfo{author}{S.~Maekawa},
\newblock \bibinfo{title}{Theory of magnon-driven spin {S}eebeck effect},
\newblock \bibinfo{journal}{Phys. Rev. B} \bibinfo{volume}{81}
  (\bibinfo{year}{2010}) \bibinfo{pages}{214418}.
  \DOIprefix\doi{10.1103/PhysRevB.81.214418}.
\bibitem[{Adachi et~al.(2011)Adachi, Ohe, Takahashi, and
  Maekawa}]{AdachiPRB2011}
\bibinfo{author}{H.~Adachi}, \bibinfo{author}{J.-i. Ohe},
  \bibinfo{author}{S.~Takahashi}, \bibinfo{author}{S.~Maekawa},
\newblock \bibinfo{title}{Linear-response theory of spin {S}eebeck effect in
  ferromagnetic insulators},
\newblock \bibinfo{journal}{Phys. Rev. B} \bibinfo{volume}{83}
  (\bibinfo{year}{2011}) \bibinfo{pages}{094410}.
  \DOIprefix\doi{10.1103/PhysRevB.83.094410}.
\bibitem[{Rezende et~al.(2014)Rezende, Rodriguez-Suarez, Lopez~Ortiz, and
  Azevedo}]{RezendePRB2014}
\bibinfo{author}{S.~M. Rezende}, \bibinfo{author}{R.~L. Rodriguez-Suarez},
  \bibinfo{author}{J.~C. Lopez~Ortiz}, \bibinfo{author}{A.~Azevedo},
\newblock \bibinfo{title}{Thermal properties of magnons and the spin {S}eebeck
  effect in yttrium iron garnet/normal metal hybrid structures},
\newblock \bibinfo{journal}{Phys. Rev. B} \bibinfo{volume}{89}
  (\bibinfo{year}{2014}) \bibinfo{pages}{134406}.
  \DOIprefix\doi{10.1103/PhysRevB.89.134406}.
\bibitem[{Rezende et~al.(2018)Rezende, Azevedo, and
  Rodr{\'{\i}}guez-Su{\'{a}}rez}]{RezendeJPhysD2018}
\bibinfo{author}{S.~M. Rezende}, \bibinfo{author}{A.~Azevedo},
  \bibinfo{author}{R.~L. Rodr{\'{\i}}guez-Su{\'{a}}rez},
\newblock \bibinfo{title}{Magnon diffusion theory for the spin {S}eebeck effect
  in ferromagnetic and antiferromagnetic insulators},
\newblock \bibinfo{journal}{Journal of Physics D: Applied Physics}
  \bibinfo{volume}{51} (\bibinfo{year}{2018}) \bibinfo{pages}{174004}.
  \DOIprefix\doi{10.1088/1361-6463/aab5f8}.
\bibitem[{Serga et~al.(2010)Serga, Chumak, and Hillebrands}]{SergaJPhysD2010}
\bibinfo{author}{A.~A. Serga}, \bibinfo{author}{A.~V. Chumak},
  \bibinfo{author}{B.~Hillebrands},
\newblock \bibinfo{title}{{YIG} magnonics},
\newblock \bibinfo{journal}{Journal of Physics D: Applied Physics}
  \bibinfo{volume}{43} (\bibinfo{year}{2010}) \bibinfo{pages}{264002}.
  \DOIprefix\doi{10.1088/0022-3727/43/26/264002}.
\bibitem[{Thiery et~al.(2018)Thiery, Naletov, Vila, Marty, Brenac, Jacquot,
  de~Loubens, Viret, Anane, Cros, Ben~Youssef, Beaulieu, Demidov, Divinskiy,
  Demokritov, and Klein}]{ThieryPRB2018charge}
\bibinfo{author}{N.~Thiery}, \bibinfo{author}{V.~V. Naletov},
  \bibinfo{author}{L.~Vila}, \bibinfo{author}{A.~Marty},
  \bibinfo{author}{A.~Brenac}, \bibinfo{author}{J.-F. Jacquot},
  \bibinfo{author}{G.~de~Loubens}, \bibinfo{author}{M.~Viret},
  \bibinfo{author}{A.~Anane}, \bibinfo{author}{V.~Cros},
  \bibinfo{author}{J.~Ben~Youssef}, \bibinfo{author}{N.~Beaulieu},
  \bibinfo{author}{V.~E. Demidov}, \bibinfo{author}{B.~Divinskiy},
  \bibinfo{author}{S.~O. Demokritov}, \bibinfo{author}{O.~Klein},
\newblock \bibinfo{title}{Electrical properties of epitaxial yttrium iron
  garnet ultrathin films at high temperatures},
\newblock \bibinfo{journal}{Phys. Rev. B} \bibinfo{volume}{97}
  (\bibinfo{year}{2018}) \bibinfo{pages}{064422}.
  \DOIprefix\doi{10.1103/PhysRevB.97.064422}.
\bibitem[{Kikkawa et~al.(2013)Kikkawa, Uchida, Shiomi, Qiu, Hou, Tian,
  Nakayama, Jin, and Saitoh}]{KikkawaPRL2013}
\bibinfo{author}{T.~Kikkawa}, \bibinfo{author}{K.~Uchida},
  \bibinfo{author}{Y.~Shiomi}, \bibinfo{author}{Z.~Qiu},
  \bibinfo{author}{D.~Hou}, \bibinfo{author}{D.~Tian},
  \bibinfo{author}{H.~Nakayama}, \bibinfo{author}{X.-F. Jin},
  \bibinfo{author}{E.~Saitoh},
\newblock \bibinfo{title}{Longitudinal spin {S}eebeck effect free from the
  proximity {N}ernst effect},
\newblock \bibinfo{journal}{Phys. Rev. Lett.} \bibinfo{volume}{110}
  (\bibinfo{year}{2013}) \bibinfo{pages}{067207}.
  \DOIprefix\doi{10.1103/PhysRevLett.110.067207}.
\bibitem[{Sola et~al.(2019)Sola, Basso, Kuepferling, Pasquale, ne~Meier, Reiss,
  Kuschel, Kikkawa, ichi Uchida, Saitoh, Jin, Watzman, Boona, Heremans,
  Jungfleisch, Zhang, Pearson, Hoffmann, and Schumacher}]{SolaIEEEtim2019}
\bibinfo{author}{A.~Sola}, \bibinfo{author}{V.~Basso},
  \bibinfo{author}{M.~Kuepferling}, \bibinfo{author}{M.~Pasquale},
  \bibinfo{author}{D.~C. ne~Meier}, \bibinfo{author}{G.~Reiss},
  \bibinfo{author}{T.~Kuschel}, \bibinfo{author}{T.~Kikkawa},
  \bibinfo{author}{K.~ichi Uchida}, \bibinfo{author}{E.~Saitoh},
  \bibinfo{author}{H.~Jin}, \bibinfo{author}{S.~J. Watzman},
  \bibinfo{author}{S.~Boona}, \bibinfo{author}{J.~Heremans},
  \bibinfo{author}{M.~B. Jungfleisch}, \bibinfo{author}{W.~Zhang},
  \bibinfo{author}{J.~E. Pearson}, \bibinfo{author}{A.~Hoffmann},
  \bibinfo{author}{H.~W. Schumacher},
\newblock \bibinfo{title}{Spincaloritronic measurements: A round robin
  comparison of the longitudinal spin {S}eebeck effect},
\newblock \bibinfo{journal}{{IEEE} Transactions on Instrumentation and
  Measurement} \bibinfo{volume}{68} (\bibinfo{year}{2019})
  \bibinfo{pages}{1765--1773}. \DOIprefix\doi{10.1109/tim.2018.2882930}.
\bibitem[{Sola et~al.(2017)Sola, Bougiatioti, Kuepferling, Meier, Reiss,
  Pasquale, Kuschel, and Basso}]{SolaSciRep2017}
\bibinfo{author}{A.~Sola}, \bibinfo{author}{P.~Bougiatioti},
  \bibinfo{author}{M.~Kuepferling}, \bibinfo{author}{D.~Meier},
  \bibinfo{author}{G.~Reiss}, \bibinfo{author}{M.~Pasquale},
  \bibinfo{author}{T.~Kuschel}, \bibinfo{author}{V.~Basso},
\newblock \bibinfo{title}{Longitudinal spin {S}eebeck coefficient: heat flux
  vs. temperature difference method},
\newblock \bibinfo{journal}{Scientific Reports} \bibinfo{volume}{7}
  (\bibinfo{year}{2017}). \DOIprefix\doi{SolaSciRep2017}.
\bibitem[{Boona et~al.(2021)Boona, Jin, and Watzman}]{BoonaJAP2021}
\bibinfo{author}{S.~R. Boona}, \bibinfo{author}{H.~Jin},
  \bibinfo{author}{S.~Watzman},
\newblock \bibinfo{title}{Transverse thermal energy conversion using spin and
  topological structures},
\newblock \bibinfo{journal}{Journal of Applied Physics} \bibinfo{volume}{130}
  (\bibinfo{year}{2021}) \bibinfo{pages}{171101}.
  \DOIprefix\doi{10.1063/5.0062559}.
\bibitem[{Wu et~al.(2015{\natexlab{a}})Wu, Pearson, and
  Bhattacharya}]{WuPRL2015}
\bibinfo{author}{S.~M. Wu}, \bibinfo{author}{J.~E. Pearson},
  \bibinfo{author}{A.~Bhattacharya},
\newblock \bibinfo{title}{Paramagnetic spin {S}eebeck effect},
\newblock \bibinfo{journal}{Phys. Rev. Lett.} \bibinfo{volume}{114}
  (\bibinfo{year}{2015}{\natexlab{a}}) \bibinfo{pages}{186602}.
  \DOIprefix\doi{10.1103/PhysRevLett.114.186602}.
\bibitem[{Wu et~al.(2015{\natexlab{b}})Wu, Fradin, Hoffman, Hoffmann, and
  Bhattacharya}]{WuJAP2015}
\bibinfo{author}{S.~M. Wu}, \bibinfo{author}{F.~Y. Fradin},
  \bibinfo{author}{J.~Hoffman}, \bibinfo{author}{A.~Hoffmann},
  \bibinfo{author}{A.~Bhattacharya},
\newblock \bibinfo{title}{Spin {S}eebeck devices using local on-chip heating},
\newblock \bibinfo{journal}{Journal of Applied Physics} \bibinfo{volume}{117}
  (\bibinfo{year}{2015}{\natexlab{b}}) \bibinfo{pages}{--}.
  \DOIprefix\doi{http://dx.doi.org/10.1063/1.4916188}.
\bibitem[{Weiler et~al.(2012)Weiler, Althammer, Czeschka, Huebl, Wagner, Opel,
  Imort, Reiss, Thomas, Gross, and Goennenwein}]{WeilerPRL2012}
\bibinfo{author}{M.~Weiler}, \bibinfo{author}{M.~Althammer},
  \bibinfo{author}{F.~D. Czeschka}, \bibinfo{author}{H.~Huebl},
  \bibinfo{author}{M.~S. Wagner}, \bibinfo{author}{M.~Opel},
  \bibinfo{author}{I.-M. Imort}, \bibinfo{author}{G.~Reiss},
  \bibinfo{author}{A.~Thomas}, \bibinfo{author}{R.~Gross},
  \bibinfo{author}{S.~T.~B. Goennenwein},
\newblock \bibinfo{title}{Local charge and spin currents in magnetothermal
  landscapes},
\newblock \bibinfo{journal}{Phys. Rev. Lett.} \bibinfo{volume}{108}
  (\bibinfo{year}{2012}) \bibinfo{pages}{106602}.
  \DOIprefix\doi{10.1103/PhysRevLett.108.106602}.
\bibitem[{Schreier et~al.(2013)Schreier, Roschewsky, Dobler, Meyer, Huebl,
  Gross, and Goennenwein}]{SchreierAPL2013}
\bibinfo{author}{M.~Schreier}, \bibinfo{author}{N.~Roschewsky},
  \bibinfo{author}{E.~Dobler}, \bibinfo{author}{S.~Meyer},
  \bibinfo{author}{H.~Huebl}, \bibinfo{author}{R.~Gross},
  \bibinfo{author}{S.~T.~B. Goennenwein},
\newblock \bibinfo{title}{Current heating induced spin {S}eebeck effect},
\newblock \bibinfo{journal}{Applied Physics Letters} \bibinfo{volume}{103}
  (\bibinfo{year}{2013}) \bibinfo{pages}{242404}.
  \DOIprefix\doi{10.1063/1.4839395}.
\bibitem[{Wang et~al.(2014)Wang, Wang, Zou, Cai, Sun, and Sun}]{WangAPL2014}
\bibinfo{author}{W.~X. Wang}, \bibinfo{author}{S.~H. Wang},
  \bibinfo{author}{L.~K. Zou}, \bibinfo{author}{J.~W. Cai},
  \bibinfo{author}{Z.~G. Sun}, \bibinfo{author}{J.~R. Sun},
\newblock \bibinfo{title}{Joule heating-induced coexisted spin {S}eebeck effect
  and spin {H}all magnetoresistance in the
  platinum/{Y}$_{3}${F}e$_{5}${O}$_{12}$ structure},
\newblock \bibinfo{journal}{Applied Physics Letters} \bibinfo{volume}{105}
  (\bibinfo{year}{2014}) \bibinfo{pages}{182403}.
  \DOIprefix\doi{10.1063/1.4901101}.
\bibitem[{Vlietstra et~al.(2014)Vlietstra, Shan, van Wees, Isasa, Casanova, and
  Ben~Youssef}]{VlietstraPRB2014}
\bibinfo{author}{N.~Vlietstra}, \bibinfo{author}{J.~Shan},
  \bibinfo{author}{B.~J. van Wees}, \bibinfo{author}{M.~Isasa},
  \bibinfo{author}{F.~Casanova}, \bibinfo{author}{J.~Ben~Youssef},
\newblock \bibinfo{title}{Simultaneous detection of the spin-{H}all
  magnetoresistance and the spin-{S}eebeck effect in platinum and tantalum on
  yttrium iron garnet},
\newblock \bibinfo{journal}{Phys. Rev. B} \bibinfo{volume}{90}
  (\bibinfo{year}{2014}) \bibinfo{pages}{174436}.
  \DOIprefix\doi{10.1103/PhysRevB.90.174436}.
\bibitem[{Barker and Bauer(2016)}]{BarkerPRL2016}
\bibinfo{author}{J.~Barker}, \bibinfo{author}{G.~E.~W. Bauer},
\newblock \bibinfo{title}{Thermal spin dynamics of yttrium iron garnet},
\newblock \bibinfo{journal}{Phys. Rev. Lett.} \bibinfo{volume}{117}
  (\bibinfo{year}{2016}) \bibinfo{pages}{217201}.
  \DOIprefix\doi{10.1103/PhysRevLett.117.217201}.
\bibitem[{Itoh et~al.(2017)Itoh, Iguchi, Daimon, Oyanagi, Uchida, and
  Saitoh}]{ItohPRB2017}
\bibinfo{author}{R.~Itoh}, \bibinfo{author}{R.~Iguchi},
  \bibinfo{author}{S.~Daimon}, \bibinfo{author}{K.~Oyanagi},
  \bibinfo{author}{K.-i. Uchida}, \bibinfo{author}{E.~Saitoh},
\newblock \bibinfo{title}{Magnetic-field-induced decrease of the spin {P}eltier
  effect in {P}t/{Y}$_{3}${F}e$_{5}${O}$_{12}$ system at room temperature},
\newblock \bibinfo{journal}{Phys. Rev. B} \bibinfo{volume}{96}
  (\bibinfo{year}{2017}) \bibinfo{pages}{184422}.
  \DOIprefix\doi{10.1103/PhysRevB.96.184422}.
\bibitem[{Kikkawa et~al.(2015)Kikkawa, Uchida, Daimon, Qiu, Shiomi, and
  Saitoh}]{KikkawaPRB2015}
\bibinfo{author}{T.~Kikkawa}, \bibinfo{author}{K.-i. Uchida},
  \bibinfo{author}{S.~Daimon}, \bibinfo{author}{Z.~Qiu},
  \bibinfo{author}{Y.~Shiomi}, \bibinfo{author}{E.~Saitoh},
\newblock \bibinfo{title}{Critical suppression of spin {S}eebeck effect by
  magnetic fields},
\newblock \bibinfo{journal}{Phys. Rev. B} \bibinfo{volume}{92}
  (\bibinfo{year}{2015}) \bibinfo{pages}{064413}.
  \DOIprefix\doi{10.1103/PhysRevB.92.064413}.
\bibitem[{Jin et~al.(2015)Jin, Boona, Yang, Myers, and Heremans}]{JinPRB2015}
\bibinfo{author}{H.~Jin}, \bibinfo{author}{S.~R. Boona},
  \bibinfo{author}{Z.~Yang}, \bibinfo{author}{R.~C. Myers},
  \bibinfo{author}{J.~P. Heremans},
\newblock \bibinfo{title}{Effect of the magnon dispersion on the longitudinal
  spin {S}eebeck effect in yttrium iron garnets},
\newblock \bibinfo{journal}{Phys. Rev. B} \bibinfo{volume}{92}
  (\bibinfo{year}{2015}) \bibinfo{pages}{054436}.
  \DOIprefix\doi{10.1103/PhysRevB.92.054436}.
\bibitem[{Guo et~al.(2016)Guo, Cramer, Kehlberger, Ferguson, MacLaren, Jakob,
  and Kl\"aui}]{GuoPRX2016}
\bibinfo{author}{E.-J. Guo}, \bibinfo{author}{J.~Cramer},
  \bibinfo{author}{A.~Kehlberger}, \bibinfo{author}{C.~A. Ferguson},
  \bibinfo{author}{D.~A. MacLaren}, \bibinfo{author}{G.~Jakob},
  \bibinfo{author}{M.~Kl\"aui},
\newblock \bibinfo{title}{Influence of thickness and interface on the
  low-temperature enhancement of the spin {S}eebeck effect in {YIG} films},
\newblock \bibinfo{journal}{Phys. Rev. X} \bibinfo{volume}{6}
  (\bibinfo{year}{2016}) \bibinfo{pages}{031012}.
  \DOIprefix\doi{10.1103/PhysRevX.6.031012}.
\bibitem[{Kikkawa et~al.(2016)Kikkawa, Shen, Flebus, Duine, Uchida, Qiu, Bauer,
  and Saitoh}]{KikkawaPRL2016}
\bibinfo{author}{T.~Kikkawa}, \bibinfo{author}{K.~Shen},
  \bibinfo{author}{B.~Flebus}, \bibinfo{author}{R.~A. Duine},
  \bibinfo{author}{K.-i. Uchida}, \bibinfo{author}{Z.~Qiu},
  \bibinfo{author}{G.~E.~W. Bauer}, \bibinfo{author}{E.~Saitoh},
\newblock \bibinfo{title}{Magnon polarons in the spin {S}eebeck effect},
\newblock \bibinfo{journal}{Phys. Rev. Lett.} \bibinfo{volume}{117}
  (\bibinfo{year}{2016}) \bibinfo{pages}{207203}.
  \DOIprefix\doi{10.1103/PhysRevLett.117.207203}.
\bibitem[{Yagmur et~al.(2018)Yagmur, Iguchi, Gepr{\"a}gs, Erb, Daimon, Saitoh,
  Gross, and Uchida}]{YagmurJPhysD2018}
\bibinfo{author}{A.~Yagmur}, \bibinfo{author}{R.~Iguchi},
  \bibinfo{author}{S.~Gepr{\"a}gs}, \bibinfo{author}{A.~Erb},
  \bibinfo{author}{S.~Daimon}, \bibinfo{author}{E.~Saitoh},
  \bibinfo{author}{R.~Gross}, \bibinfo{author}{K.~Uchida},
\newblock \bibinfo{title}{Lock-in thermography measurements of the spin
  {P}eltier effect in a compensated ferrimagnet and its comparison to the spin
  {S}eebeck effect},
\newblock \bibinfo{journal}{Journal of Physics D: Applied Physics}
  \bibinfo{volume}{51} (\bibinfo{year}{2018}) \bibinfo{pages}{194002}.
  \DOIprefix\doi{10.1088/1361-6463/aabc75}.
\bibitem[{Flipse et~al.(2014)Flipse, Dejene, Wagenaar, Bauer, Youssef, and van
  Wees}]{FlipsePRL2014}
\bibinfo{author}{J.~Flipse}, \bibinfo{author}{F.~K. Dejene},
  \bibinfo{author}{D.~Wagenaar}, \bibinfo{author}{G.~E.~W. Bauer},
  \bibinfo{author}{J.~B. Youssef}, \bibinfo{author}{B.~J. van Wees},
\newblock \bibinfo{title}{Observation of the spin {P}eltier effect for magnetic
  insulators},
\newblock \bibinfo{journal}{Phys. Rev. Lett.} \bibinfo{volume}{113}
  (\bibinfo{year}{2014}) \bibinfo{pages}{027601}.
  \DOIprefix\doi{10.1103/PhysRevLett.113.027601}.
\bibitem[{Daimon et~al.(2016)Daimon, Iguchi, Hioki, Saitoh, and ichi
  Uchida}]{DaimonNatComms2016}
\bibinfo{author}{S.~Daimon}, \bibinfo{author}{R.~Iguchi},
  \bibinfo{author}{T.~Hioki}, \bibinfo{author}{E.~Saitoh},
  \bibinfo{author}{K.~ichi Uchida},
\newblock \bibinfo{title}{Thermal imaging of spin {P}eltier effect},
\newblock \bibinfo{journal}{Nature Communications} \bibinfo{volume}{7}
  (\bibinfo{year}{2016}). \DOIprefix\doi{10.1038/ncomms13754}.
\bibitem[{Yahiro et~al.(2020)Yahiro, Kikkawa, Ramos, Oyanagi, Hioki, Daimon,
  and Saitoh}]{YahiroPRB2020}
\bibinfo{author}{R.~Yahiro}, \bibinfo{author}{T.~Kikkawa},
  \bibinfo{author}{R.~Ramos}, \bibinfo{author}{K.~Oyanagi},
  \bibinfo{author}{T.~Hioki}, \bibinfo{author}{S.~Daimon},
  \bibinfo{author}{E.~Saitoh},
\newblock \bibinfo{title}{Magnon polarons in the spin {P}eltier effect},
\newblock \bibinfo{journal}{Phys. Rev. B} \bibinfo{volume}{101}
  (\bibinfo{year}{2020}) \bibinfo{pages}{024407}.
  \DOIprefix\doi{10.1103/PhysRevB.101.024407}.
\bibitem[{Uchida et~al.(2008)Uchida, Takahashi, Harii, Ieda, Koshibae, Ando,
  Maekawa, and Saitoh}]{UchidaNature08}
\bibinfo{author}{K.~Uchida}, \bibinfo{author}{S.~Takahashi},
  \bibinfo{author}{K.~Harii}, \bibinfo{author}{J.~Ieda},
  \bibinfo{author}{W.~Koshibae}, \bibinfo{author}{K.~Ando},
  \bibinfo{author}{S.~Maekawa}, \bibinfo{author}{E.~Saitoh},
\newblock \bibinfo{title}{Observation of the spin {S}eebeck effect},
\newblock \bibinfo{journal}{Nature} \bibinfo{volume}{455}
  (\bibinfo{year}{2008}) \bibinfo{pages}{778}.
\bibitem[{Jaworski et~al.(2010)Jaworski, Yang, Mack, Awschalom, Heremans, and
  Myers}]{MyersNatMater2010}
\bibinfo{author}{C.~M. Jaworski}, \bibinfo{author}{J.~Yang},
  \bibinfo{author}{S.~Mack}, \bibinfo{author}{D.~D. Awschalom},
  \bibinfo{author}{J.~P. Heremans}, \bibinfo{author}{R.~C. Myers},
\newblock \bibinfo{title}{Observation of the spin-{S}eebeck effect in a
  ferromagnetic semiconductor},
\newblock \bibinfo{journal}{Nature Materials} \bibinfo{volume}{9}
  (\bibinfo{year}{2010}) \bibinfo{pages}{898}.
\bibitem[{Uchida et~al.(2010)Uchida, Xiao, Adachi, Ohe, Takahashi, Ieda, Ota,
  Kajiwara, Umezawa, Kawai, Bauer, Maekawa, and Saitoh}]{XiaoNatMater2010}
\bibinfo{author}{K.~Uchida}, \bibinfo{author}{J.~Xiao},
  \bibinfo{author}{H.~Adachi}, \bibinfo{author}{J.~Ohe},
  \bibinfo{author}{S.~Takahashi}, \bibinfo{author}{J.~Ieda},
  \bibinfo{author}{T.~Ota}, \bibinfo{author}{Y.~Kajiwara},
  \bibinfo{author}{H.~Umezawa}, \bibinfo{author}{H.~Kawai},
  \bibinfo{author}{G.~E.~W. Bauer}, \bibinfo{author}{S.~Maekawa},
  \bibinfo{author}{E.~Saitoh},
\newblock \bibinfo{title}{Spin {S}eebeck insulator},
\newblock \bibinfo{journal}{Nature Materials} \bibinfo{volume}{9}
  (\bibinfo{year}{2010}) \bibinfo{pages}{894}.
\bibitem[{Jaworski et~al.(2012)Jaworski, Myers, Johnston-Halperin, and
  Heremans}]{JaworskiNature2012}
\bibinfo{author}{C.~M. Jaworski}, \bibinfo{author}{R.~C. Myers},
  \bibinfo{author}{E.~Johnston-Halperin}, \bibinfo{author}{J.~P. Heremans},
\newblock \bibinfo{title}{Giant spin {S}eebeck effect in a non-magnetic
  material},
\newblock \bibinfo{journal}{Nature} \bibinfo{volume}{487}
  (\bibinfo{year}{2012}) \bibinfo{pages}{210}.
  \DOIprefix\doi{10.1038/nature11221}.
\bibitem[{Srichandan et~al.(2018)Srichandan, Kronseder, Vogel, Back, and
  Strunk}]{SrichandanJPhysD2018}
\bibinfo{author}{S.~Srichandan}, \bibinfo{author}{M.~Kronseder},
  \bibinfo{author}{M.~Vogel}, \bibinfo{author}{C.~H. Back},
  \bibinfo{author}{C.~Strunk},
\newblock \bibinfo{title}{A microcalorimeter for simultaneous measurement of
  the electric and thermal transport coefficients in ferromagnetic thin films},
\newblock \bibinfo{journal}{Journal of Physics D: Applied Physics}
  \bibinfo{volume}{51} (\bibinfo{year}{2018}) \bibinfo{pages}{294006}.
  \DOIprefix\doi{10.1088/1361-6463/aacace}.
\bibitem[{Yin et~al.(2013)Yin, Mao, Meng, Li, and Zhao}]{YinPRB2013}
\bibinfo{author}{S.~L. Yin}, \bibinfo{author}{Q.~Mao}, \bibinfo{author}{Q.~Y.
  Meng}, \bibinfo{author}{D.~Li}, \bibinfo{author}{H.~W. Zhao},
\newblock \bibinfo{title}{Hybrid anomalous and planar {N}ernst effect in
  permalloy thin films},
\newblock \bibinfo{journal}{Phys. Rev. B} \bibinfo{volume}{88}
  (\bibinfo{year}{2013}) \bibinfo{pages}{064410}.
  \DOIprefix\doi{10.1103/PhysRevB.88.064410}.
\bibitem[{Wegrowe et~al.(2014)Wegrowe, Drouhin, and Lacour}]{WegrowePRB2014}
\bibinfo{author}{J.-E. Wegrowe}, \bibinfo{author}{H.-J. Drouhin},
  \bibinfo{author}{D.~Lacour},
\newblock \bibinfo{title}{Anisotropic magnetothermal transport and spin
  {S}eebeck effect},
\newblock \bibinfo{journal}{Phys. Rev. B} \bibinfo{volume}{89}
  (\bibinfo{year}{2014}) \bibinfo{pages}{094409}.
  \DOIprefix\doi{10.1103/PhysRevB.89.094409}.
\bibitem[{Soldatov et~al.(2014)Soldatov, Panarina, Hess, Schultz, and
  Sch\"afer}]{SoldatovPRB2014}
\bibinfo{author}{I.~V. Soldatov}, \bibinfo{author}{N.~Panarina},
  \bibinfo{author}{C.~Hess}, \bibinfo{author}{L.~Schultz},
  \bibinfo{author}{R.~Sch\"afer},
\newblock \bibinfo{title}{Thermoelectric effects and magnetic anisotropy of
  {G}a$_{1-x}${M}n$_{x}${A}s thin films},
\newblock \bibinfo{journal}{Phys. Rev. B} \bibinfo{volume}{90}
  (\bibinfo{year}{2014}) \bibinfo{pages}{104423}.
  \DOIprefix\doi{10.1103/PhysRevB.90.104423}.
\bibitem[{Jayathilaka et~al.(2015)Jayathilaka, Belyea, Fawcett, and
  Miller}]{JayathilakaJMMM2015}
\bibinfo{author}{P.~Jayathilaka}, \bibinfo{author}{D.~Belyea},
  \bibinfo{author}{T.~Fawcett}, \bibinfo{author}{C.~W. Miller},
\newblock \bibinfo{title}{Anisotropic magnetothermopower in ferromagnetic thin
  films grown on macroscopic substrates},
\newblock \bibinfo{journal}{Journal of Magnetism and Magnetic Materials}
  \bibinfo{volume}{382} (\bibinfo{year}{2015}) \bibinfo{pages}{376 -- 379}.
  \DOIprefix\doi{http://dx.doi.org/10.1016/j.jmmm.2015.02.006}.
\bibitem[{Cao et~al.(2016)Cao, Feng, Liu, Wang, Yang, Zhang, Zhao, Jiang, Liu,
  Yang, Zelalem, and Yu}]{CaoAIPadv2016}
\bibinfo{author}{Y.~Cao}, \bibinfo{author}{C.~Feng}, \bibinfo{author}{D.~X.
  Liu}, \bibinfo{author}{L.~J. Wang}, \bibinfo{author}{G.~Yang},
  \bibinfo{author}{J.~Y. Zhang}, \bibinfo{author}{B.~Zhao},
  \bibinfo{author}{S.~L. Jiang}, \bibinfo{author}{Q.~Q. Liu},
  \bibinfo{author}{K.~Yang}, \bibinfo{author}{A.~B. Zelalem},
  \bibinfo{author}{G.~H. Yu},
\newblock \bibinfo{title}{Observation of a thermally enhanced magnetoresistance
  in {NiFe}},
\newblock \bibinfo{journal}{{AIP} Advances} \bibinfo{volume}{6}
  (\bibinfo{year}{2016}) \bibinfo{pages}{045314}.
  \DOIprefix\doi{10.1063/1.4948310}.
\bibitem[{Kimling and Kuschel(2019)}]{KimlingPRB2019}
\bibinfo{author}{J.~Kimling}, \bibinfo{author}{T.~Kuschel},
\newblock \bibinfo{title}{Comment on ``{O}ptical detection of transverse
  spin-{S}eebeck effect in permalloy film using {S}agnac interferometer
  microscopy''},
\newblock \bibinfo{journal}{Phys. Rev. B} \bibinfo{volume}{99}
  (\bibinfo{year}{2019}) \bibinfo{pages}{106401}.
  \DOIprefix\doi{10.1103/PhysRevB.99.106401}.
\bibitem[{Takahashi et~al.(2012)Takahashi, Kasai, Hirayama, Mitani, and
  Hono}]{TakahashiAPL2012}
\bibinfo{author}{Y.~K. Takahashi}, \bibinfo{author}{S.~Kasai},
  \bibinfo{author}{S.~Hirayama}, \bibinfo{author}{S.~Mitani},
  \bibinfo{author}{K.~Hono},
\newblock \bibinfo{title}{All-metallic lateral spin valves using
  {C}o$_{2}${F}e({G}e$_{0.5}${G}a$_{0.5}$) {H}eusler alloy with a large spin
  signal},
\newblock \bibinfo{journal}{Applied Physics Letters} \bibinfo{volume}{100}
  (\bibinfo{year}{2012}) \bibinfo{pages}{052405}.
  \DOIprefix\doi{10.1063/1.3681804}.
\bibitem[{Yamada et~al.(2013)Yamada, Sato, Yoshida, Sato, Meguro, and
  Ogawa}]{YamadaIEEETransMag2013}
\bibinfo{author}{M.~Yamada}, \bibinfo{author}{D.~Sato},
  \bibinfo{author}{N.~Yoshida}, \bibinfo{author}{M.~Sato},
  \bibinfo{author}{K.~Meguro}, \bibinfo{author}{S.~Ogawa},
\newblock \bibinfo{title}{Scalability of spin accumulation sensor},
\newblock \bibinfo{journal}{Magnetics, IEEE Transactions on}
  \bibinfo{volume}{49} (\bibinfo{year}{2013}) \bibinfo{pages}{713--717}.
  \DOIprefix\doi{10.1109/TMAG.2012.2226871}.
\bibitem[{Vedyayev et~al.(2018)Vedyayev, Ryzhanova, Strelkov, Andrianov,
  Lobachev, and Dieny}]{VedyayevPRApp2018}
\bibinfo{author}{A.~Vedyayev}, \bibinfo{author}{N.~Ryzhanova},
  \bibinfo{author}{N.~Strelkov}, \bibinfo{author}{T.~Andrianov},
  \bibinfo{author}{A.~Lobachev}, \bibinfo{author}{B.~Dieny},
\newblock \bibinfo{title}{Nonlocal signal and noise in $\mathsf{T}$-shaped
  lateral spin-valve structures},
\newblock \bibinfo{journal}{Phys. Rev. Applied} \bibinfo{volume}{10}
  (\bibinfo{year}{2018}) \bibinfo{pages}{064047}.
  \DOIprefix\doi{10.1103/PhysRevApplied.10.064047}.
\bibitem[{Wright et~al.(2021)Wright, Erickson, Bromley, Crowell, Leighton, and
  O'Brien}]{WrightPRB2021}
\bibinfo{author}{A.~J. Wright}, \bibinfo{author}{M.~J. Erickson},
  \bibinfo{author}{D.~Bromley}, \bibinfo{author}{P.~A. Crowell},
  \bibinfo{author}{C.~Leighton}, \bibinfo{author}{L.~O'Brien},
\newblock \bibinfo{title}{Origin of the magnetic field enhancement of the spin
  signal in metallic nonlocal spin transport devices},
\newblock \bibinfo{journal}{Phys. Rev. B} \bibinfo{volume}{104}
  (\bibinfo{year}{2021}) \bibinfo{pages}{014423}.
  \DOIprefix\doi{10.1103/PhysRevB.104.014423}.
\bibitem[{Watts et~al.(2019)Watts, O'Brien, Jeong, Mkhoyan, Crowell, and
  Leighton}]{WattsPRM2019}
\bibinfo{author}{J.~D. Watts}, \bibinfo{author}{L.~O'Brien},
  \bibinfo{author}{J.~S. Jeong}, \bibinfo{author}{K.~A. Mkhoyan},
  \bibinfo{author}{P.~A. Crowell}, \bibinfo{author}{C.~Leighton},
\newblock \bibinfo{title}{Magnetic impurities as the origin of the variability
  in spin relaxation rates in {C}u-based spin transport devices},
\newblock \bibinfo{journal}{Phys. Rev. Materials} \bibinfo{volume}{3}
  (\bibinfo{year}{2019}) \bibinfo{pages}{124409}.
  \DOIprefix\doi{10.1103/PhysRevMaterials.3.124409}.
\bibitem[{Watts et~al.(2022)Watts, Batley, Rabideau, Hoch, O'Brien, Crowell,
  and Leighton}]{WattsPRL2022}
\bibinfo{author}{J.~D. Watts}, \bibinfo{author}{J.~T. Batley},
  \bibinfo{author}{N.~A. Rabideau}, \bibinfo{author}{J.~P. Hoch},
  \bibinfo{author}{L.~O'Brien}, \bibinfo{author}{P.~A. Crowell},
  \bibinfo{author}{C.~Leighton},
\newblock \bibinfo{title}{Finite-size effect in phonon-induced
  {E}lliott-{Y}afet spin relaxation in {A}l},
\newblock \bibinfo{journal}{Phys. Rev. Lett.} \bibinfo{volume}{128}
  (\bibinfo{year}{2022}) \bibinfo{pages}{207201}.
  \DOIprefix\doi{10.1103/PhysRevLett.128.207201}.
\bibitem[{O'Brien et~al.(2016)O'Brien, Spivak, Jeong, Mkhoyan, Crowell, and
  Leighton}]{OBrienPRB2016}
\bibinfo{author}{L.~O'Brien}, \bibinfo{author}{D.~Spivak},
  \bibinfo{author}{J.~S. Jeong}, \bibinfo{author}{K.~A. Mkhoyan},
  \bibinfo{author}{P.~A. Crowell}, \bibinfo{author}{C.~Leighton},
\newblock \bibinfo{title}{Interdiffusion-controlled {K}ondo suppression of
  injection efficiency in metallic nonlocal spin valves},
\newblock \bibinfo{journal}{Phys. Rev. B} \bibinfo{volume}{93}
  (\bibinfo{year}{2016}) \bibinfo{pages}{014413}.
  \DOIprefix\doi{10.1103/PhysRevB.93.014413}.
\bibitem[{O'Brien et~al.(2014)O'Brien, Erickson, Spivak, Ambaye, Goyette,
  Lauter, Crowell, and Leighton}]{ObrienNatComm2014}
\bibinfo{author}{L.~O'Brien}, \bibinfo{author}{M.~J. Erickson},
  \bibinfo{author}{D.~Spivak}, \bibinfo{author}{H.~Ambaye},
  \bibinfo{author}{R.~J. Goyette}, \bibinfo{author}{V.~Lauter},
  \bibinfo{author}{P.~A. Crowell}, \bibinfo{author}{C.~Leighton},
\newblock \bibinfo{title}{Kondo physics in non-local metallic spin transport
  devices},
\newblock \bibinfo{journal}{Nature Communications} \bibinfo{volume}{5}
  (\bibinfo{year}{2014}) \bibinfo{pages}{3927}.
\bibitem[{Wakamura et~al.(2015)Wakamura, Akaike, Omori, Niimi, Takahashi,
  Fujimaki, Maekawa, and Otani}]{WakamuraNatMater2015}
\bibinfo{author}{T.~Wakamura}, \bibinfo{author}{H.~Akaike},
  \bibinfo{author}{Y.~Omori}, \bibinfo{author}{Y.~Niimi},
  \bibinfo{author}{S.~Takahashi}, \bibinfo{author}{A.~Fujimaki},
  \bibinfo{author}{S.~Maekawa}, \bibinfo{author}{Y.~Otani},
\newblock \bibinfo{title}{Quasiparticle-mediated spin {H}all effect
  in~a~superconductor},
\newblock \bibinfo{journal}{Nature Materials} \bibinfo{volume}{14}
  (\bibinfo{year}{2015}) \bibinfo{pages}{675--678}.
  \DOIprefix\doi{10.1038/nmat4276}.
\bibitem[{Niimi et~al.(2014)Niimi, Suzuki, Kawanishi, Omori, Valet, Fert, and
  Otani}]{NiimiPRB2014}
\bibinfo{author}{Y.~Niimi}, \bibinfo{author}{H.~Suzuki},
  \bibinfo{author}{Y.~Kawanishi}, \bibinfo{author}{Y.~Omori},
  \bibinfo{author}{T.~Valet}, \bibinfo{author}{A.~Fert},
  \bibinfo{author}{Y.~Otani},
\newblock \bibinfo{title}{Extrinsic spin {H}all effects measured with lateral
  spin valve structures},
\newblock \bibinfo{journal}{Phys. Rev. B} \bibinfo{volume}{89}
  (\bibinfo{year}{2014}) \bibinfo{pages}{054401}.
  \DOIprefix\doi{10.1103/PhysRevB.89.054401}.
\bibitem[{Niimi et~al.(2012)Niimi, Kawanishi, Wei, Deranlot, Yang, Chshiev,
  Valet, Fert, and Otani}]{NiimiPRL2012}
\bibinfo{author}{Y.~Niimi}, \bibinfo{author}{Y.~Kawanishi},
  \bibinfo{author}{D.~H. Wei}, \bibinfo{author}{C.~Deranlot},
  \bibinfo{author}{H.~X. Yang}, \bibinfo{author}{M.~Chshiev},
  \bibinfo{author}{T.~Valet}, \bibinfo{author}{A.~Fert},
  \bibinfo{author}{Y.~Otani},
\newblock \bibinfo{title}{Giant spin {H}all effect induced by skew scattering
  from bismuth impurities inside thin film {C}u{B}i alloys},
\newblock \bibinfo{journal}{Phys. Rev. Lett.} \bibinfo{volume}{109}
  (\bibinfo{year}{2012}) \bibinfo{pages}{156602}.
  \DOIprefix\doi{10.1103/PhysRevLett.109.156602}.
\bibitem[{Kimura et~al.(2007)Kimura, Otani, Sato, Takahashi, and
  Maekawa}]{KimuraPRL2007}
\bibinfo{author}{T.~Kimura}, \bibinfo{author}{Y.~Otani},
  \bibinfo{author}{T.~Sato}, \bibinfo{author}{S.~Takahashi},
  \bibinfo{author}{S.~Maekawa},
\newblock \bibinfo{title}{Room-temperature reversible spin {H}all effect},
\newblock \bibinfo{journal}{Phys. Rev. Lett.} \bibinfo{volume}{98}
  (\bibinfo{year}{2007}) \bibinfo{pages}{156601}.
  \DOIprefix\doi{10.1103/PhysRevLett.98.156601}.
\bibitem[{Isshiki et~al.(2022)Isshiki, Zhu, Mizuno, Uesugi, Higo, Nakatsuji,
  and Otani}]{IsshikiPRM2022}
\bibinfo{author}{H.~Isshiki}, \bibinfo{author}{Z.~Zhu},
  \bibinfo{author}{H.~Mizuno}, \bibinfo{author}{R.~Uesugi},
  \bibinfo{author}{T.~Higo}, \bibinfo{author}{S.~Nakatsuji},
  \bibinfo{author}{Y.~Otani},
\newblock \bibinfo{title}{Determination of spin {H}all angle in the {W}eyl
  ferromagnet {C}o$_{2}${M}n{G}a by taking into account the thermoelectric
  contributions},
\newblock \bibinfo{journal}{Phys. Rev. Materials} \bibinfo{volume}{6}
  (\bibinfo{year}{2022}) \bibinfo{pages}{084411}.
  \DOIprefix\doi{10.1103/PhysRevMaterials.6.084411}.
\bibitem[{Bakker et~al.(2010)Bakker, Slachter, Adam, and van
  Wees}]{BakkerPRL2010}
\bibinfo{author}{F.~L. Bakker}, \bibinfo{author}{A.~Slachter},
  \bibinfo{author}{J.-P. Adam}, \bibinfo{author}{B.~J. van Wees},
\newblock \bibinfo{title}{Interplay of {P}eltier and {S}eebeck effects in
  nanoscale nonlocal spin valves},
\newblock \bibinfo{journal}{Phys. Rev. Lett.} \bibinfo{volume}{105}
  (\bibinfo{year}{2010}) \bibinfo{pages}{136601}.
  \DOIprefix\doi{10.1103/PhysRevLett.105.136601}.
\bibitem[{Casanova et~al.(2009)Casanova, Sharoni, Erekhinsky, and
  Schuller}]{CasanovaPRB2009}
\bibinfo{author}{F.~Casanova}, \bibinfo{author}{A.~Sharoni},
  \bibinfo{author}{M.~Erekhinsky}, \bibinfo{author}{I.~K. Schuller},
\newblock \bibinfo{title}{Control of spin injection by direct current in
  lateral spin valves},
\newblock \bibinfo{journal}{Phys. Rev. B} \bibinfo{volume}{79}
  (\bibinfo{year}{2009}) \bibinfo{pages}{184415}.
  \DOIprefix\doi{10.1103/PhysRevB.79.184415}.
\bibitem[{Hu and Kimura(2013)}]{HuPRB2013}
\bibinfo{author}{S.~Hu}, \bibinfo{author}{T.~Kimura},
\newblock \bibinfo{title}{Anomalous {N}ernst-{E}ttingshausen effect in nonlocal
  spin valve measurement under high-bias current injection},
\newblock \bibinfo{journal}{Phys. Rev. B} \bibinfo{volume}{87}
  (\bibinfo{year}{2013}) \bibinfo{pages}{014424}.
\bibitem[{Kasai et~al.(2014)Kasai, Hirayama, Takahashi, Mitani, Hono, Adachi,
  Ieda, and Maekawa}]{KasaiAPL2014}
\bibinfo{author}{S.~Kasai}, \bibinfo{author}{S.~Hirayama},
  \bibinfo{author}{Y.~K. Takahashi}, \bibinfo{author}{S.~Mitani},
  \bibinfo{author}{K.~Hono}, \bibinfo{author}{H.~Adachi},
  \bibinfo{author}{J.~Ieda}, \bibinfo{author}{S.~Maekawa},
\newblock \bibinfo{title}{Thermal engineering of non-local resistance in
  lateral spin valves},
\newblock \bibinfo{journal}{Applied Physics Letters} \bibinfo{volume}{104}
  (\bibinfo{year}{2014}) \bibinfo{pages}{--}.
  \DOIprefix\doi{http://dx.doi.org/10.1063/1.4873687}.
\bibitem[{Hojem(2016)}]{HojemThesis}
\bibinfo{author}{A.~Hojem}, \bibinfo{title}{Thermal effects on spin currents in
  non-local metallic spin valves}, Ph.D. thesis, University of Denver,
  \bibinfo{year}{2016}.
\bibitem[{Yamashita et~al.(2018)Yamashita, Ando, Koike, Miwa, Suzuki, and
  Shiraishi}]{YamashitaPRApp2018}
\bibinfo{author}{N.~Yamashita}, \bibinfo{author}{Y.~Ando},
  \bibinfo{author}{H.~Koike}, \bibinfo{author}{S.~Miwa},
  \bibinfo{author}{Y.~Suzuki}, \bibinfo{author}{M.~Shiraishi},
\newblock \bibinfo{title}{Thermally generated spin signals in a nondegenerate
  silicon spin valve},
\newblock \bibinfo{journal}{Phys. Rev. Applied} \bibinfo{volume}{9}
  (\bibinfo{year}{2018}) \bibinfo{pages}{054002}.
  \DOIprefix\doi{10.1103/PhysRevApplied.9.054002}.
\bibitem[{Kuschel et~al.(2019)Kuschel, Czerner, Walowski, Thomas, Schumacher,
  Reiss, Heiliger, and M{\"u}nzenberg}]{KuschelJPhysD2019}
\bibinfo{author}{T.~Kuschel}, \bibinfo{author}{M.~Czerner},
  \bibinfo{author}{J.~Walowski}, \bibinfo{author}{A.~Thomas},
  \bibinfo{author}{H.~W. Schumacher}, \bibinfo{author}{G.~Reiss},
  \bibinfo{author}{C.~Heiliger}, \bibinfo{author}{M.~M{\"u}nzenberg},
\newblock \bibinfo{title}{Tunnel magneto-{S}eebeck effect},
\newblock \bibinfo{journal}{Journal of Physics D: Applied Physics}
  \bibinfo{volume}{52} (\bibinfo{year}{2019}) \bibinfo{pages}{133001}.
  \DOIprefix\doi{10.1088/1361-6463/aafa5f}.
\bibitem[{Le~Breton et~al.(2011)Le~Breton, S., H., S., and
  R.}]{LeBretonNature2011}
\bibinfo{author}{J.-C. Le~Breton}, \bibinfo{author}{S.~S.},
  \bibinfo{author}{S.~H.}, \bibinfo{author}{Y.~S.}, \bibinfo{author}{J.~R.},
\newblock \bibinfo{title}{Thermal spin current from a ferromagnet to silicon by
  {S}eebeck spin tunnelling},
\newblock \bibinfo{journal}{Nature} \bibinfo{volume}{475}
  (\bibinfo{year}{2011}) \bibinfo{pages}{82--85}.
\bibitem[{Jang et~al.(2020)Jang, Marnitz, Huebner, Kimling, Kuschel, and
  Cahill}]{JangPRA2020}
\bibinfo{author}{H.~Jang}, \bibinfo{author}{L.~Marnitz},
  \bibinfo{author}{T.~Huebner}, \bibinfo{author}{J.~Kimling},
  \bibinfo{author}{T.~Kuschel}, \bibinfo{author}{D.~G. Cahill},
\newblock \bibinfo{title}{Thermal conductivity of oxide tunnel barriers in
  magnetic tunnel junctions measured by ultrafast thermoreflectance and
  magneto-optic {K}err effect thermometry},
\newblock \bibinfo{journal}{Phys. Rev. Applied} \bibinfo{volume}{13}
  (\bibinfo{year}{2020}) \bibinfo{pages}{024007}.
  \DOIprefix\doi{10.1103/PhysRevApplied.13.024007}.
\bibitem[{Taniguchi and Imamura(2011)}]{TaniguchiPRB2011}
\bibinfo{author}{T.~Taniguchi}, \bibinfo{author}{H.~Imamura},
\newblock \bibinfo{title}{Thermally assisted spin transfer torque switching in
  synthetic free layers},
\newblock \bibinfo{journal}{Phys. Rev. B} \bibinfo{volume}{83}
  (\bibinfo{year}{2011}) \bibinfo{pages}{054432}.
  \DOIprefix\doi{10.1103/PhysRevB.83.054432}.
\bibitem[{Prejbeanu et~al.(2007)Prejbeanu, Kerekes, Sousa, Sibuet, Redon,
  Dieny, and Nozi{\`e}res}]{PrejbeanuJPCM2007}
\bibinfo{author}{I.~Prejbeanu}, \bibinfo{author}{M.~Kerekes},
  \bibinfo{author}{R.~C. Sousa}, \bibinfo{author}{H.~Sibuet},
  \bibinfo{author}{O.~Redon}, \bibinfo{author}{B.~Dieny},
  \bibinfo{author}{J.~Nozi{\`e}res},
\newblock \bibinfo{title}{Thermally assisted {MRAM}},
\newblock \bibinfo{journal}{Journal of Physics: Condensed Matter}
  \bibinfo{volume}{19} (\bibinfo{year}{2007}) \bibinfo{pages}{165218}.
\bibitem[{Jungfleisch et~al.(2018)Jungfleisch, Zhang, and
  Hoffmann}]{JungfleischPhysLettA2018}
\bibinfo{author}{M.~B. Jungfleisch}, \bibinfo{author}{W.~Zhang},
  \bibinfo{author}{A.~Hoffmann},
\newblock \bibinfo{title}{Perspectives of antiferromagnetic spintronics},
\newblock \bibinfo{journal}{Physics Letters A} \bibinfo{volume}{382}
  (\bibinfo{year}{2018}) \bibinfo{pages}{865--871}.
  \DOIprefix\doi{10.1016/j.physleta.2018.01.008}.
\bibitem[{Baltz et~al.(2018)Baltz, Manchon, Tsoi, Moriyama, Ono, and
  Tserkovnyak}]{BaltzRMP2018}
\bibinfo{author}{V.~Baltz}, \bibinfo{author}{A.~Manchon},
  \bibinfo{author}{M.~Tsoi}, \bibinfo{author}{T.~Moriyama},
  \bibinfo{author}{T.~Ono}, \bibinfo{author}{Y.~Tserkovnyak},
\newblock \bibinfo{title}{Antiferromagnetic spintronics},
\newblock \bibinfo{journal}{Rev. Mod. Phys.} \bibinfo{volume}{90}
  (\bibinfo{year}{2018}) \bibinfo{pages}{015005}.
  \DOIprefix\doi{10.1103/RevModPhys.90.015005}.
\bibitem[{Jungwirth et~al.(2016)Jungwirth, Marti, Wadley, and
  Wunderlich}]{JungwirthNatNano2016}
\bibinfo{author}{T.~Jungwirth}, \bibinfo{author}{X.~Marti},
  \bibinfo{author}{P.~Wadley}, \bibinfo{author}{J.~Wunderlich},
\newblock \bibinfo{title}{Antiferromagnetic spintronics},
\newblock \bibinfo{journal}{Nature Nanotechnology} \bibinfo{volume}{11}
  (\bibinfo{year}{2016}) \bibinfo{pages}{231--241}.
  \DOIprefix\doi{10.1038/nnano.2016.18}.
\bibitem[{Marti et~al.(2015)Marti, Fina, and Jungwirth}]{MartiIEEETransMag2015}
\bibinfo{author}{X.~Marti}, \bibinfo{author}{I.~Fina},
  \bibinfo{author}{T.~Jungwirth},
\newblock \bibinfo{title}{Prospect for antiferromagnetic spintronics},
\newblock \bibinfo{journal}{Magnetics, IEEE Transactions on}
  \bibinfo{volume}{51} (\bibinfo{year}{2015}) \bibinfo{pages}{1--4}.
  \DOIprefix\doi{10.1109/TMAG.2014.2358939}.
\bibitem[{Kim et~al.(2022)Kim, Beach, Lee, Ono, Rasing, and
  Yang}]{KimNatMater2022}
\bibinfo{author}{S.~K. Kim}, \bibinfo{author}{G.~S. Beach},
  \bibinfo{author}{K.-J. Lee}, \bibinfo{author}{T.~Ono},
  \bibinfo{author}{T.~Rasing}, \bibinfo{author}{H.~Yang},
\newblock \bibinfo{title}{Ferrimagnetic spintronics},
\newblock \bibinfo{journal}{Nature Materials} \bibinfo{volume}{21}
  (\bibinfo{year}{2022}) \bibinfo{pages}{24--34}.
\bibitem[{He et~al.(2022)He, Hughes, Armitage, Tokura, and
  Wang}]{HeNatMater2022}
\bibinfo{author}{Q.~L. He}, \bibinfo{author}{T.~L. Hughes},
  \bibinfo{author}{N.~P. Armitage}, \bibinfo{author}{Y.~Tokura},
  \bibinfo{author}{K.~L. Wang},
\newblock \bibinfo{title}{Topological spintronics and magnetoelectronics},
\newblock \bibinfo{journal}{Nature Materials} \bibinfo{volume}{21}
  (\bibinfo{year}{2022}) \bibinfo{pages}{15--23}.
\bibitem[{{\v{S}}mejkal et~al.(2018){\v{S}}mejkal, Mokrousov, Yan, and
  MacDonald}]{SmejkalNatPhys2018}
\bibinfo{author}{L.~{\v{S}}mejkal}, \bibinfo{author}{Y.~Mokrousov},
  \bibinfo{author}{B.~Yan}, \bibinfo{author}{A.~H. MacDonald},
\newblock \bibinfo{title}{Topological antiferromagnetic spintronics},
\newblock \bibinfo{journal}{Nature Physics} \bibinfo{volume}{14}
  (\bibinfo{year}{2018}) \bibinfo{pages}{242--251}.
  \DOIprefix\doi{10.1038/s41567-018-0064-5}.
\bibitem[{Sakai et~al.(2018)Sakai, Mizuta, Nugroho, Sihombing, Koretsune,
  Suzuki, Takemori, Ishii, Nishio-Hamane, Arita et~al.}]{SakaiNatPhys2018}
\bibinfo{author}{A.~Sakai}, \bibinfo{author}{Y.~P. Mizuta},
  \bibinfo{author}{A.~A. Nugroho}, \bibinfo{author}{R.~Sihombing},
  \bibinfo{author}{T.~Koretsune}, \bibinfo{author}{M.-T. Suzuki},
  \bibinfo{author}{N.~Takemori}, \bibinfo{author}{R.~Ishii},
  \bibinfo{author}{D.~Nishio-Hamane}, \bibinfo{author}{R.~Arita}, et~al.,
\newblock \bibinfo{title}{Giant anomalous {N}ernst effect and quantum-critical
  scaling in a ferromagnetic semimetal},
\newblock \bibinfo{journal}{Nature Physics} \bibinfo{volume}{14}
  (\bibinfo{year}{2018}) \bibinfo{pages}{1119--1124}.
\bibitem[{Chen et~al.(2021)Chen, Tomita, Minami, Fu, Koretsune, Kitatani,
  Muhammad, Nishio-Hamane, Ishii, Ishii et~al.}]{ChenNatComms2021}
\bibinfo{author}{T.~Chen}, \bibinfo{author}{T.~Tomita},
  \bibinfo{author}{S.~Minami}, \bibinfo{author}{M.~Fu},
  \bibinfo{author}{T.~Koretsune}, \bibinfo{author}{M.~Kitatani},
  \bibinfo{author}{I.~Muhammad}, \bibinfo{author}{D.~Nishio-Hamane},
  \bibinfo{author}{R.~Ishii}, \bibinfo{author}{F.~Ishii}, et~al.,
\newblock \bibinfo{title}{Anomalous transport due to {W}eyl fermions in the
  chiral antiferromagnets {M}n$_{3}${X}, {X}= {S}n, {G}e},
\newblock \bibinfo{journal}{Nature Communications} \bibinfo{volume}{12}
  (\bibinfo{year}{2021}) \bibinfo{pages}{1--14}.
\bibitem[{Leiva et~al.(2022)Leiva, Granville, Zhang, Dushenko, Shigematsu,
  Ohshima, Ando, and Shiraishi}]{LeivaPRM2022}
\bibinfo{author}{L.~Leiva}, \bibinfo{author}{S.~Granville},
  \bibinfo{author}{Y.~Zhang}, \bibinfo{author}{S.~Dushenko},
  \bibinfo{author}{E.~Shigematsu}, \bibinfo{author}{R.~Ohshima},
  \bibinfo{author}{Y.~Ando}, \bibinfo{author}{M.~Shiraishi},
\newblock \bibinfo{title}{Efficient room-temperature magnetization direction
  detection by means of the enhanced anomalous {N}ernst effect in a {W}eyl
  ferromagnet},
\newblock \bibinfo{journal}{Phys. Rev. Materials} \bibinfo{volume}{6}
  (\bibinfo{year}{2022}) \bibinfo{pages}{064201}.
  \DOIprefix\doi{10.1103/PhysRevMaterials.6.064201}.
\bibitem[{Siddiqui et~al.(2020)Siddiqui, Sklenar, Kang, Gilbert, Schleife,
  Mason, and Hoffmann}]{SiddiquiJAP2020}
\bibinfo{author}{S.~A. Siddiqui}, \bibinfo{author}{J.~Sklenar},
  \bibinfo{author}{K.~Kang}, \bibinfo{author}{M.~J. Gilbert},
  \bibinfo{author}{A.~Schleife}, \bibinfo{author}{N.~Mason},
  \bibinfo{author}{A.~Hoffmann},
\newblock \bibinfo{title}{Metallic antiferromagnets},
\newblock \bibinfo{journal}{Journal of Applied Physics} \bibinfo{volume}{128}
  (\bibinfo{year}{2020}) \bibinfo{pages}{040904}.
  \DOIprefix\doi{10.1063/5.0009445}.
\bibitem[{Seki et~al.(2015)Seki, Ideue, Kubota, Kozuka, Takagi, Nakamura,
  Kaneko, Kawasaki, and Tokura}]{SekiPRL2015}
\bibinfo{author}{S.~Seki}, \bibinfo{author}{T.~Ideue},
  \bibinfo{author}{M.~Kubota}, \bibinfo{author}{Y.~Kozuka},
  \bibinfo{author}{R.~Takagi}, \bibinfo{author}{M.~Nakamura},
  \bibinfo{author}{Y.~Kaneko}, \bibinfo{author}{M.~Kawasaki},
  \bibinfo{author}{Y.~Tokura},
\newblock \bibinfo{title}{Thermal generation of spin current in an
  antiferromagnet},
\newblock \bibinfo{journal}{Phys. Rev. Lett.} \bibinfo{volume}{115}
  (\bibinfo{year}{2015}) \bibinfo{pages}{266601}.
  \DOIprefix\doi{10.1103/PhysRevLett.115.266601}.
\bibitem[{Wu et~al.(2016)Wu, Zhang, KC, Borisov, Pearson, Jiang, Lederman,
  Hoffmann, and Bhattacharya}]{WuPRL2016}
\bibinfo{author}{S.~M. Wu}, \bibinfo{author}{W.~Zhang},
  \bibinfo{author}{A.~KC}, \bibinfo{author}{P.~Borisov}, \bibinfo{author}{J.~E.
  Pearson}, \bibinfo{author}{J.~S. Jiang}, \bibinfo{author}{D.~Lederman},
  \bibinfo{author}{A.~Hoffmann}, \bibinfo{author}{A.~Bhattacharya},
\newblock \bibinfo{title}{Antiferromagnetic spin {S}eebeck effect},
\newblock \bibinfo{journal}{Phys. Rev. Lett.} \bibinfo{volume}{116}
  (\bibinfo{year}{2016}) \bibinfo{pages}{097204}.
  \DOIprefix\doi{10.1103/PhysRevLett.116.097204}.
\bibitem[{Rezende et~al.(2016)Rezende, Rodr\'{\i}guez-Su\'arez, and
  Azevedo}]{RezendePRB2016b}
\bibinfo{author}{S.~M. Rezende}, \bibinfo{author}{R.~L.
  Rodr\'{\i}guez-Su\'arez}, \bibinfo{author}{A.~Azevedo},
\newblock \bibinfo{title}{Theory of the spin {S}eebeck effect in
  antiferromagnets},
\newblock \bibinfo{journal}{Phys. Rev. B} \bibinfo{volume}{93}
  (\bibinfo{year}{2016}) \bibinfo{pages}{014425}.
  \DOIprefix\doi{10.1103/PhysRevB.93.014425}.
\bibitem[{Reitz et~al.(2020)Reitz, Li, Yuan, Shi, and
  Tserkovnyak}]{ReitzPRB2020}
\bibinfo{author}{D.~Reitz}, \bibinfo{author}{J.~Li}, \bibinfo{author}{W.~Yuan},
  \bibinfo{author}{J.~Shi}, \bibinfo{author}{Y.~Tserkovnyak},
\newblock \bibinfo{title}{Spin {S}eebeck effect near the antiferromagnetic
  spin-flop transition},
\newblock \bibinfo{journal}{Phys. Rev. B} \bibinfo{volume}{102}
  (\bibinfo{year}{2020}) \bibinfo{pages}{020408}.
  \DOIprefix\doi{10.1103/PhysRevB.102.020408}.
\bibitem[{Lin et~al.(2016)Lin, Chen, Zhang, and Chien}]{LinPRL2016}
\bibinfo{author}{W.~Lin}, \bibinfo{author}{K.~Chen},
  \bibinfo{author}{S.~Zhang}, \bibinfo{author}{C.~L. Chien},
\newblock \bibinfo{title}{Enhancement of thermally injected spin current
  through an antiferromagnetic insulator},
\newblock \bibinfo{journal}{Phys. Rev. Lett.} \bibinfo{volume}{116}
  (\bibinfo{year}{2016}) \bibinfo{pages}{186601}.
  \DOIprefix\doi{10.1103/PhysRevLett.116.186601}.
\bibitem[{Cramer et~al.(2018)Cramer, Ritzmann, Dong, Jaiswal, Qiu, Saitoh,
  Nowak, and Kl{\"a}ui}]{CramerJPhysD2018}
\bibinfo{author}{J.~Cramer}, \bibinfo{author}{U.~Ritzmann},
  \bibinfo{author}{B.-W. Dong}, \bibinfo{author}{S.~Jaiswal},
  \bibinfo{author}{Z.~Qiu}, \bibinfo{author}{E.~Saitoh},
  \bibinfo{author}{U.~Nowak}, \bibinfo{author}{M.~Kl{\"a}ui},
\newblock \bibinfo{title}{Spin transport across antiferromagnets induced by the
  spin {S}eebeck effect},
\newblock \bibinfo{journal}{Journal of Physics D: Applied Physics}
  \bibinfo{volume}{51} (\bibinfo{year}{2018}) \bibinfo{pages}{144004}.
  \DOIprefix\doi{10.1088/1361-6463/aab223}.
\bibitem[{Bia{\l}ek et~al.(2018)Bia{\l}ek, Br{\'{e}}chet, and
  Ansermet}]{BialekJPhysD2018}
\bibinfo{author}{M.~Bia{\l}ek}, \bibinfo{author}{S.~Br{\'{e}}chet},
  \bibinfo{author}{J.-P. Ansermet},
\newblock \bibinfo{title}{Heat-driven spin torques in antiferromagnets},
\newblock \bibinfo{journal}{Journal of Physics D: Applied Physics}
  \bibinfo{volume}{51} (\bibinfo{year}{2018}) \bibinfo{pages}{164001}.
  \DOIprefix\doi{10.1088/1361-6463/aab2f7}.
\bibitem[{Ikhlas et~al.(2017)Ikhlas, Tomita, Koretsune, Suzuki, Nishio-Hamane,
  Arita, Otani, and Nakatsuji}]{IkhlasNatPhys2017}
\bibinfo{author}{M.~Ikhlas}, \bibinfo{author}{T.~Tomita},
  \bibinfo{author}{T.~Koretsune}, \bibinfo{author}{M.-T. Suzuki},
  \bibinfo{author}{D.~Nishio-Hamane}, \bibinfo{author}{R.~Arita},
  \bibinfo{author}{Y.~Otani}, \bibinfo{author}{S.~Nakatsuji},
\newblock \bibinfo{title}{Large anomalous {N}ernst effect at room temperature
  in a chiral antiferromagnet},
\newblock \bibinfo{journal}{Nature Physics} \bibinfo{volume}{13}
  (\bibinfo{year}{2017}) \bibinfo{pages}{1085--1090}.
\bibitem[{Li et~al.(2017)Li, Xu, Ding, Wang, Shen, Lu, Zhu, and
  Behnia}]{LiPRL2017}
\bibinfo{author}{X.~Li}, \bibinfo{author}{L.~Xu}, \bibinfo{author}{L.~Ding},
  \bibinfo{author}{J.~Wang}, \bibinfo{author}{M.~Shen},
  \bibinfo{author}{X.~Lu}, \bibinfo{author}{Z.~Zhu},
  \bibinfo{author}{K.~Behnia},
\newblock \bibinfo{title}{Anomalous {N}ernst and {R}ighi-{L}educ effects in
  {M}n$_{3}${S}n: Berry curvature and entropy flow},
\newblock \bibinfo{journal}{Phys. Rev. Lett.} \bibinfo{volume}{119}
  (\bibinfo{year}{2017}) \bibinfo{pages}{056601}.
  \DOIprefix\doi{10.1103/PhysRevLett.119.056601}.
\bibitem[{Shao et~al.(2021)Shao, Li, Liu, Yang, Fukami, Razavi, Wu, Wang,
  Freimuth, Mokrousov, Stiles, Emori, Hoffmann, Akerman, Roy, Wang, Yang,
  Garello, and Zhang}]{ShaoIEEETransMag2021}
\bibinfo{author}{Q.~Shao}, \bibinfo{author}{P.~Li}, \bibinfo{author}{L.~Liu},
  \bibinfo{author}{H.~Yang}, \bibinfo{author}{S.~Fukami},
  \bibinfo{author}{A.~Razavi}, \bibinfo{author}{H.~Wu},
  \bibinfo{author}{K.~Wang}, \bibinfo{author}{F.~Freimuth},
  \bibinfo{author}{Y.~Mokrousov}, \bibinfo{author}{M.~D. Stiles},
  \bibinfo{author}{S.~Emori}, \bibinfo{author}{A.~Hoffmann},
  \bibinfo{author}{J.~Akerman}, \bibinfo{author}{K.~Roy},
  \bibinfo{author}{J.-P. Wang}, \bibinfo{author}{S.-H. Yang},
  \bibinfo{author}{K.~Garello}, \bibinfo{author}{W.~Zhang},
\newblock \bibinfo{title}{Roadmap of spin-orbit torques},
\newblock \bibinfo{journal}{{IEEE} Transactions on Magnetics}
  \bibinfo{volume}{57} (\bibinfo{year}{2021}) \bibinfo{pages}{1--39}.
  \DOIprefix\doi{10.1109/tmag.2021.3078583}.
\bibitem[{Manchon et~al.(2019)Manchon, \ifmmode~\check{Z}\else
  \v{Z}\fi{}elezn\'y, Miron, Jungwirth, Sinova, Thiaville, Garello, and
  Gambardella}]{ManchonRMP2019}
\bibinfo{author}{A.~Manchon}, \bibinfo{author}{J.~\ifmmode~\check{Z}\else
  \v{Z}\fi{}elezn\'y}, \bibinfo{author}{I.~M. Miron},
  \bibinfo{author}{T.~Jungwirth}, \bibinfo{author}{J.~Sinova},
  \bibinfo{author}{A.~Thiaville}, \bibinfo{author}{K.~Garello},
  \bibinfo{author}{P.~Gambardella},
\newblock \bibinfo{title}{Current-induced spin-orbit torques in ferromagnetic
  and antiferromagnetic systems},
\newblock \bibinfo{journal}{Rev. Mod. Phys.} \bibinfo{volume}{91}
  (\bibinfo{year}{2019}) \bibinfo{pages}{035004}.
  \DOIprefix\doi{10.1103/RevModPhys.91.035004}.
\bibitem[{Ramaswamy et~al.(2018)Ramaswamy, Lee, Cai, and
  Yang}]{RamaswamyAPR2018}
\bibinfo{author}{R.~Ramaswamy}, \bibinfo{author}{J.~M. Lee},
  \bibinfo{author}{K.~Cai}, \bibinfo{author}{H.~Yang},
\newblock \bibinfo{title}{Recent advances in spin-orbit torques: Moving towards
  device applications},
\newblock \bibinfo{journal}{Applied Physics Reviews} \bibinfo{volume}{5}
  (\bibinfo{year}{2018}) \bibinfo{pages}{031107}.
  \DOIprefix\doi{10.1063/1.5041793}.
\bibitem[{Wadley et~al.(2016)Wadley, Howells, {\v Z}elezn{\'y}, Andrews, Hills,
  Campion, Nov{\'a}k, Olejn{\'\i}k, Maccherozzi, Dhesi, Martin, Wagner,
  Wunderlich, Freimuth, Mokrousov, Kune{\v s}, Chauhan, Grzybowski, Rushforth,
  Edmonds, Gallagher, and Jungwirth}]{WadleyScience2016}
\bibinfo{author}{P.~Wadley}, \bibinfo{author}{B.~Howells},
  \bibinfo{author}{J.~{\v Z}elezn{\'y}}, \bibinfo{author}{C.~Andrews},
  \bibinfo{author}{V.~Hills}, \bibinfo{author}{R.~P. Campion},
  \bibinfo{author}{V.~Nov{\'a}k}, \bibinfo{author}{K.~Olejn{\'\i}k},
  \bibinfo{author}{F.~Maccherozzi}, \bibinfo{author}{S.~S. Dhesi},
  \bibinfo{author}{S.~Y. Martin}, \bibinfo{author}{T.~Wagner},
  \bibinfo{author}{J.~Wunderlich}, \bibinfo{author}{F.~Freimuth},
  \bibinfo{author}{Y.~Mokrousov}, \bibinfo{author}{J.~Kune{\v s}},
  \bibinfo{author}{J.~S. Chauhan}, \bibinfo{author}{M.~J. Grzybowski},
  \bibinfo{author}{A.~W. Rushforth}, \bibinfo{author}{K.~W. Edmonds},
  \bibinfo{author}{B.~L. Gallagher}, \bibinfo{author}{T.~Jungwirth},
\newblock \bibinfo{title}{Electrical switching of an antiferromagnet},
\newblock \bibinfo{journal}{Science} \bibinfo{volume}{351}
  (\bibinfo{year}{2016}) \bibinfo{pages}{587--590}.
  \DOIprefix\doi{10.1126/science.aab1031}.
\bibitem[{Olejn{\'{\i}}k et~al.(2017)Olejn{\'{\i}}k, Schuler, Marti,
  Nov{\'{a}}k, Ka{\v{s}}par, Wadley, Campion, Edmonds, Gallagher, Garces,
  Baumgartner, Gambardella, and Jungwirth}]{OlejnikNatComms2017}
\bibinfo{author}{K.~Olejn{\'{\i}}k}, \bibinfo{author}{V.~Schuler},
  \bibinfo{author}{X.~Marti}, \bibinfo{author}{V.~Nov{\'{a}}k},
  \bibinfo{author}{Z.~Ka{\v{s}}par}, \bibinfo{author}{P.~Wadley},
  \bibinfo{author}{R.~P. Campion}, \bibinfo{author}{K.~W. Edmonds},
  \bibinfo{author}{B.~L. Gallagher}, \bibinfo{author}{J.~Garces},
  \bibinfo{author}{M.~Baumgartner}, \bibinfo{author}{P.~Gambardella},
  \bibinfo{author}{T.~Jungwirth},
\newblock \bibinfo{title}{Antiferromagnetic {CuMnAs} multi-level memory cell
  with microelectronic compatibility},
\newblock \bibinfo{journal}{Nature Communications} \bibinfo{volume}{8}
  (\bibinfo{year}{2017}). \DOIprefix\doi{10.1038/ncomms15434}.
\bibitem[{Chen et~al.(2018)Chen, Zarzuela, Zhang, Song, Zhou, Shi, Li, Zhou,
  Jiang, Pan, and Tserkovnyak}]{ChenPRL2018}
\bibinfo{author}{X.~Z. Chen}, \bibinfo{author}{R.~Zarzuela},
  \bibinfo{author}{J.~Zhang}, \bibinfo{author}{C.~Song}, \bibinfo{author}{X.~F.
  Zhou}, \bibinfo{author}{G.~Y. Shi}, \bibinfo{author}{F.~Li},
  \bibinfo{author}{H.~A. Zhou}, \bibinfo{author}{W.~J. Jiang},
  \bibinfo{author}{F.~Pan}, \bibinfo{author}{Y.~Tserkovnyak},
\newblock \bibinfo{title}{Antidamping-torque-induced switching in biaxial
  antiferromagnetic insulators},
\newblock \bibinfo{journal}{Phys. Rev. Lett.} \bibinfo{volume}{120}
  (\bibinfo{year}{2018}) \bibinfo{pages}{207204}.
  \DOIprefix\doi{10.1103/PhysRevLett.120.207204}.
\bibitem[{Moriyama et~al.(2018)Moriyama, Oda, Ohkochi, Kimata, and
  Ono}]{MoriyamaSciRep2018}
\bibinfo{author}{T.~Moriyama}, \bibinfo{author}{K.~Oda},
  \bibinfo{author}{T.~Ohkochi}, \bibinfo{author}{M.~Kimata},
  \bibinfo{author}{T.~Ono},
\newblock \bibinfo{title}{Spin torque control of antiferromagnetic moments in
  {NiO}},
\newblock \bibinfo{journal}{Scientific Reports} \bibinfo{volume}{8}
  (\bibinfo{year}{2018}) \bibinfo{pages}{14167}.
  \DOIprefix\doi{10.1038/s41598-018-32508-w}.
\bibitem[{Baldrati et~al.(2019)Baldrati, Gomonay, Ross, Filianina, Lebrun,
  Ramos, Leveille, Fuhrmann, Forrest, Maccherozzi, Valencia, Kronast, Saitoh,
  Sinova, and Kl\"aui}]{BaldratiPRL2019}
\bibinfo{author}{L.~Baldrati}, \bibinfo{author}{O.~Gomonay},
  \bibinfo{author}{A.~Ross}, \bibinfo{author}{M.~Filianina},
  \bibinfo{author}{R.~Lebrun}, \bibinfo{author}{R.~Ramos},
  \bibinfo{author}{C.~Leveille}, \bibinfo{author}{F.~Fuhrmann},
  \bibinfo{author}{T.~R. Forrest}, \bibinfo{author}{F.~Maccherozzi},
  \bibinfo{author}{S.~Valencia}, \bibinfo{author}{F.~Kronast},
  \bibinfo{author}{E.~Saitoh}, \bibinfo{author}{J.~Sinova},
  \bibinfo{author}{M.~Kl\"aui},
\newblock \bibinfo{title}{Mechanism of {N}\'eel order switching in
  antiferromagnetic thin films revealed by magnetotransport and direct
  imaging},
\newblock \bibinfo{journal}{Phys. Rev. Lett.} \bibinfo{volume}{123}
  (\bibinfo{year}{2019}) \bibinfo{pages}{177201}.
  \DOIprefix\doi{10.1103/PhysRevLett.123.177201}.
\bibitem[{Chiang et~al.(2019)Chiang, Huang, Qu, Wu, and Chien}]{ChiangPRL2019}
\bibinfo{author}{C.~C. Chiang}, \bibinfo{author}{S.~Y. Huang},
  \bibinfo{author}{D.~Qu}, \bibinfo{author}{P.~H. Wu}, \bibinfo{author}{C.~L.
  Chien},
\newblock \bibinfo{title}{Absence of evidence of electrical switching of the
  antiferromagnetic {N}\'eel vector},
\newblock \bibinfo{journal}{Phys. Rev. Lett.} \bibinfo{volume}{123}
  (\bibinfo{year}{2019}) \bibinfo{pages}{227203}.
  \DOIprefix\doi{10.1103/PhysRevLett.123.227203}.
\bibitem[{Grzybowski et~al.(2017)Grzybowski, Wadley, Edmonds, Beardsley, Hills,
  Campion, Gallagher, Chauhan, Novak, Jungwirth, Maccherozzi, and
  Dhesi}]{GrzybowskiPRL2017}
\bibinfo{author}{M.~J. Grzybowski}, \bibinfo{author}{P.~Wadley},
  \bibinfo{author}{K.~W. Edmonds}, \bibinfo{author}{R.~Beardsley},
  \bibinfo{author}{V.~Hills}, \bibinfo{author}{R.~P. Campion},
  \bibinfo{author}{B.~L. Gallagher}, \bibinfo{author}{J.~S. Chauhan},
  \bibinfo{author}{V.~Novak}, \bibinfo{author}{T.~Jungwirth},
  \bibinfo{author}{F.~Maccherozzi}, \bibinfo{author}{S.~S. Dhesi},
\newblock \bibinfo{title}{Imaging current-induced switching of
  antiferromagnetic domains in {CuMnAs}},
\newblock \bibinfo{journal}{Phys. Rev. Lett.} \bibinfo{volume}{118}
  (\bibinfo{year}{2017}) \bibinfo{pages}{057701}.
  \DOIprefix\doi{10.1103/PhysRevLett.118.057701}.
\bibitem[{Wadley et~al.(2018)Wadley, Reimers, Grzybowski, Andrews, Wang,
  Chauhan, Gallagher, Campion, Edmonds, Dhesi, Maccherozzi, Novak, Wunderlich,
  and Jungwirth}]{WadleyNatNano2018}
\bibinfo{author}{P.~Wadley}, \bibinfo{author}{S.~Reimers},
  \bibinfo{author}{M.~J. Grzybowski}, \bibinfo{author}{C.~Andrews},
  \bibinfo{author}{M.~Wang}, \bibinfo{author}{J.~S. Chauhan},
  \bibinfo{author}{B.~L. Gallagher}, \bibinfo{author}{R.~P. Campion},
  \bibinfo{author}{K.~W. Edmonds}, \bibinfo{author}{S.~S. Dhesi},
  \bibinfo{author}{F.~Maccherozzi}, \bibinfo{author}{V.~Novak},
  \bibinfo{author}{J.~Wunderlich}, \bibinfo{author}{T.~Jungwirth},
\newblock \bibinfo{title}{Current polarity-dependent manipulation of
  antiferromagnetic domains},
\newblock \bibinfo{journal}{Nature Nanotechnology} \bibinfo{volume}{13}
  (\bibinfo{year}{2018}) \bibinfo{pages}{362--365}.
  \DOIprefix\doi{10.1038/s41565-018-0079-1}.
\bibitem[{Bodnar et~al.(2019)Bodnar, Filianina, Bommanaboyena, Forrest,
  Maccherozzi, Sapozhnik, Skourski, Kl\"{a}ui, and Jourdan}]{BodnarPRB2019}
\bibinfo{author}{S.~Y. Bodnar}, \bibinfo{author}{M.~Filianina},
  \bibinfo{author}{S.~P. Bommanaboyena}, \bibinfo{author}{T.~Forrest},
  \bibinfo{author}{F.~Maccherozzi}, \bibinfo{author}{A.~A. Sapozhnik},
  \bibinfo{author}{Y.~Skourski}, \bibinfo{author}{M.~Kl\"{a}ui},
  \bibinfo{author}{M.~Jourdan},
\newblock \bibinfo{title}{Imaging of current induced {N}{\'{e}}el vector
  switching in antiferromagnetic {Mn}$_{2}${Au}},
\newblock \bibinfo{journal}{Physical Review B} \bibinfo{volume}{99}
  (\bibinfo{year}{2019}) \bibinfo{pages}{140409}.
  \DOIprefix\doi{10.1103/physrevb.99.140409}.
\bibitem[{Gray et~al.(2019)Gray, Moriyama, Sivadas, Stiehl, Heron, Need, Kirby,
  Low, Nowack, Schlom, Ralph, Ono, and Fuchs}]{GrayPRX2019}
\bibinfo{author}{I.~Gray}, \bibinfo{author}{T.~Moriyama},
  \bibinfo{author}{N.~Sivadas}, \bibinfo{author}{G.~M. Stiehl},
  \bibinfo{author}{J.~T. Heron}, \bibinfo{author}{R.~Need},
  \bibinfo{author}{B.~J. Kirby}, \bibinfo{author}{D.~H. Low},
  \bibinfo{author}{K.~C. Nowack}, \bibinfo{author}{D.~G. Schlom},
  \bibinfo{author}{D.~C. Ralph}, \bibinfo{author}{T.~Ono},
  \bibinfo{author}{G.~D. Fuchs},
\newblock \bibinfo{title}{Spin {S}eebeck imaging of spin-torque switching in
  antiferromagnetic $\mathrm{Pt}/\mathrm{NiO}$ heterostructures},
\newblock \bibinfo{journal}{Phys. Rev. X} \bibinfo{volume}{9}
  (\bibinfo{year}{2019}) \bibinfo{pages}{041016}.
  \DOIprefix\doi{10.1103/PhysRevX.9.041016}.
\bibitem[{Zhang et~al.(2019)Zhang, Finley, Safi, and Liu}]{ZhangPRL2019}
\bibinfo{author}{P.~Zhang}, \bibinfo{author}{J.~Finley},
  \bibinfo{author}{T.~Safi}, \bibinfo{author}{L.~Liu},
\newblock \bibinfo{title}{Quantitative study on current-induced effect in an
  antiferromagnet insulator/{P}t bilayer film},
\newblock \bibinfo{journal}{Phys. Rev. Lett.} \bibinfo{volume}{123}
  (\bibinfo{year}{2019}) \bibinfo{pages}{247206}.
  \DOIprefix\doi{10.1103/PhysRevLett.123.247206}.
\bibitem[{Cogulu et~al.(2022)Cogulu, Zhang, Statuto, Cheng, Yang, Cheng, and
  Kent}]{CoguluPRL2022}
\bibinfo{author}{E.~Cogulu}, \bibinfo{author}{H.~Zhang}, \bibinfo{author}{N.~N.
  Statuto}, \bibinfo{author}{Y.~Cheng}, \bibinfo{author}{F.~Yang},
  \bibinfo{author}{R.~Cheng}, \bibinfo{author}{A.~D. Kent},
\newblock \bibinfo{title}{Quantifying spin-orbit torques in
  antiferromagnet--heavy-metal heterostructures},
\newblock \bibinfo{journal}{Phys. Rev. Lett.} \bibinfo{volume}{128}
  (\bibinfo{year}{2022}) \bibinfo{pages}{247204}.
  \DOIprefix\doi{10.1103/PhysRevLett.128.247204}.
\bibitem[{Meer et~al.(2021)Meer, Schreiber, Schmitt, Ramos, Saitoh, Gomonay,
  Sinova, Baldrati, and Kl\"{a}ui}]{MeerNanoLetts2021}
\bibinfo{author}{H.~Meer}, \bibinfo{author}{F.~Schreiber},
  \bibinfo{author}{C.~Schmitt}, \bibinfo{author}{R.~Ramos},
  \bibinfo{author}{E.~Saitoh}, \bibinfo{author}{O.~Gomonay},
  \bibinfo{author}{J.~Sinova}, \bibinfo{author}{L.~Baldrati},
  \bibinfo{author}{M.~Kl\"{a}ui},
\newblock \bibinfo{title}{Direct imaging of current-induced antiferromagnetic
  switching revealing a pure thermomagnetoelastic switching mechanism in
  {N}i{O}},
\newblock \bibinfo{journal}{Nano Letters} \bibinfo{volume}{21}
  (\bibinfo{year}{2021}) \bibinfo{pages}{114--119}.
  \DOIprefix\doi{10.1021/acs.nanolett.0c03367}, \bibinfo{note}{pMID: 33306407}.
\bibitem[{Miao et~al.(2014)Miao, Huang, Qu, and Chien}]{MiaoPRL2014}
\bibinfo{author}{B.~F. Miao}, \bibinfo{author}{S.~Y. Huang},
  \bibinfo{author}{D.~Qu}, \bibinfo{author}{C.~L. Chien},
\newblock \bibinfo{title}{Physical origins of the new magnetoresistance in
  {P}t/{YIG}},
\newblock \bibinfo{journal}{Phys. Rev. Lett.} \bibinfo{volume}{112}
  (\bibinfo{year}{2014}) \bibinfo{pages}{236601}.
  \DOIprefix\doi{10.1103/PhysRevLett.112.236601}.
\bibitem[{Althammer et~al.(2013)Althammer, Meyer, Nakayama, Schreier,
  Altmannshofer, Weiler, Huebl, Gepr\"ags, Opel, Gross, Meier, Klewe, Kuschel,
  Schmalhorst, Reiss, Shen, Gupta, Chen, Bauer, Saitoh, and
  Goennenwein}]{AlthammerPRB2013}
\bibinfo{author}{M.~Althammer}, \bibinfo{author}{S.~Meyer},
  \bibinfo{author}{H.~Nakayama}, \bibinfo{author}{M.~Schreier},
  \bibinfo{author}{S.~Altmannshofer}, \bibinfo{author}{M.~Weiler},
  \bibinfo{author}{H.~Huebl}, \bibinfo{author}{S.~Gepr\"ags},
  \bibinfo{author}{M.~Opel}, \bibinfo{author}{R.~Gross},
  \bibinfo{author}{D.~Meier}, \bibinfo{author}{C.~Klewe},
  \bibinfo{author}{T.~Kuschel}, \bibinfo{author}{J.-M. Schmalhorst},
  \bibinfo{author}{G.~Reiss}, \bibinfo{author}{L.~Shen},
  \bibinfo{author}{A.~Gupta}, \bibinfo{author}{Y.-T. Chen},
  \bibinfo{author}{G.~Bauer}, \bibinfo{author}{E.~Saitoh},
  \bibinfo{author}{S.~Goennenwein},
\newblock \bibinfo{title}{Quantitative study of the spin {H}all
  magnetoresistance in ferromagnetic insulator/normal metal hybrids},
\newblock \bibinfo{journal}{Phys. Rev. B} \bibinfo{volume}{87}
  (\bibinfo{year}{2013}) \bibinfo{pages}{224401}.
  \DOIprefix\doi{10.1103/PhysRevB.87.224401}.
\bibitem[{Nakayama et~al.(2013)Nakayama, Althammer, Chen, Uchida, Kajiwara,
  Kikuchi, Ohtani, Gepr\"ags, Opel, Takahashi, Gross, Bauer, Goennenwein, and
  Saitoh}]{NakayamaPRL2013}
\bibinfo{author}{H.~Nakayama}, \bibinfo{author}{M.~Althammer},
  \bibinfo{author}{Y.-T. Chen}, \bibinfo{author}{K.~Uchida},
  \bibinfo{author}{Y.~Kajiwara}, \bibinfo{author}{D.~Kikuchi},
  \bibinfo{author}{T.~Ohtani}, \bibinfo{author}{S.~Gepr\"ags},
  \bibinfo{author}{M.~Opel}, \bibinfo{author}{S.~Takahashi},
  \bibinfo{author}{R.~Gross}, \bibinfo{author}{G.~E.~W. Bauer},
  \bibinfo{author}{S.~T.~B. Goennenwein}, \bibinfo{author}{E.~Saitoh},
\newblock \bibinfo{title}{Spin {H}all magnetoresistance induced by a
  nonequilibrium proximity effect},
\newblock \bibinfo{journal}{Phys. Rev. Lett.} \bibinfo{volume}{110}
  (\bibinfo{year}{2013}) \bibinfo{pages}{206601}.
  \DOIprefix\doi{10.1103/PhysRevLett.110.206601}.
\bibitem[{Liu et~al.(2020)Liu, Peiro, de~Wal, Leutenantsmeyer, Guimar\~aes, and
  van Wees}]{TianLiuPRB2020}
\bibinfo{author}{T.~Liu}, \bibinfo{author}{J.~Peiro}, \bibinfo{author}{D.~K.
  de~Wal}, \bibinfo{author}{J.~C. Leutenantsmeyer}, \bibinfo{author}{M.~H.~D.
  Guimar\~aes}, \bibinfo{author}{B.~J. van Wees},
\newblock \bibinfo{title}{Spin caloritronics in a {C}r{B}r$_{3}$-based magnetic
  van der {W}aals heterostructure},
\newblock \bibinfo{journal}{Phys. Rev. B} \bibinfo{volume}{101}
  (\bibinfo{year}{2020}) \bibinfo{pages}{205407}.
  \DOIprefix\doi{10.1103/PhysRevB.101.205407}.
\bibitem[{Sierra et~al.(2017)Sierra, Neumann, Cuppens, Raes, Costache, and
  Valenzuela}]{SierraNatNano2017}
\bibinfo{author}{J.~F. Sierra}, \bibinfo{author}{I.~Neumann},
  \bibinfo{author}{J.~Cuppens}, \bibinfo{author}{B.~Raes},
  \bibinfo{author}{M.~V. Costache}, \bibinfo{author}{S.~O. Valenzuela},
\newblock \bibinfo{title}{Thermoelectric spin voltage in graphene},
\newblock \bibinfo{journal}{Nature Nanotechnology} \bibinfo{volume}{13}
  (\bibinfo{year}{2017}) \bibinfo{pages}{107--111}.
  \DOIprefix\doi{10.1038/s41565-017-0015-9}.
\bibitem[{Vera-Marun et~al.(2016)Vera-Marun, van~den Berg, Dejene, and van
  Wees}]{VeraMarunNatComms2016}
\bibinfo{author}{I.~J. Vera-Marun}, \bibinfo{author}{J.~J. van~den Berg},
  \bibinfo{author}{F.~K. Dejene}, \bibinfo{author}{B.~J. van Wees},
\newblock \bibinfo{title}{Direct electronic measurement of {P}eltier cooling
  and heating in graphene},
\newblock \bibinfo{journal}{Nature Communications} \bibinfo{volume}{7}
  (\bibinfo{year}{2016}) \bibinfo{pages}{11525}.
  \DOIprefix\doi{10.1038/ncomms11525}.
\bibitem[{Kief and Victora(2018)}]{KiefMRS2018}
\bibinfo{author}{M.~Kief}, \bibinfo{author}{R.~Victora},
\newblock \bibinfo{title}{Materials for heat-assisted magnetic recording},
\newblock \bibinfo{journal}{MRS Bulletin} \bibinfo{volume}{43}
  (\bibinfo{year}{2018}) \bibinfo{pages}{87--92}.
  \DOIprefix\doi{10.1557/mrs.2018.2}.
\bibitem[{Weller et~al.(2014)Weller, Parker, Mosendz, Champion, Stipe, Wang,
  Klemmer, Ju, and Ajan}]{WellerIEEETransMag2014}
\bibinfo{author}{D.~Weller}, \bibinfo{author}{G.~Parker},
  \bibinfo{author}{O.~Mosendz}, \bibinfo{author}{E.~Champion},
  \bibinfo{author}{B.~Stipe}, \bibinfo{author}{X.~Wang},
  \bibinfo{author}{T.~Klemmer}, \bibinfo{author}{G.~Ju},
  \bibinfo{author}{A.~Ajan},
\newblock \bibinfo{title}{A {HAMR} media technology roadmap to an areal density
  of 4 {T}b/in},
\newblock \bibinfo{journal}{{IEEE} Trans. Magn.} \bibinfo{volume}{50}
  (\bibinfo{year}{2014}) \bibinfo{pages}{1--8}.
  \DOIprefix\doi{10.1109/tmag.2013.2281027}.
\bibitem[{Xu et~al.(2012)Xu, Liu, Ji, Toh, Hu, Li, Zhang, Ye, and
  Chia}]{XuJAP2012}
\bibinfo{author}{B.~X. Xu}, \bibinfo{author}{Z.~J. Liu},
  \bibinfo{author}{R.~Ji}, \bibinfo{author}{Y.~T. Toh}, \bibinfo{author}{J.~F.
  Hu}, \bibinfo{author}{J.~M. Li}, \bibinfo{author}{J.~Zhang},
  \bibinfo{author}{K.~D. Ye}, \bibinfo{author}{C.~W. Chia},
\newblock \bibinfo{title}{Thermal issues and their effects on heat-assisted
  magnetic recording system (invited)},
\newblock \bibinfo{journal}{Journal of Applied Physics} \bibinfo{volume}{111}
  (\bibinfo{year}{2012}) \bibinfo{pages}{07B701}.
  \DOIprefix\doi{10.1063/1.3671421}.
\bibitem[{Del~Valle et~al.(2020)Del~Valle, Salev, Kalcheim, and
  Schuller}]{DelValleSciRep2020}
\bibinfo{author}{J.~Del~Valle}, \bibinfo{author}{P.~Salev},
  \bibinfo{author}{Y.~Kalcheim}, \bibinfo{author}{I.~K. Schuller},
\newblock \bibinfo{title}{A caloritronics-based {M}ott neuristor},
\newblock \bibinfo{journal}{Scientific reports} \bibinfo{volume}{10}
  (\bibinfo{year}{2020}) \bibinfo{pages}{1--10}.
\bibitem[{Lucassen et~al.(2011)Lucassen, Wong, Duine, and
  Tserkovnyak}]{LucassenAPL2011}
\bibinfo{author}{M.~E. Lucassen}, \bibinfo{author}{C.~H. Wong},
  \bibinfo{author}{R.~A. Duine}, \bibinfo{author}{Y.~Tserkovnyak},
\newblock \bibinfo{title}{Spin-transfer mechanism for magnon-drag thermopower},
\newblock \bibinfo{journal}{Applied Physics Letters} \bibinfo{volume}{99}
  (\bibinfo{year}{2011}) \bibinfo{pages}{--}.
  \DOIprefix\doi{http://dx.doi.org/10.1063/1.3672207}.
\bibitem[{Watzman et~al.(2016)Watzman, Duine, Tserkovnyak, Boona, Jin, Prakash,
  Zheng, and Heremans}]{WatzmanPRB2016}
\bibinfo{author}{S.~J. Watzman}, \bibinfo{author}{R.~A. Duine},
  \bibinfo{author}{Y.~Tserkovnyak}, \bibinfo{author}{S.~R. Boona},
  \bibinfo{author}{H.~Jin}, \bibinfo{author}{A.~Prakash},
  \bibinfo{author}{Y.~Zheng}, \bibinfo{author}{J.~P. Heremans},
\newblock \bibinfo{title}{Magnon-drag thermopower and {N}ernst coefficient in
  {F}e, {C}o, and {N}i},
\newblock \bibinfo{journal}{Phys. Rev. B} \bibinfo{volume}{94}
  (\bibinfo{year}{2016}) \bibinfo{pages}{144407}.
  \DOIprefix\doi{10.1103/PhysRevB.94.144407}.
\bibitem[{Tserkovnyak et~al.(2016)Tserkovnyak, Bender, Duine, and
  Flebus}]{TserkovnyakPRB2016}
\bibinfo{author}{Y.~Tserkovnyak}, \bibinfo{author}{S.~A. Bender},
  \bibinfo{author}{R.~A. Duine}, \bibinfo{author}{B.~Flebus},
\newblock \bibinfo{title}{Bose-{E}instein condensation of magnons pumped by the
  bulk spin {S}eebeck effect},
\newblock \bibinfo{journal}{Phys. Rev. B} \bibinfo{volume}{93}
  (\bibinfo{year}{2016}) \bibinfo{pages}{100402}.
  \DOIprefix\doi{10.1103/PhysRevB.93.100402}.
\bibitem[{Flebus et~al.(2016)Flebus, Duine, and Tserkovnyak}]{FlebusEPL2016}
\bibinfo{author}{B.~Flebus}, \bibinfo{author}{R.~A. Duine},
  \bibinfo{author}{Y.~Tserkovnyak},
\newblock \bibinfo{title}{Landau-{L}ifshitz theory of the magnon-drag
  thermopower},
\newblock \bibinfo{journal}{{EPL} (Europhysics Letters)} \bibinfo{volume}{115}
  (\bibinfo{year}{2016}) \bibinfo{pages}{57004}.
  \DOIprefix\doi{10.1209/0295-5075/115/57004}.
\bibitem[{Natale et~al.(2021)Natale, Wesenberg, Edwards, Nembach, Shaw, and
  Zink}]{NatalePRM2021}
\bibinfo{author}{M.~R. Natale}, \bibinfo{author}{D.~J. Wesenberg},
  \bibinfo{author}{E.~R.~J. Edwards}, \bibinfo{author}{H.~T. Nembach},
  \bibinfo{author}{J.~M. Shaw}, \bibinfo{author}{B.~L. Zink},
\newblock \bibinfo{title}{Field-dependent nonelectronic contributions to
  thermal conductivity in a metallic ferromagnet with low {G}ilbert damping},
\newblock \bibinfo{journal}{Phys. Rev. Materials} \bibinfo{volume}{5}
  (\bibinfo{year}{2021}) \bibinfo{pages}{L111401}.
  \DOIprefix\doi{10.1103/PhysRevMaterials.5.L111401}.
\bibitem[{Gallo and Sebastian(2020)}]{LeGalloJPhysD2020}
\bibinfo{author}{M.~L. Gallo}, \bibinfo{author}{A.~Sebastian},
\newblock \bibinfo{title}{An overview of phase-change memory device physics},
\newblock \bibinfo{journal}{Journal of Physics D: Applied Physics}
  \bibinfo{volume}{53} (\bibinfo{year}{2020}) \bibinfo{pages}{213002}.
  \DOIprefix\doi{10.1088/1361-6463/ab7794}.
\bibitem[{Zhang et~al.(2021)Zhang, Bartell, Karsch, Gray, and
  Fuchs}]{ZhangNanoLett2021}
\bibinfo{author}{C.~Zhang}, \bibinfo{author}{J.~M. Bartell},
  \bibinfo{author}{J.~C. Karsch}, \bibinfo{author}{I.~Gray},
  \bibinfo{author}{G.~D. Fuchs},
\newblock \bibinfo{title}{Nanoscale magnetization and current imaging using
  time-resolved scanning-probe magnetothermal microscopy},
\newblock \bibinfo{journal}{Nano Letters} \bibinfo{volume}{21}
  (\bibinfo{year}{2021}) \bibinfo{pages}{4966--4972}.
  \DOIprefix\doi{10.1021/acs.nanolett.1c00704}.
\bibitem[{Althammer(2021)}]{AlthammerPSS2021}
\bibinfo{author}{M.~Althammer},
\newblock \bibinfo{title}{All-electrical magnon transport experiments in
  magnetically ordered insulators},
\newblock \bibinfo{journal}{physica status solidi (RRL)} \bibinfo{volume}{15}
  (\bibinfo{year}{2021}) \bibinfo{pages}{2100130}.
  \DOIprefix\doi{https://doi.org/10.1002/pssr.202100130}.
\bibitem[{Choi et~al.(2014)Choi, Min, Lee, and Cahill}]{ChoiNatComms2014}
\bibinfo{author}{G.-M. Choi}, \bibinfo{author}{B.-C. Min},
  \bibinfo{author}{K.-J. Lee}, \bibinfo{author}{D.~G. Cahill},
\newblock \bibinfo{title}{Spin current generated by thermally driven ultrafast
  demagnetization},
\newblock \bibinfo{journal}{Nature communications} \bibinfo{volume}{5}
  (\bibinfo{year}{2014}) \bibinfo{pages}{4334}.
\bibitem[{Ostler et~al.(2012)Ostler, Barker, Evans, Chantrell, Atxitia,
  Chubykalo-Fesenko, Moussaoui, Guyader, Mengotti, Heyderman, Nolting,
  Tsukamoto, Itoh, Afanasiev, Ivanov, Kalashnikova, Vahaplar, Mentink,
  Kirilyuk, Rasing, and Kimel}]{OstlerNatComms2012}
\bibinfo{author}{T.~Ostler}, \bibinfo{author}{J.~Barker},
  \bibinfo{author}{R.~Evans}, \bibinfo{author}{R.~Chantrell},
  \bibinfo{author}{U.~Atxitia}, \bibinfo{author}{O.~Chubykalo-Fesenko},
  \bibinfo{author}{S.~E. Moussaoui}, \bibinfo{author}{L.~L. Guyader},
  \bibinfo{author}{E.~Mengotti}, \bibinfo{author}{L.~Heyderman},
  \bibinfo{author}{F.~Nolting}, \bibinfo{author}{A.~Tsukamoto},
  \bibinfo{author}{A.~Itoh}, \bibinfo{author}{D.~Afanasiev},
  \bibinfo{author}{B.~Ivanov}, \bibinfo{author}{A.~Kalashnikova},
  \bibinfo{author}{K.~Vahaplar}, \bibinfo{author}{J.~Mentink},
  \bibinfo{author}{A.~Kirilyuk}, \bibinfo{author}{T.~Rasing},
  \bibinfo{author}{A.~Kimel},
\newblock \bibinfo{title}{Ultrafast heating as a sufficient stimulus for
  magnetization reversal in a ferrimagnet},
\newblock \bibinfo{journal}{Nature Communications} \bibinfo{volume}{3}
  (\bibinfo{year}{2012}) \bibinfo{pages}{666}.
  \DOIprefix\doi{10.1038/ncomms1666}.
\bibitem[{Kirilyuk et~al.(2010)Kirilyuk, Kimel, and Rasing}]{KirilyukRMP2010}
\bibinfo{author}{A.~Kirilyuk}, \bibinfo{author}{A.~V. Kimel},
  \bibinfo{author}{T.~Rasing},
\newblock \bibinfo{title}{Ultrafast optical manipulation of magnetic order},
\newblock \bibinfo{journal}{Rev. Mod. Phys.} \bibinfo{volume}{82}
  (\bibinfo{year}{2010}) \bibinfo{pages}{2731--2784}.
  \DOIprefix\doi{10.1103/RevModPhys.82.2731}.
\bibitem[{Kimling et~al.(2017)Kimling, Choi, Brangham, Matalla-Wagner, Huebner,
  Kuschel, Yang, and Cahill}]{KimlingPRL2017}
\bibinfo{author}{J.~Kimling}, \bibinfo{author}{G.-M. Choi},
  \bibinfo{author}{J.~T. Brangham}, \bibinfo{author}{T.~Matalla-Wagner},
  \bibinfo{author}{T.~Huebner}, \bibinfo{author}{T.~Kuschel},
  \bibinfo{author}{F.~Yang}, \bibinfo{author}{D.~G. Cahill},
\newblock \bibinfo{title}{Picosecond spin {S}eebeck effect},
\newblock \bibinfo{journal}{Phys. Rev. Lett.} \bibinfo{volume}{118}
  (\bibinfo{year}{2017}) \bibinfo{pages}{057201}.
  \DOIprefix\doi{10.1103/PhysRevLett.118.057201}.
\bibitem[{Seifert et~al.(2018)Seifert, Jaiswal, Barker, Weber, Razdolski,
  Cramer, Gueckstock, Maehrlein, Nadvornik, Watanabe, Ciccarelli, Melnikov,
  Jakob, M\"{u}nzenberg, Goennenwein, Woltersdorf, Rethfeld, Brouwer, Wolf,
  Kl\"{a}ui, and Kampfrath}]{SeifertNatComms2018}
\bibinfo{author}{T.~S. Seifert}, \bibinfo{author}{S.~Jaiswal},
  \bibinfo{author}{J.~Barker}, \bibinfo{author}{S.~T. Weber},
  \bibinfo{author}{I.~Razdolski}, \bibinfo{author}{J.~Cramer},
  \bibinfo{author}{O.~Gueckstock}, \bibinfo{author}{S.~F. Maehrlein},
  \bibinfo{author}{L.~Nadvornik}, \bibinfo{author}{S.~Watanabe},
  \bibinfo{author}{C.~Ciccarelli}, \bibinfo{author}{A.~Melnikov},
  \bibinfo{author}{G.~Jakob}, \bibinfo{author}{M.~M\"{u}nzenberg},
  \bibinfo{author}{S.~T.~B. Goennenwein}, \bibinfo{author}{G.~Woltersdorf},
  \bibinfo{author}{B.~Rethfeld}, \bibinfo{author}{P.~W. Brouwer},
  \bibinfo{author}{M.~Wolf}, \bibinfo{author}{M.~Kl\"{a}ui},
  \bibinfo{author}{T.~Kampfrath},
\newblock \bibinfo{title}{Femtosecond formation dynamics of the spin {S}eebeck
  effect revealed by terahertz spectroscopy},
\newblock \bibinfo{journal}{Nature Communications} \bibinfo{volume}{9}
  (\bibinfo{year}{2018}). \DOIprefix\doi{10.1038/s41467-018-05135-2}.

\end{thebibliography}

\end{document}